\documentclass[nofootinbib]{revtex4-1}  %% !!! arxiv, oldaltörés!

\draft % marks overfull lines with a black rule on the right

\usepackage[]{graphicx}
\graphicspath{{figs/}}
\usepackage{float}
\usepackage{amsmath,amsfonts,amssymb,nicefrac,esint,bbm}
\DeclareMathAlphabet{\mathpzc}{OT1}{pzc}{m}{it}
\usepackage{xcolor}

\usepackage{multirow}
\usepackage[bbgreekl]{mathbbol}  %% void \bbalpha ..., also has a \mathbb{},
\usepackage{hyperref}

\newcommand\nc{\newcommand}

\nc\tensorr\mathbf
\nc\Tensor\boldsymbol

\nc\wrt{w.r.t.\ }  \nc\cf{cf.\ }
\nc\lat\textit
\nc\ie{\lat{i.e.,\ }}  \nc\etal{\lat{et al.\ }}  \nc\eg{\lat{e.g.,\ }}
\nc\zero{\tensorr{0}}

\nc\re[1]{(\ref{#1})}

\nc\dd{{\rm d}}
\nc\pd{\partial}

\nc\pder[3]{\left. \frac{\pd #1}{\pd #2} \right|_{#3}}
\nc\dt[1]{\frac{\dd #1}{\dd t}}
\nc\dtt[1]{\frac{\dd^2 #1}{\dd t^2}}
\nc\pdt[1]{\frac{\pd #1}{\pd t}}
\nc\pdct[1]{\frac{\pd #1}{\pd \check{t}}}
\nc\pdcta[1]{\frac{\pd #1}{\pd \check{t}_a}}
\nc\pdctd[1]{\frac{\pd #1}{\pd \check{t}_d}}
\nc\pdx[1]{\frac{\pd #1}{\pd x}}
\nc\pdxi[1]{\frac{\pd #1}{\pd \xi}}
\nc\pdz[1]{\frac{\pd #1}{\pd \zeta}}
\nc\pdr[1]{\frac{\pd #1}{\pd r}}
\nc\ppd[2]{\frac{\pd^2 #1}{\pd #2^2}}

\nc\ee{\mathrm{e}}

\nc\qrho{\varrho}

\DeclareMathOperator{\erf}{erf}
\DeclareMathOperator{\erfc}{erfc}
\DeclareMathOperator{\ierfc}{ierfc}
\DeclareMathOperator{\erfcx}{erfcx}

\nc\htheta{\hat{\vartheta}}
\nc\hTheta{\hat{\Theta}}

\nc\ct{\check{t}}
\nc\cx{\check{x}}
\nc\cz{\check{\zeta}}
\nc\cTheta{\check{\Theta}}

\nc\qB{\mathcal{B}}

\nc\qqr{\tensorr{r}}
\nc\qqv{\tensorr{v}}
\nc\qas{\mathsf{a}_s}

\DeclareMathAlphabet{\mathpzc}{OT1}{pzc}{m}{it}
\nc\qPe{\mathpzc{Pe}}

\begin{document}

\title{Exact analytical solutions for the piston effect in supercritical fluids under post-acoustic approximation -- Short-time asymptotics, thermal penetration depth and comparison with the Spacelab D-2 experiments}
\author{Mátyás Szücs}
\email[]{szucs.matyas@gpk.bme.hu}
\affiliation{Department of Energy Engineering, Faculty of Mechanical Engineering, Budapest University of Technology and Economics, Műegyetem rkp.\ 3., H-1111 Budapest, Hungary}
\affiliation{Department of Theoretical Physics, Institute for Particle and Nuclear Physics, HUN-REN Wigner Research Centre for Physics, Konkoly-Thege Miklós út 29--33., H-1121 Budapest, Hungary}
\affiliation{Montavid Thermodynamic Research Group, Society for the Unity of Science and Technology, Lovas út 18., H-1012 Budapest, Hungary}

\date{\today}

\begin{abstract}
    Near the liquid-vapor critical point, fluids become highly compressible, giving rise to a special, strongly coupled thermo-mechanical process: the piston effect. In this phenomenon, a thin thermal boundary layer develops near a heated wall; owing to strong thermal expansion, this layer acts like a piston, compressing the bulk fluid adiabatically and resulting in a seemingly accelerated thermal response. Although the piston effect is a thermoacoustic process, the characteristic time scale of the boundary perturbation is typically orders of magnitude larger than the acoustic time scale of the setup. Consequently, rapid acoustic propagation can be neglected, justifying a post-acoustic approximation with a spatially uniform but time-dependent bulk pressure. Within the linear regime, the temporal evolution of pressure can be directly connected to the heat flux entering through the boundaries. As a result, the problem reduces to a diffusion equation governed by a spatially homogeneous source term that depends explicitly on the boundary conditions. Exact, closed-form analytical solutions are derived for effectively one-dimensional problems in both Cartesian and spherical coordinates, considering boundary conditions of the first and second kinds. Short-time asymptotic behavior and thermal penetration depth are analyzed for all four cases. By incorporating the heat capacity of a container via a homogeneous model, an effective boundary condition coupling the wall heat flux and the time derivative of the wall temperature is derived, allowing for a direct comparison with experimental data from the Spacelab D-2 mission. The analytical predictions show good agreement with the experimental results without relying on any numerical simulations.  
\end{abstract}

\maketitle %\maketitle must follow title, authors, abstract and \pacs

\section{Introduction}

The thermodynamic distinction between the liquid and vapor phases vanishes at the critical point, where the coexistence curve terminates and the latent heat of vaporization becomes zero. Above the critical point, the fluid exhibits both liquid-like and gas-like properties at the same time, combining liquid-magnitude densities with gas-like, extremely high compressibility and thermal expansion \cite{carles2010brief,imre2019anomalous}, the latter two being strongly correlated through the Imre ellipse \cite{takacs2024leading}. The emerging high compressibility---supplemented by the strong state dependence of thermophysical properties---gives rise to strongly coupled thermo-mechanical phenomena \cite{carles2006thermoacoustic,zappoli2003nearcritical,zappoli2015heat,hasan2012thermoacoustic}.

The efficiency of thermodynamic cycles operating above the critical point drastically increases \cite{daniarta2022thermodynamic}, a feature that is increasingly exploited by next-generation energy technologies, such as Supercritical Water-Cooled Reactors \cite{rahman2020design,wu2022review} and Enhanced Geothermal Systems \cite{dobson2017supercritical,reinsch2017utilizing}. By applying supercritical fluid technologies, the risk of a conventional, subcritical boiling crisis \cite{theofanous2002boiling1,theofanous2002boiling2} is entirely eliminated. However, this engineering advantage introduces a further thermal challenge: the phenomenon of heat transfer deterioration \cite{longmire2022onset}, where a localized, sudden drop in the heat transfer coefficient leads to hazardous wall-temperature spikes that threaten structural integrity. Ensuring safe system operation in these applications fundamentally requires a deep understanding of the underlying strongly coupled thermo-mechanical interactions.

To clearly understand how different physical parameters affect these processes, exact analytical solutions provide the most powerful tool, offering direct physical insights that are often obscured in numerical formulations. However, the derivation of exact analytical solutions is severely limited, particularly for partial differential equations governing complex thermal systems. In these cases, the mathematical constraints imposed by both the boundary conditions and the non-trivial source terms often make closed-form solutions impossible to obtain. Consequently, explicit analytical expressions are rarely found in the literature for highly coupled processes.

The anomalous heat conduction behavior observed in the vicinity of the liquid–vapor critical point can be explained via a specific thermo-mechanical coupling \cite{onuki1990fast,boukari1990critical,zappoli1990anomalous}. Near the critical point, thermal diffusivity tends to zero, meaning that thermal equilibration is expected to slow down---a phenomenon referred to in the literature as ``critical slowing down.'' In contrast, microgravity experiments, where buoyancy-driven convection is eliminated, have demonstrated unexpectedly fast temperature equilibration, termed ``critical speeding up'' \cite{garrabos1998relaxation,straub1995process}. This contradiction is resolved by thermal expansion. Within a closed tank filled with a supercritical fluid, a heated wall induces a thermal boundary layer that expands intensively due to the large value of the thermal expansion coefficient. This expanding boundary layer compresses the bulk fluid like a piston, thereby causing an adiabatic-like temperature increase throughout the entire volume. This phenomenon, known as the piston effect, was originally described by Onuki \etal \cite{onuki1990fast} through the mechanism of initial isentropic acoustic-wave propagation, and its characteristic time-scale structure has been discussed in detail in subsequent studies \cite{carles2005two}. The associated thermoacoustic-wave dynamics, including acoustic emission, reflection, and resonance near the critical point, have also been investigated in subsequent thermoacoustic formulations \cite{onuki2007thermoacoustic,shen2011thermoacoustic}.

Regarding the piston effect---a classical benchmark problem that has been investigated both theoretically and experimentally---analytical solutions have been reported in the literature \cite{onuki1990fast,straub1995process}. However, most classical analytical treatments are commonly restricted to idealized constant-temperature boundary conditions and one-dimensional Cartesian geometries. Moreover, these solutions are approximate in the sense that the rising bulk temperature is typically incorporated as a parametric correction in the thermal boundary layer rather than being obtained as a fully coupled state variable of the transient problem \cite{onuki1990fast}. As a consequence, they do not provide an explicit closed-form description of the mutual evolution of the bulk temperature and the transient thermal boundary layer throughout the entire process. To the best of our knowledge, a fully coupled exact closed-form solution accounting for such thermal interactions over the complete transient evolution has not yet been reported for the configuration considered here.

To bridge this gap, the objective of this paper is to derive exact, closed-form analytical solutions for the post-acoustic approximation of the piston effect for effectively one-dimensional problems. The main scientific contributions of the paper can be summarized as follows:
\begin{itemize}
    \item The post-acoustic approximation is derived systematically from the linearized thermo-hydrodynamic balance equations by means of a two-time-scale asymptotic expansion, thereby providing a hydrodynamic justification for the thermal equilibration model proposed by Boukari \etal \cite{boukari1990critical}.
    \item The post-acoustic piston-effect model is reformulated as a boundary-coupled diffusion equation driven by a spatially homogeneous source term originating directly from the boundary conditions, establishing the equivalence between a local diffusion equation and the integro-differential formulation commonly used in the literature.
    \item Exact closed-form analytical solutions are derived for four fundamental configurations, namely Cartesian and spherical geometries subject to Dirichlet (prescribed wall temperature) and Neumann (prescribed wall heat flux) boundary conditions.
    \item Short-time asymptotic approximations and thermal penetration-depth expressions for all configurations are derived and analyzed, providing direct physical insight into the interplay between boundary-layer diffusion and bulk heating by the piston effect.
    \item The boundary-coupled diffusion equation is solved analytically for a dynamic conjugate boundary condition accounting for the thermal inertia of the container wall. Accordingly, the resulting solution shows good agreement with the available Spacelab D-2 microgravity experimental data without relying on numerical simulations, while also providing physical insight into the roles of wall thermal inertia, interfacial thermal coupling, and critical-point proximity in the observed transient response \cite{straub1995dynamic}.
\end{itemize}
The remainder of the paper is organized as follows. Section~\ref{sec:post-ac} presents the derivation of the post-acoustic approximation. Exact analytical solutions are derived in Sec.~\ref{sec:an-sol}, while their asymptotic behavior and thermal penetration depths are investigated in Sec.~\ref{sec:asy}. The dynamic conjugate boundary condition and the corresponding analysis are presented in Sec.~\ref{sec:spacelab}.

\section{The post-acoustic approximation of the piston effect} \label{sec:post-ac}

The coupled thermo-mechanical process describing the piston effect can be formulated via the balance equations of mass, linear momentum, and internal energy for a heat-conducting inviscid fluid\footnote{Internal friction of the fluid, characterized by its shear and bulk viscosities, also plays an important role in transport processes near the critical point \cite{carles1998effect}; however, for the purposes of this study, these do not make a significant contribution.}. When the volumetric force density is omitted under microgravity conditions, these conservation laws are expressed as
\begin{align}
    \label{eq:bal-m}
    \mathcal{D}_\qqv \qrho + \qrho \nabla \cdot \qqv &= 0 , \\
    \label{eq:bal-v}
    \qrho \mathcal{D}_\qqv \qqv &= - \nabla p , \\
    \label{eq:bal-e}
    \qrho \mathcal{D}_\qqv e &= - \nabla \cdot \dot{\tensorr{q}} - p \nabla \cdot \qqv ,
\end{align}
where $ \qrho $, $ \qqv $, $ p $, $ e $ and $ \dot{\tensorr{q}} $ denote the (mass) density, velocity, pressure, (mass) specific internal energy, and heat current density fields, respectively \cite{degroot1962nonequilibrium,gyarmati1970nonequilibrium}. All these physical quantities are functions of the time $ t $ and spatial coordinates $ \qqr $. The material time derivative, illustrated here on the density field as
\begin{align}
    \mathcal{D}_\qqv \qrho = \pdt{\qrho} + \qqv \cdot \nabla \qrho ,
\end{align}
characterizes the rate of change of a physical property following a specific material particle moving with velocity $ \qqv $. Throughout this study, $ \pdt{} $ denotes the partial time derivative, while $ \nabla $ is the nabla operator representing the gradient or the divergence. The heat current density is given by Fourier's law
\begin{align}
    \dot{\tensorr{q}} = - \lambda \nabla T ,
\end{align}
where $ T $ represents the temperature and $ \lambda $ is the thermal conductivity. The thermodynamic state variables are coupled through the thermal and caloric equations of state
\begin{align}
    p &= p ( T , \qrho ) , & 
    e &= e ( T , \qrho ) .
\end{align}
In a single-phase, single-component fluid, all specific extensive and intensive thermodynamic state variables can be expressed in terms of any two independent state variables. Assuming the required invertibility between the selected state variables, the thermal and caloric equations of state are frequently expressed in differential form as
\begin{align}
    \label{eq:th-eos}
    \frac{1}{\qrho} \dd \qrho &= - \beta_p \dd T + \chi_T^{} \dd p , \\
    \label{eq:cal-eos}
    \dd e &= c_v \dd T + \pder{e}{\qrho}{T} \dd \qrho,
\end{align}
with
\begin{align}
    \label{eq:bp-def}
    &\bullet \ \text{the volumetric isobaric thermal expansion coefficient}
    & \beta_p &= - \frac{1}{\qrho} \pder{\qrho}{T}{p} , \\
    \label{eq:kT-def}
    & \bullet \ \text{the isothermal compressibility}
    & \chi_T^{} &= \frac{1}{\qrho} \pder{\qrho}{p}{T} > 0 , \\
    \label{eq:cv-def}
    & \bullet \ \text{and the isochoric specific heat capacity}
    & c_v &= \pder{e}{T}{\qrho} > 0 .
\end{align}
The positivity of the latter two coefficients reflects the thermodynamic material stability conditions \cite{grigull1964prinzip,matolcsi2004ordinary}.

Although the thermal and caloric equations of state may appear to be independent, the existence of the specific entropy $ s $ imposes a thermodynamic consistency condition between them, which is expressed by the Gibbs relation
\begin{align}
    \label{eq:Gibbs}
    \dd e = T \dd s - p \dd v = T \dd s + \frac{p}{\qrho^2} \dd \qrho ,
\end{align}
where $ v = \frac{1}{\qrho} $ denotes the specific volume. Consequently, $ \dd v = - \frac{1}{\qrho^2} \dd \qrho $, and differentiation with respect to the specific volume can be expressed as $ \frac{\pd}{\pd v} = - \qrho^2 \frac{\pd}{\pd \qrho} $. According to \re{eq:Gibbs}, the temperature and pressure are obtained as the partial derivatives
\begin{align}
    \label{eq:T,p-conj}
    T ( s , v ) &= \pder{e}{s}{v} , &
    p ( s , v ) &= - \pder{e}{v}{s} .
\end{align}
To determine the partial derivative of the specific internal energy with respect to the specific volume in an isothermal process, we consider the composite function $ e ( T , v ) = e \big( s ( T , v ) , v \big) $, yielding
\begin{align}
    \label{eq:de_v_T}
    \pder{e}{v}{T} = \pder{e}{s}{v} \pder{s}{v}{T} + \pder{e}{v}{s} \overset{\re{eq:T,p-conj}}{=} T \pder{s}{v}{T} - p .
\end{align}
Introducing the specific Helmholtz free energy $ \mathsf{f} = e - T s $, the Gibbs relation \re{eq:Gibbs} becomes
\begin{align}
    \dd \mathsf{f} = - s \dd T - p \dd v ,
\end{align}
and therefore
\begin{align}
    s ( T , v ) &= - \pder{\mathsf{f}}{T}{v} , &
    p ( T , v ) &= - \pder{\mathsf{f}}{v}{T} .
\end{align}
Provided that the Helmholtz free energy is twice continuously differentiable, the equality of its mixed second derivatives yields the Maxwell relation
\begin{align}
    \pder{s}{v}{T} = \pder{p}{T}{v} .
\end{align}
Substitution of this Maxwell relation into \re{eq:de_v_T} gives
\begin{align}
    \pder{e}{v}{T} = T \pder{p}{T}{v} - p = - T \frac{\pder{v}{T}{p}}{\pder{v}{p}{T}} - p,
\end{align}
where we have used the cyclic identity for partial derivatives, $ -1 = \pder{p}{T}{v} \pder{T}{v}{p} \pder{v}{p}{T} $, together with the inverse-function relation. From the definitions \re{eq:bp-def} and \re{eq:kT-def}, it follows that
\begin{align}
    \pder{v}{T}{p} = \pder{\frac{1}{\qrho}}{T}{p} = - \frac{1}{\qrho^2} \pder{\qrho}{T}{p} = v \beta_p
    \qquad \text{and} \qquad
    \pder{v}{p}{T} = - v \chi_T^{} .
\end{align}
Consequently, $ \pder{e}{v}{T} = T \frac{\beta_p}{\chi_T^{}} - p $, and the differential form of the caloric equation of state is
\begin{align}
    \label{eq:cal-eos-1}
    \dd e &= c_v \dd T - \frac{1}{\qrho^2} \left( T \frac{\beta_p}{\chi_T^{}} - p \right) \dd \qrho .
\end{align}

A further consequence of thermodynamic consistency is that only three material properties are mutually independent, allowing all other thermodynamic quantities to be calculated from them. In this study, we utilize
\begin{align}
    \label{eq:cp-def}
    &\bullet \ \text{the isobaric specific heat capacity}
    & c_p &= c_v + \frac{T}{\qrho} \frac{\beta_p^2}{\chi_T^{}} , \\
    \label{eq:gam-def}
    & \bullet \ \text{the heat capacity ratio}
    & \gamma &= \frac{c_p}{c_v} = 1 + \frac{T}{\qrho} \frac{\beta_p^2}{c_v \chi_T^{}} , \\
    \label{eq:as-def}
    & \bullet \ \text{and the isentropic speed of sound}
    & \qas &= \sqrt{\pder{p}{\qrho}{s}} = \sqrt{ \frac{\gamma}{\qrho \chi_T^{}} } .
\end{align}
Based on this definition of the speed of sound, the heat capacity ratio can be equivalently rewritten as 
\begin{align}
    \gamma = 1 + T \frac{\beta_p^2 \qas^2}{c_p}.
\end{align}
By substituting \re{eq:cal-eos-1} and \re{eq:as-def} into \re{eq:bal-e} and using the mass continuity equation \re{eq:bal-m}, the balance equation of internal energy \re{eq:bal-e} can be reformulated in terms of the temperature as
\begin{align}
    \label{eq:heat-eq}
    \qrho c_v \mathcal{D}_\qqv T &= \nabla \cdot \left( \lambda \nabla T \right) - T \qrho \frac{\beta_p \qas^2}{\gamma} \nabla \cdot \qqv .
\end{align}

Let us now focus on small perturbations around a homogeneous static equilibrium state characterized by its temperature, density, and velocity $ \left( T_0 , \qrho_0 , \ \text{and} \ \qqv_0 \equiv \zero \right) $. Under these conditions, the system of governing equations \re{eq:bal-m}, \re{eq:bal-v} and \re{eq:heat-eq} can be linearized, which read as
\begin{align}
    \label{eq:bal-m-lin}
    \pdt{P} &= - \qrho_0 \nabla \cdot \Tensor{\Upsilon} , \\
    \label{eq:bal-v-lin}
    \pdt{\Tensor{\Upsilon}} &= - \frac{1}{\qrho_0} \nabla \Pi , \\
    \label{eq:heat-eq-lin}
    \pdt{\Theta} &= \gamma a \nabla^2 \Theta - T_0 \frac{\beta_p \qas^2}{c_p} \nabla \cdot \Tensor{\Upsilon} ,
\end{align}
where $ P = \qrho - \qrho_0 $, $ \Tensor{\Upsilon} = \qqv - \qqv_0 = \qqv $, $ \Theta = T - T_0 $ and $ \Pi = p - p_0 $ are the deviations of the density, velocity, temperature, and pressure fields measured from their respective equilibrium values, furthermore, $ a = \frac{\lambda}{\qrho c_p} $ denotes the thermal diffusivity, and $ \nabla^2 $ is the Laplace operator. Note that $ p_0 = p ( T_0 , \qrho_0 ) $, and all thermophysical parameters are evaluated in the equilibrium state; however, for better transparency of the formulas, these are not marked separately. The linearized thermal equation of state can be derived directly from \re{eq:th-eos}, since the pressure differential can be expressed as 
\begin{align}
    \dd p = \frac{\beta_p}{\chi_T^{}} \dd T + \frac{1}{\qrho \chi_T^{}} \dd \qrho \overset{\re{eq:as-def}}{=} \qrho \frac{\beta_p \qas^2}{\gamma} \dd T + \frac{\qas^2}{\gamma} \dd \qrho .
\end{align}
In accordance with the linearization procedure, the resulting constant-coefficient relation is integrated from the equilibrium state $ ( T_0 , \qrho_0 , p_0 ) $ to the perturbed state $ ( T , \qrho , p ) $, yielding the linearized thermal equation of state
\begin{align}
    \label{eq:eos}
    \Pi ( \Theta , P ) = \qrho_0 \frac{\beta_p \qas^2}{\gamma} \Theta + \frac{\qas^2}{\gamma} P .
\end{align}

\subsection{Asymptotic separation of acoustic and thermal phenomena}

Asymptotic descriptions of near-critical thermoacoustic and boundary-layer processes have been developed in various previous studies \cite{chen2022asymptotic,zhang2023comparison}. In particular, such analyses have identified the coexistence of wave-like behavior in the bulk and diffusion-dominated dynamics in the boundary layer. However, the present objective is different: here, the multiple-scale analysis \cite{bender1999advanced} is applied to systematically resolve the coupled thermoacoustic phenomena. As will be demonstrated, the reciprocal of the Péclet number corresponds exactly to the ratio of the acoustic and thermal diffusion time scales, which serves as the small scale-separation parameter. By combining this two-scale asymptotic method with a systematic averaging over the fast acoustic processes, the post-acoustic approximation of the piston effect is obtained, which proves to be equivalent to the fundamental model proposed by Boukari \etal \cite{boukari1990critical}. While their formulation relies on the heuristic neglect of fluid flow, the present derivation provides a multi-scale hydrodynamic validation.

First, we introduce the dimensionless space and time coordinates as
\begin{align}
    \check{\qqr} &= \frac{\qqr}{\ell} , &
    \ct &= \frac{t}{\nicefrac{\ell}{\qas}} ,
\end{align}
where $\ell$ denotes a macroscopic characteristic length scale (for instance, typically a geometric dimension of an experimental container). Although this scale can be chosen with some flexibility, its optimal definition can be suggested by the appearing coefficients, as will be fixed in \re{eq:char-length}. The dimensionless time is defined via the acoustic time scale $ \tau_{\rm a} = \frac{\ell}{\qas} $, which characterizes the travel time of a perturbation propagating at the speed of sound. Correspondingly, the derivatives are given by
\begin{align}
    \nabla &= \frac{1}{\ell} \check \nabla , &
    \pdt{} &= \frac{\qas}{\ell} \pdct{} ,
\end{align}
and the linearized governing equations \re{eq:bal-m-lin}, \re{eq:bal-v-lin}, and \re{eq:heat-eq-lin} can be formulated in terms of the dimensionless variables as
\begin{align}
    \label{eq:bal-m-lin-dimless}
    \pdct{P} &= - \frac{\qrho_0}{\qas} \check \nabla \cdot \Tensor{\Upsilon} , \\
    \label{eq:bal-v-lin-dimless}
    \pdct{\Tensor{\Upsilon}} &= - \frac{1}{\qrho_0 \qas} \check \nabla \Pi , \\
    \label{eq:heat-eq-lin-dimless-0}
    \pdct{\Theta} &= \gamma \frac{a}{\qas \ell} \check \nabla^2 \Theta - T_0 \frac{\beta_p \qas}{c_p} \check \nabla \cdot \Tensor{\Upsilon} .
\end{align}
The (acoustic) Péclet number $ \qPe = \frac{\qas \ell}{a} $ naturally emerges in the coefficient of the diffusion term in \re{eq:heat-eq-lin-dimless-0}, and thus \re{eq:heat-eq-lin-dimless-0} can be equivalently formulated as
\begin{align}
    \label{eq:heat-eq-lin-dimless}
    \pdct{\Theta} &= \frac{\gamma}{\qPe} \check \nabla^2 \Theta - T_0 \frac{\beta_p \qas}{c_p} \check \nabla \cdot \Tensor{\Upsilon} .
\end{align}
The Péclet number expresses the ratio of the thermal diffusion time scale $ \tau_{\rm d} = \frac{\ell^2}{a}$ to the acoustic time scale, \ie
\begin{align}
    \qPe = \frac{\ell}{\nicefrac{a}{\qas}} = \frac{\nicefrac{\ell^2}{a}}{\nicefrac{\ell}{\qas}} = \frac{\tau_{\rm d}}{\tau_{\rm a}} .
\end{align}
This definition inherently introduces a material-specific characteristic length scale, $ \ell_{\rm mat} = \frac{a}{\qas} $. Near the liquid–vapor critical point, this length scale remains exceptionally small, \eg for carbon dioxide at the order of $ 10^{-9} $~m \cite{toth2025initial}. In any realistic experiment or industrial process, the condition $ \ell_{\rm mat} \ll \ell $ strictly holds, ensuring that $ \tau_{\rm a} \ll \tau_{\rm d} $. Consequently, the reciprocal of the Péclet number, $ \frac{1}{\qPe} \ll 1 $, emerges as a natural small parameter, providing a clear physical basis for analyzing the rescaled governing equations with a two-time-scale asymptotic expansion.

To separate the simultaneously occurring fast acoustic and slow diffusion phenomena, an artificial slow time scale $ \ct_{\rm d} = \frac{1}{\qPe} \ct $ is introduced. Therefore, the time-dependent fields are extended into an artificial, higher-dimensional variable space and are considered as formal asymptotic power series in terms of the small parameter $ \frac{1}{\qPe} $ up to the first order, which, illustrated on the field $ P $, reads as
\begin{align}
    P ( \ct , \check{\qqr} ) &= \tilde{P}^{(0)} ( \ct , \ct_{\rm d} , \check{\qqr} ) + \frac{1}{\qPe} \tilde{P}^{(1)} ( \ct , \ct_{\rm d} , \check{\qqr} ) + \mathcal{O} \left( \qPe^{-2} \right) . %, \\
    %\Tensor{\Upsilon} ( \ct , \check{\qqr} ) &= \tilde{\Tensor{\Upsilon}}^{(0)} ( \ct , \ct_{\rm d} , \check{\qqr} ) + \frac{1}{\qPe} \tilde{\Tensor{\Upsilon}}^{(1)} ( \ct , \ct_{\rm d} , \check{\qqr} ) + \mathcal{O} \left( \qPe^{-2} \right) , \\
    %\Theta ( \ct , \check{\qqr} ) &= \tilde{\Theta}^{(0)} ( \ct , \ct_{\rm d} , \check{\qqr} ) + \frac{1}{\qPe} \tilde{\Theta}^{(1)} ( \ct , \ct_{\rm d} , \check{\qqr} ) + \mathcal{O} \left( \qPe^{-2} \right) , \\
    %\Pi ( \ct , \check{\qqr} ) &= \tilde{\Pi}^{(0)} ( \ct , \ct_{\rm d} , \check{\qqr} ) + \frac{1}{\qPe} \tilde{\Pi}^{(1)} ( \ct , \ct_{\rm d} , \check{\qqr} ) + \mathcal{O} \left( \qPe^{-2} \right) .
\end{align}
Correspondingly, since the extended fields are composite functions, for instance, $ \tilde{P}^{(0)} ( \ct , \ct_{\rm d} , \check{\qqr} ) = \tilde{P}^{(0)} ( \ct , \qPe^{-1} \ct , \check{\qqr} ) $, the dimensionless partial time derivative, illustrated also on the field $P$, can be formulated via the chain rule as
\begin{align}
    \pdct{P} = \left( \pdct{\tilde{P}^{(0)}} + \pdctd{\tilde{P}^{(0)}} \frac{\dd \ct_{\rm d}}{\dd \ct} \right) + \frac{1}{\qPe} \left( \pdct{\tilde{P}^{(1)}} + \pdctd{\tilde{P}^{(1)}} \frac{\dd \ct_{\rm d}}{\dd \ct} \right) = \pdct{\tilde{P}^{(0)}} + \frac{1}{\qPe} \left( \pdctd{\tilde{P}^{(0)}} + \pdct{\tilde{P}^{(1)}} \right) + \mathcal{O} \left( \qPe^{-2} \right) .
\end{align}

Substituting these asymptotic representations into the dimensionless balance equations \re{eq:bal-m-lin-dimless}--\re{eq:heat-eq-lin-dimless} and collecting the terms proportional to the zeroth and first power of the small parameter $\frac{1}{\qPe}$ reveals the complex thermoacoustic interactions in a decoupled form, as formulated by
\begin{align}
    \label{eq:Pe0-1}
    \qPe^0 &: &
    \pdct{\tilde P^{(0)}} &= - \frac{\qrho_0}{\qas} \check \nabla \cdot \tilde{\Tensor{\Upsilon}}^{(0)} , &
    \\
    \label{eq:Pe0-2}
    &&
    \pdct{\tilde{\Tensor{\Upsilon}}^{(0)}} &= - \frac{1}{\qrho_0 \qas} \check \nabla \tilde \Pi^{(0)} , \\
    \label{eq:Pe0-3}
    &&
    \pdct{\tilde \Theta^{(0)}} &= - T_0 \frac{\beta_p \qas}{c_p} \check \nabla \cdot \tilde{\Tensor{\Upsilon}}^{(0)} , \\
    \label{eq:Pe1-1}
    \qPe^{-1} &: &
    \pdctd{\tilde P^{(0)}} + \pdct{\tilde P^{(1)}} &= - \frac{\qrho_0}{\qas} \check \nabla \cdot \tilde{\Tensor{\Upsilon}}^{(1)} , \\
    \label{eq:Pe1-2}
    &&
    \pdctd{\tilde{\Tensor{\Upsilon}}^{(0)}} + \pdct{\tilde{\Tensor{\Upsilon}}^{(1)}} &= - \frac{1}{\qrho_0 \qas} \check \nabla \tilde \Pi^{(1)} , \\
    \label{eq:Pe1-3}
    &&
    \pdctd{\tilde \Theta^{(0)}} + \pdct{\tilde \Theta^{(1)}} &= \gamma \check \nabla^2 \tilde \Theta^{(0)} - T_0 \frac{\beta_p \qas}{c_p} \check \nabla \cdot \tilde{\Tensor{\Upsilon}}^{(1)} .
\end{align}
The leading-order dynamics ($ \qPe^0 $) characterizes isentropic acoustic wave propagation, where the localized pressure gradients drive high-frequency oscillations across the fluid domain before any significant thermal diffusion can take place. The first-order correction ($ \qPe^{-1} $) captures the formal coupling between the rapid acoustic oscillations and the long-term thermal transients. On the left-hand side of these equations, the mixed time derivatives directly illustrate how the leading-order fields evolve under the slow time scale $ \ct_{\rm d} $. Note that thermal conduction manifests itself explicitly only at this perturbation order.

Typically, in experiments, the characteristic time scale of the thermal boundary perturbation is orders of magnitude longer than the acoustic time scale but remains only a small fraction of the diffusion time scale. Therefore, an intermediate asymptotic regime is established, implying that a detailed knowledge of the short-time acoustic transients is not necessary to determine the long-term thermal response. Assuming that all physical fields undergo bounded deviations, the rapid time dependence can be systematically averaged out. Let us introduce an intermediate time scale $ \tau_{\rm int} $, which satisfies the separation condition $ \tau_{\rm a} \ll \tau_{\rm int} \ll \tau_{\rm d} $. Note that this intermediate time scale is not necessarily the time scale of the boundary perturbation. The fast-time average of any field, illustrated here for $ \tilde P^{(0)} ( \ct , \ct_{\rm d} , \check{\qqr} ) $, is denoted and defined as
\begin{align}
    \left\langle \tilde P^{(0)} ( \ct , \ct_{\rm d} , \check{\qqr} ) \right\rangle = \frac{1}{\nicefrac{\tau_{\rm int}}{\tau_{\rm a}}} \int_0^{\nicefrac{\tau_{\rm int}}{\tau_{\rm a}}} \tilde P^{(0)} ( \ct , \ct_{\rm d} , \check{\qqr} ) \dd \ct .
\end{align}
Over this intermediate time window, the slow-scale variations remain virtually unchanged, yielding the averaged behavior (\eg at $\check{t}_d = 0$)
\begin{align}
    \left\langle \pdctd{\tilde P^{(0)}} \right\rangle = \frac{1}{\nicefrac{\tau_{\rm int}}{\tau_{\rm a}}} \int_0^{\nicefrac{\tau_{\rm int}}{\tau_{\rm a}}} \pdctd{\tilde P^{(0)}} \dd \ct \approx \pdctd{\tilde P^{(0)}} ,
\end{align}
whereas the fast-scale oscillations average out over this time window, resulting in
\begin{align}
    \left\langle \pdct{\tilde P^{(0)}} \right\rangle = \frac{1}{\nicefrac{\tau_{\rm int}}{\tau_{\rm a}}} \int_0^{\nicefrac{\tau_{\rm int}}{\tau_{\rm a}}} \pdct{\tilde P^{(0)}} \dd \ct = \frac{\tilde P^{(0)} \left( \nicefrac{\tau_{\rm int}}{\tau_{\rm a}} \right) - \tilde P^{(0)}(0)}{\nicefrac{\tau_{\rm int}}{\tau_{\rm a}}} \approx 0 ,
\end{align}
and similarly for the other fields. In this asymptotic limit, the continuous expansion at the boundary effectively suppresses the acoustic wave reflections, smoothing out the spatial pressure fluctuations, just as previously demonstrated numerically in \cite{takacs2025piston}. Correspondingly, the fast-scale time derivatives vanish, meaning that the leading-order equations reduce to a state where spatial pressure gradients become negligible, and \re{eq:Pe0-2} reduces to
\begin{align}
    \label{eq:piston-hom-p}
    \check \nabla \tilde \Pi^{(0)} = 0 .
\end{align}
Over the slow time scale, \re{eq:Pe0-1} reduces to $ \check \nabla \cdot \tilde{\Tensor{\Upsilon}}^{(0)} = 0 $, characterizing the emergence of an effective incompressibility. Accordingly, \re{eq:Pe0-3} serves as a compatibility condition. The first-order correction system \re{eq:Pe1-1}--\re{eq:Pe1-3} then simplifies to
\begin{align}
    \label{eq:piston-1}
    \pdctd{\tilde P^{(0)}} &= - \frac{\qrho_0}{\qas} \check \nabla \cdot \tilde{\Tensor{\Upsilon}}^{(1)} , \\
    \label{eq:piston-2}
    \pdctd{\tilde{\Tensor{\Upsilon}}^{(0)}} &= - \frac{1}{\qrho_0 \qas} \check \nabla \tilde \Pi^{(1)} , \\
    \label{eq:piston-3}
    \pdctd{\tilde \Theta^{(0)}} &= \gamma \check \nabla^2 \tilde \Theta^{(0)} - T_0 \frac{\beta_p \qas}{c_p} \check \nabla \cdot \tilde{\Tensor{\Upsilon}}^{(1)} \stackrel{\re{eq:piston-1}}{=} \gamma \check \nabla^2 \tilde \Theta^{(0)} + \frac{T_0}{\qrho_0} \frac{\beta_p \qas^2}{c_p} \pdctd{\tilde P^{(0)}} .
\end{align}
Utilizing the linearized thermal equation of state \re{eq:eos}, the resulting slow time scale heat conduction equation reads as
\begin{align}
    \label{eq:piston-1-fast-bou}
    \left( 1 + T_0 \frac{\beta_p^2 \qas^2}{c_p} \right) \pdctd{\tilde \Theta^{(0)}} = \gamma \check \nabla^2 \tilde \Theta^{(0)} + \gamma T_0 \frac{\beta_p}{\qrho_0 c_p} \frac{\dd \tilde \Pi^{(0)}}{\dd \ct_{\rm d}} .
\end{align}
Recognizing that the dimensionless coefficient on the left-hand side is the specific heat ratio $ \gamma $ in leading order, this factor cancels out from all terms. By utilizing the fact that the leading-order pressure deviation is spatially homogeneous [cf.\ \re{eq:piston-hom-p}], the pressure field loses its spatial dependence and becomes a function of time alone. Consequently, its time derivative reduces to an ordinary derivative, completing the post-acoustic reduction of the thermodynamic problem. This directly reveals that within a closed container, the local thermal expansion near the heated wall cannot escape, thereby inducing a global bulk pressure rise. 

An examination of the first-order momentum equation \re{eq:piston-2} reveals an important feature of the post-acoustic coupling. At the slow time scale, the acceleration of the leading-order velocity field $ \tilde{\Tensor{\Upsilon}}^{(0)} $ is driven exclusively by the gradient of the first-order pressure correction $ \tilde \Pi^{(1)} $. This relation demonstrates that macroscopic fluid motion during the thermal transient is strictly a weak secondary effect. This clear separation of driving forces provides a justification for the post-acoustic approximation, as the fluid mechanics are governed by minor pressure fluctuations while the thermodynamic state is dominated by a spatially uniform bulk pressure. 

The resulting system of equations formed by \re{eq:piston-hom-p} and \re{eq:piston-1-fast-bou} turns out to be equivalent to the fundamental model proposed by Boukari \etal \cite{boukari1990critical}. Remarkably, while their formulation relies on the simple neglect of fluid flow, the presented multi-scale asymptotic derivation provides a systematic hydrodynamic justification for this reduction, proving that the induced slow flow is indeed a secondary effect that cancels out from the leading-order internal energy balance.

\subsection{Boundary-coupled formulation and emergence of the piston time scale}

Having established the validity of the post-acoustic approximation, we restrict our subsequent analysis exclusively to the slow thermal time scale. By directly substituting the slow time scale relation $ \ct_{\rm d} = \frac{1}{\qPe} \ct $ and  reverting from the dimensionless variables back to the dimensional ones, the piston effect is found to be governed by the dimensional equations
\begin{align}
    \nabla \Pi &= 0 , \\
    \label{eq:heat-eq-piston}
    \pdt{\Theta} &= a \nabla^2 \Theta + \frac{T_0 \beta_p}{\qrho_0 c_p} \frac{\dd \Pi}{\dd t} ,
\end{align}
the dimensional versions of equations \re{eq:piston-hom-p} and \re{eq:piston-1-fast-bou}. Evidently, this local system of differential equations is underdetermined on its own, as it contains two unknown fields---the temperature $ \Theta( t , \qqr ) $ and the spatially uniform pressure $ \Pi ( t ) $---but provides no explicit equation for the temporal evolution of the pressure. To achieve mathematical closure, the global mass balance provides assistance. Let $ \Omega $ denote the spatial domain filled with the supercritical fluid, which is typically a tank with volume $ \mathcal{V} $ at rest with respect to a reference system; therefore,
\begin{align}
    0 = \dt{} \int\limits_{\Omega} \qrho \dd V = \int\limits_{\Omega} \pdt{P} \dd V \stackrel{\re{eq:eos}}{=} \int\limits_{\Omega} \left( \frac{\gamma}{\qas^2} \dt{\Pi} - \qrho_0 \beta_p \pdt{\Theta} \right) \dd V = \mathcal{V} \frac{\gamma}{\qas^2} \dt{\Pi} - \qrho_0 \beta_p \int\limits_{\Omega} \pdt{\Theta} \dd V ;
\end{align}
correspondingly, the time evolution of the pressure connects directly to the instantaneous average temperature, \ie
\begin{align}
    \label{eq:dpdt}
    \dt{\Pi} = \qrho_0 \frac{\beta_p \qas^2}{\gamma} \frac{1}{\mathcal{V}} \int\limits_{\Omega} \pdt{\Theta} \dd V .
\end{align}
Substituting \re{eq:dpdt} into \re{eq:heat-eq-piston} and applying the relationship $ \gamma = 1 + T_0 \frac{\beta_p^2 \qas^2}{c_p} $, the integro-differential equation
\begin{align}
    \pdt{\Theta} = a \nabla^2 \Theta + \frac{\gamma - 1}{\gamma} \frac{1}{\mathcal{V}} \int\limits_{\Omega} \pdt{\Theta} \dd V
\end{align}
on the temperature field is obtained \cite{onuki1990fast}. However, still limited to linearity, \re{eq:dpdt} can be reformulated as
\begin{align}
    \dt{\Pi} = \frac{\beta_p \qas^2}{c_p} \frac{1}{\mathcal{V}} \int\limits\limits_{\Omega} \qrho_0 \pdt{e} \dd V = - \frac{\beta_p \qas^2}{c_p} \frac{1}{\mathcal{V}} \int\limits_{\pd \Omega} \dot{\tensorr{q}} \cdot \dd \tensorr{A} .
\end{align}
Note that we have utilized that heat transfer occurs only at the boundary of the region, denoted by $ \pd \Omega $, and $ \dd \tensorr{A} $ is its outward-pointing surface element. When the surface heating is homogeneous over the entire boundary with total surface area $ \mathcal{A} $, via Fourier's law of heat conduction the pressure evolution can be directly coupled to the temperature gradient at the boundary, \ie
\begin{align}
    \label{eq:dpdt-bc}
    \dt{\Pi} = \lambda \frac{\beta_p \qas^2}{c_p} \frac{\mathcal{A}}{\mathcal{V}} \left. \nabla \Theta \right|_{\pd \Omega} \cdot \tensorr{n} ,
\end{align}
where $ \tensorr{n} $ is the local outward-pointing normal vector. Substituting \re{eq:dpdt-bc} into \re{eq:heat-eq-piston}, the classical diffusion equation is obtained
\begin{align}
    \label{eq:heat-eq-piston-bound-term}
    \pdt{\Theta} = a \nabla^2 \Theta + ( \gamma - 1 ) a \frac{\mathcal{A}}{\mathcal{V}} \left. \nabla \Theta \right|_{\pd \Omega} \cdot \tensorr{n} ,
\end{align}
where a homogeneous volumetric heat source emerges, originating from the boundary heating.

Let us observe that in \re{eq:heat-eq-piston-bound-term}, the ratio of the heated surface area to the fluid volume automatically introduces a characteristic geometric length scale, hence now we fix $ \ell $ as
\begin{align}
    \label{eq:char-length}
    \ell = f \frac{\mathcal{V}}{\mathcal{A}} ,
\end{align}
where $ f $ denotes a dimensionless geometric shape factor reflecting the dimensionality of the domain:
\begin{align}
    \label{formfactor}
    \text{plane wall: } f = 1,
    \quad \text{cylinder: } f = 2,
    \quad \text{sphere: } f = 3.
\end{align}
Alongside the classical diffusion time scale $ \tau_{\rm d} = \frac{\ell^2}{a} $, a further time scale characterizing the apparent bulk heating appears, defined respectively as
\begin{align}
    \tau_{\rm p} &= \frac{\tau_{\rm d}}{f \left( \gamma - 1 \right)} ,
\end{align}
thus reducing the governing equation \re{eq:heat-eq-piston-bound-term} to
\begin{align}
    \label{eq:piston}
    \pdt{\Theta} = a \nabla^2 \Theta + \frac{\ell}{\tau_{\rm p}} \left. \nabla \Theta \right|_{\pd \Omega} \cdot \mathbf{n} .
\end{align}
Although the acoustically originated spatial pressure gradients have vanished, compressibility manifests itself in the energy balance through the specific heat ratio $ \gamma $ embedded within $ \tau_{\rm p} $. Approaching the critical point, heat capacity ratio diverges, hence $ \tau_{\rm p} \ll \tau_{\rm d} $, but in parallel, $ \tau_{\rm a} \ll \tau_{\rm p} $, therefore, $ \tau_{\rm p} $ emerges as an intermediate time scale reflecting the permanent footprint of the suppressed fast time scale acoustic dynamics on the slow diffusion time scale.

To characterize how the localized conduction and the global compression interact, \re{eq:piston} can be rearranged as
\begin{align}
    \label{eq:piston-ratio-form}
    \pdt{\Theta} = \frac{\ell^2}{\tau_{\rm d}} \left( \nabla^2 \Theta + \frac{1}{\ell} \frac{\tau_{\rm d}}{\tau_{\rm p}} \left. \nabla \Theta \right|_{\pd \Omega} \cdot \mathbf{n} \right) .
\end{align}
The explicit appearance of the time-scale ratio $ \frac{\tau_{\rm p}}{\tau_{\rm d}} \ll 1 $ in \re{eq:piston-ratio-form} provides a clear physical basis to establish distinct transport regimes. Introducing $ \zeta $ as the wall-normal coordinate measured directly from the heated boundary, the domain naturally splits into two asymptotic regions:
\begin{itemize}
    \item Near the boundary, \ie when $ \zeta \ll \ell \frac{\tau_{\rm p}}{\tau_{\rm d}} $, local heat conduction dominates the process, establishing a steep thermal gradient, nevertheless modulated by the global compression.
    \item Throughout the vast majority of the bulk fluid, \ie when $ \zeta \gg \ell \frac{\tau_{\rm p}}{\tau_{\rm d}} $, the piston effect drives the dynamics. The rapid adiabatic compression entirely dictates the instantaneous temperature rise, mathematically explaining the famously observed critical speeding up of the system.
\end{itemize}

From a thermoacoustic perspective, this separation of regimes reflects the underlying coupling between fast acoustic propagation and slow thermal diffusion. In particular, the rapid adiabatic compression of the bulk---mediated by acoustic waves---appears in the post-acoustic limit as an effective volumetric heat source, while the thermal diffusion remains confined to a thin boundary layer. Therefore, the classical thermoacoustic description---typically formulated in terms of coupled wave and diffusion equations---reduces in the present asymptotic limit to a boundary-driven heat transfer problem, where the piston effect manifests as a homogeneous source term in the diffusion equation. 

\section{Analytical determination of the temperature field for the piston effect near a heated wall} \label{sec:an-sol}

In this section, exact analytical solutions for the governing equation of the post-acoustic approximation \re{eq:piston} are derived. The transient thermal response to specific boundary perturbations is investigated, assuming an initially homogeneous equilibrium temperature $ T_0 $. We consider strictly one-dimensional heat propagation, presenting solutions in both planar geometry (\eg, for "long" insulated pipes of length $ L $ and cross-sectional area $A$) and spherical geometry of radius $ R $. The latter configuration more closely mimics realistic experimental setups, such as those detailed in \cite{straub1995dynamic}. Accordingly, the characteristic geometric length scale yields $ \ell = L $ for the planar case and $ \ell = R $ for the spherical case. Denoting the generalized spatial coordinate by $ \xi $ (where $ \xi $ represents the axial coordinate $ x $ in planar geometry and the radial coordinate $ r $ in spherical geometry), the temperature field is sought in the additive form
\begin{align}
    \label{eq:bulk+diff}
    \Theta ( t , \xi ) = \Theta_{\rm b} ( t ) + \Theta_{\rm d} ( t , \xi ) ,
\end{align}
where $ \Theta_{\rm b} ( t ) $ denotes the spatially uniform bulk (hereafter also termed volumetric) temperature rise induced by the piston effect---recalling that the volumetric source term in \re{eq:piston} is homogeneous---and $ \Theta_{\rm d} ( t, \xi ) $ represents the localized contribution of thermal diffusion. As a reminder, these temperature contributions, as well as the corresponding boundary conditions, are defined as offsets relative to the initial equilibrium temperature $ T_0 $. Consequently, \re{eq:piston} can be reformulated using the generalized one-dimensional Laplace operator as
\begin{align}
    \label{eq:piston-1D}
    \pdt{\Theta_{\rm d}} = a \frac{1}{\xi^{f-1}} \pdxi{} \left( \xi^{f-1} \pdxi{\Theta_{\rm d}} \right) + \left( \frac{\ell}{\tau_{\rm p}} \left. \pdxi{\Theta_{\rm d}} \right|_{\xi \in \pd \Omega} \cdot \tensorr{n} - \dt{\Theta_{\rm b}} \right) ,
\end{align}
where $ f = 1 $ corresponds to the planar geometry and $ f = 3 $ to the spherical configuration [cf.\ \re{formfactor}]. Since the bulk heating is driven entirely by thermal expansion resulting from surface heat flux, \re{eq:piston-1D} can be decoupled into an ordinary differential equation governing the bulk temperature rise,
\begin{align}
    \label{eq:piston-ode}
    \dt{\Theta_{\rm b}} &= \frac{\ell}{\tau_{\rm p}} \left. \pdxi{\Theta_{\rm d}} \right|_{\xi \in \pd \Omega} \cdot \tensorr{n} , \\
    \Theta_{\rm b} ( 0 ) &= 0 ,
\end{align}
and a classical, unforced diffusion equation for the localized thermal perturbation,
\begin{align}
    \label{eq:piston-pde}
    \pdt{\Theta_{\rm d}} &= a \frac{1}{\xi^{f-1}} \pdxi{} \left( \xi^{f-1} \pdxi{\Theta_{\rm d}} \right) , \\
    \Theta_{\rm d} ( 0 , \xi ) &= 0 .
\end{align}
The dynamic coupling between the bulk temperature rise \re{eq:piston-ode} and the diffusive field \re{eq:piston-pde} is established through the temperature gradient at the boundary.

Within both geometrical configurations, boundary conditions of the first (\ie Dirichlet) and second (\ie Neumann) kind are investigated. For the Dirichlet case, a constant wall temperature $ T_{\rm w} $ is prescribed at the boundary, \ie
\begin{align}
    \Theta ( t , \xi ) \big|_{\xi \in \pd \Omega} = \Theta_{\rm w} ,
\end{align}
where $ \Theta_{\rm w} = T_{\rm w} - T_0 $, which yields the time-dependent boundary condition for the diffusive component:
\begin{align}
    \Theta_{\rm d} ( t , \xi ) \big|_{\xi \in \pd \Omega} = \Theta_{\rm w} - \Theta_{\rm b} ( t ) .
\end{align}
For the Neumann case, a constant wall heat current density $\dot{q}_{\rm w}$ is prescribed at the boundary, \ie
\begin{align}
    - \dot{\tensorr{q}} ( t , \xi ) \big|_{\xi \in \pd \Omega} \cdot \tensorr{n} = \dot{q}_{\rm w} ,
\end{align}
which relates to the temperature gradient via Fourier's law as
\begin{align}
    \left. \pdxi{\Theta_{\rm d}} \right|_{\xi \in \pd \Omega} \cdot \tensorr{n} = \frac{\dot{q}_{\rm w}}{\lambda} .
\end{align}
By convention, $ \dot{q}_{\rm w} > 0 $ corresponds to a heated wall. Assuming that the diffusive penetration depth remains far from the core and opposite boundaries, semi-infinite solutions are sought in terms of the wall-normal coordinate $ \zeta $, measured inward from the heated surface, subject to the asymptotic condition
\begin{align}
    \lim_{\zeta \rightarrow \infty} \Theta_{\rm d} ( t , \zeta ) = 0 .
\end{align}
Consequently, the validity of the solutions presented below is restricted to the short-time regime, $ t \ll \tau_{\rm d} $. Table~\ref{tab:parameters} summarizes the variables, parameters, and boundary conditions for both configurations.

\begin{table}[!ht]
    \begin{center}
        \caption{Applied variables, parameters and boundary conditions of the diffusive field for the planar and spherical geometry.}
        \label{tab:parameters}
        \begin{tabular}{ c | c | c }
            & Planar geometry & Spherical geometry \\
            \hline
            $ \xi $ & $ x $ & $ r $ \\
            $ f $ & $ 1 $ & $ 3 $ \\
            $ \ell $ & $ L $ & $ R $ \\
            BC of the $ 1^{\rm st} $ kind & $ \Theta_{\rm d} ( t , 0 ) = \Theta_{\rm w} - \Theta_{\rm b} ( t ) $ & $ \Theta_{\rm d} ( t , R ) = \Theta_{\rm w} - \Theta_{\rm b} ( t ) $ \\
            BC of the $ 2^{\rm nd} $ kind & $ \left. \pdx{\Theta_{\rm d}} \right|_{ x = 0 } = - \frac{\dot{q}_{\rm w}}{\lambda} $ & $ \left. \pdr{\Theta_{\rm d}} \right|_{ r = R } = \frac{\dot{q}_{\rm w}}{\lambda}$
        \end{tabular}
    \end{center}
\end{table}

Usually, analytical solutions are first presented for boundary conditions of the first kind, followed by those of the second kind. In the present case, however, due to the bulk heating effect, a constant wall temperature yields a time-dependent boundary condition of the first kind for the diffusive temperature contribution, which demands significantly more complex calculations. Therefore, for didactic reasons, we begin with the boundary condition of the second kind, thereby moving from the mathematically simpler case toward the more complex ones.

\subsection{Constant heat flux on the wall}

For a prescribed constant wall heat flux, the bulk-temperature evolution equation \re{eq:piston-ode} reduces to
\begin{align}
    \label{eq:piston-ode-BC2}
    \dt{\Theta_{\rm b}} &= \frac{\ell}{\tau_{\rm p}} \frac{ \dot{q}_{\rm w} }{\lambda} ,
\end{align}
which can be directly integrated subject to the initial condition $ \Theta_{\rm b} ( 0 ) = 0 $. Note that \re{eq:piston-ode-BC2} depends on the geometry only through the characteristic length $ \ell $. Correspondingly, for the planar configuration (\ie $ \ell = L $), this yields
\begin{align}
    \label{eq:sol-x-2-bulk}
    \Theta_{\rm b} ( t ) &= L \frac{\dot{q}_{\rm w}}{\lambda} \frac{t}{\tau_{\rm p}} ,
\end{align}
whereas for the spherical configuration (\ie $ \ell = R $), it gives
\begin{align}
    \label{eq:sol-r-2-bulk}
    \Theta_{\rm b} ( t ) &= R \frac{\dot{q}_{\rm w}}{\lambda} \frac{t}{\tau_{\rm p}} .
\end{align}
Consequently, in both geometries, the bulk temperature increases or decreases strictly linearly with time.

\subsubsection{Planar geometry}

The resulting subproblem corresponds to classical diffusion with a boundary condition of the second kind in a semi-infinite domain, the solution of which is expressed as
\begin{align}
    \label{eq:sol-x-2-diff}
    \Theta_{\rm d} ( t , x ) &= \frac{\dot{q}_{\rm w}}{\lambda} \sqrt{4 a t} \ierfc \frac{x}{\sqrt{4 a t}} ,
\end{align}
where $\ierfc \eta = \frac{\exp \left( -\eta^2 \right)}{\sqrt{\pi}} - \eta \erfc \eta$ denotes the first integral of the complementary error function, and $\erfc \eta$ is the standard complementary error function \cite{carslaw1959conduction}. To facilitate a direct comparison between the diffusive and bulk contributions, it is advantageous to rewrite the pre-factor of $\ierfc$ in \re{eq:sol-x-2-diff} into a form analogous to \re{eq:sol-x-2-bulk}, yielding
\begin{align}
    \label{eq:sol-x-2-diff-mod}
    \Theta_{\rm d} ( t , x ) &= L \frac{\dot{q}_{\rm w}}{\lambda} \sqrt{4 \frac{t}{\tau_{\rm d}}} \ierfc \frac{x}{\sqrt{4 a t}}.
\end{align}

\subsubsection{Spherical geometry}

In spherical coordinates, the governing equation can be reduced to a planar form with modified boundary conditions by employing the standard transformation
\begin{align}
    \label{eq:transf-r-2-x}
    \Theta_{\rm d} ( t , r ) = \frac{\vartheta ( t , r )}{r} .
\end{align}
The corresponding spatial derivatives take the form
\begin{align}
    \pdr{\Theta_{\rm d}} &= - \frac{1}{r^2} \vartheta + \frac{1}{r} \pdr{\vartheta} , &
    \frac{1}{r^2} \pdr{} \left( r^2 \pdr{\Theta_{\rm d}} \right) &= \frac{1}{r} \ppd{\vartheta}{r} .
\end{align}
Furthermore, we introduce a localized spatial coordinate $\zeta$, measured from the outer surface of the sphere pointing inward along the negative radial direction, defined as
\begin{align}
    \label{eq:transf-r-2-zeta}
    \zeta &= R - r , &
    \pdr{} &= - \pdz{} , &
    \ppd{}{r} &= \ppd{}{\zeta} .
\end{align}
Substituting these transformations yields the following initial-boundary value problem for the auxiliary variable $\vartheta$:
\begin{align}
    \pdt{\vartheta} &= a \ppd{\vartheta}{\zeta} , \\
    \vartheta( 0 , \zeta ) &= 0 , \\
    \frac{1}{R^2} \vartheta ( t , 0 ) + \frac{1}{R} \left. \pdz{\vartheta} \right|_{ \zeta = 0 } &= - \frac{\dot{q}_{\rm w}}{\lambda} .
\end{align}
This formulation represents a classical diffusion problem with a boundary condition of the third kind (Robin condition) in a semi-infinite region \cite{carslaw1959conduction}. Solving this system and transforming the obtained solution via \re{eq:transf-r-2-x} yields the diffusive temperature field:
\begin{align}
    \label{eq:sol-r-2-diff}
    \Theta_{\rm d} ( t , \zeta ) &= R \frac{\dot{q}_{\rm w}}{\lambda} \frac{R \exp \left( - \frac{\zeta^2}{4 a t}\right)}{R - \zeta} \left[ \erfcx \left( \frac{\zeta}{\sqrt{4 a t}} - \sqrt{\frac{t}{\tau_{\rm d}}} \right) - \erfcx \frac{\zeta}{\sqrt{4 a t}} \right] ,
\end{align}
where $ \erfcx \eta = \exp \left( \eta^2 \right) \erfc \eta $ denotes the scaled complementary error function.

\subsection{Constant wall temperature}

A constant non-zero temperature difference prescribed at the wall results in a time-dependent boundary condition of the first kind for the diffusion problem, which is treated here using the Laplace transform method. The inverse Laplace transforms of the resulting non-trivial expressions are summarized in App.~\ref{sec:app-A}.

\subsubsection{Planar geometry}

Applying the Laplace transform with respect to the time coordinate to the governing equations \re{eq:piston-ode}, \re{eq:piston-pde} and the boundary condition yields
\begin{align}
    \label{eq:piston-pde-s-bc1}
    s \hTheta_{\rm d} ( s , x ) &= a \ppd{\hTheta_{\rm d}}{x} , \\
    \label{eq:piston-ode-s-bc1}
    s \hTheta_{\rm b} ( s ) &= - \frac{L}{\tau_{\rm p}} \left. \pdx{\hTheta_{\rm d}} \right|_{ x = 0 } , \\
    \label{eq:piston-bc-s-bc1}
    \hTheta_{\rm d} ( s , 0 ) &= \frac{\Theta_{\rm w}}{s} - \hTheta_{\rm b} (s) ,
\end{align}
where $ s $ denotes the complex frequency (\ie the Laplace-domain variable). The negative sign on the right-hand side of \re{eq:piston-ode-s-bc1} originates from the outward-pointing normal vector. Enforcing the physical requirement that the temperature field must remain bounded as $ x \rightarrow \infty $, the solution of \re{eq:piston-pde-s-bc1} is given by
\begin{align}
    \label{eq:piston-pde-s-bc1-sol}
    \hTheta_{\rm d} ( s , x ) &= A ( s ) \exp \left( - \sqrt{\frac{s}{a}} x \right) ,
\end{align}
where $ A ( s ) $ represents the complex frequency-dependent amplitude. Correspondingly, the boundary values are $ \hTheta_{\rm d} ( s , 0 ) = A ( s ) $ and $ \left. \pdx{\hTheta_{\rm d}} \right|_{ x = 0 } = - \sqrt{\frac{s}{a}} A ( s )$. Substituting these expressions into \re{eq:piston-ode-s-bc1} and \re{eq:piston-bc-s-bc1} leads to the system
\begin{align}
    s \hTheta_{\rm b} &= \frac{L}{\tau_{\rm p}} \sqrt{\frac{s}{a}} A ( s ) = \frac{\sqrt{\tau_{\rm d}}}{\tau_{\rm p}} \sqrt{s} A ( s ) , &
    A ( s ) &= \frac{\Theta_{\rm w}}{s} - \hTheta_{\rm b} (s) ,
\end{align}
which, via partial fraction decomposition, yields
\begin{align}
    \label{eq:piston-Tb-s-bc1-sol}
    \hTheta_{\rm b} ( s ) = \Theta_{\rm w} \frac{\frac{\sqrt{\tau_{\rm d}}}{\tau_{\rm p}}}{s \left( \sqrt{s} + \frac{\sqrt{\tau_{\rm d}}}{\tau_{\rm p}} \right)} = \frac{\Theta_{\rm w}}{\frac{\sqrt{\tau_{\rm d}}}{\tau_{\rm p}}} \left( - \frac{1}{\sqrt{s}} + \frac{\frac{\sqrt{\tau_{\rm d}}}{\tau_{\rm p}}}{s} + \frac{1}{\sqrt{s} + \frac{\sqrt{\tau_{\rm d}}}{\tau_{\rm p}}} \right) ,
\end{align}
and, in virtue of \re{eq:piston-pde-s-bc1-sol} 
\begin{align}
    \label{eq:piston-Td-s-bc1-sol}
    \hTheta_{\rm d} ( s , x ) = \Theta_{\rm w} \frac{\exp \left( - \sqrt{\frac{s}{a}} x \right)}{\sqrt{s} \left( \sqrt{s} + \frac{\sqrt{\tau_{\rm d}}}{\tau_{\rm p}} \right)} = \frac{\Theta_{\rm w}}{\frac{\sqrt{\tau_{\rm d}}}{\tau_{\rm p}}} \exp \left( - \sqrt{\frac{s}{a}} x \right) \left( \frac{1}{\sqrt{s}} - \frac{1}{\sqrt{s} + \frac{\sqrt{\tau_{\rm d}}}{\tau_{\rm p}}} \right).
\end{align}
The time-domain temperature contributions are determined by performing the inverse Laplace transformation on \re{eq:piston-Tb-s-bc1-sol} and \re{eq:piston-Td-s-bc1-sol}, giving
\begin{align}
    \Theta_{\rm b} ( t ) &= \Theta_{\rm w} \left[ 1 - \erfcx \left( \sqrt{\frac{\tau_{\rm d}}{\tau_{\rm p}} \frac{t}{\tau_{\rm p}}} \right) \right] , \\
    \Theta_{\rm d} ( t , x ) &= \Theta_{\rm w} \exp \left( - \frac{x^2}{4 a t} \right) \erfcx \left( \frac{x}{\sqrt{4 a t}} + \sqrt{\frac{\tau_{\rm d}}{\tau_{\rm p}} \frac{t}{\tau_{\rm p}}} \right) .
\end{align}

\subsubsection{Spherical geometry}

First, the governing equations \re{eq:piston-ode}, \re{eq:piston-pde} and the associated boundary condition are converted from spherical coordinates to the localized Cartesian inward radial coordinate via \re{eq:transf-r-2-x} and \re{eq:transf-r-2-zeta}. Then, performing the Laplace transform with respect to time yields
\begin{align}
    \label{eq:piston-pde-s-bc1-r}
    s \htheta ( s , \zeta ) &= a \ppd{\htheta}{\zeta} , \\
    \label{eq:piston-ode-s-bc1-r}
    s \hTheta_{\rm b} ( s ) &= - \frac{R}{\tau_{\rm p}} \Bigg( \frac{1}{R^2} \htheta ( s , 0 ) + \frac{1}{R} \left. \pdz{\htheta} \right|_{ \zeta = 0 } \Bigg) , \\
    \label{eq:piston-bc-s-bc1-r}
    \htheta ( s , 0 ) &= R \left( \frac{\Theta_{\rm w}}{s} - \hTheta_{\rm b} ( s ) \right) .
\end{align}
Enforcing the physical requirement of a bounded temperature field as $ \zeta \to \infty $ leads to the solution of \re{eq:piston-pde-s-bc1-r} in the form
\begin{align}
    \label{eq:piston-pde-s-bc1-sol-r}
    \htheta ( s , \zeta ) &= A ( s ) \exp \left( - \sqrt{\frac{s}{a}} \zeta \right) .
\end{align}
Substituting \re{eq:piston-pde-s-bc1-sol-r} into \re{eq:piston-ode-s-bc1-r} and \re{eq:piston-bc-s-bc1-r} yields the transformed bulk temperature rise,
\begin{align}
    \label{eq:piston-Tb-s-bc1-sol-r}
    \hTheta_{\rm b} ( s ) = \Theta_{\rm w} \frac{ \frac{\sqrt{\tau_{\rm d}}}{\tau_{\rm p}} \sqrt{s} - \frac{1}{\tau_{\rm p}}}{s \left( s + \frac{\sqrt{\tau_{\rm d}}}{\tau_{\rm p}} \sqrt{s} - \frac{1}{\tau_{\rm p}} \right)} = \Theta_{\rm w} \left[ \frac{1}{s} + \frac{1}{\mu_- - \mu_+} \left( \frac{1}{\sqrt{s} - \mu_+} - \frac{1}{\sqrt{s} - \mu_-} \right) \right] ,
\end{align}
where the final expression is obtained via partial fraction decomposition using the roots $ \mu_\pm $ of the characteristic quadratic polynomial in $ \sqrt{s} $ found in the denominator:
\begin{align}
    \label{eq:piston-Tb-s-bc1-mu-r}
    \mu_\pm = - \frac{1}{2} \frac{\sqrt{\tau_{\rm d}}}{\tau_{\rm p}} \left( 1 \pm \sqrt{ 1 + 4 \frac{\tau_{\rm p}}{\tau_{\rm d}}} \right) .
\end{align}
Accordingly, the transformed diffusive field is given by
\begin{align}
    \label{eq:theta-sph-diff}
    \htheta ( s , \zeta ) = - \Theta_{\rm w} \frac{R \exp \left( - \sqrt{\frac{s}{a}} \zeta \right)}{\mu_- - \mu_+} \left( \frac{1}{\sqrt{s} - \mu_+} - \frac{1}{\sqrt{s} - \mu_-} \right) .
\end{align}
Finally, applying the inverse Laplace transformation to \re{eq:theta-sph-diff} and transforming back to the original diffusive temperature field via \re{eq:transf-r-2-x} yields the analytical solution in the time domain:
\begin{align}
    \label{eq:sol-sph-bulk}
    \Theta_{\rm b} ( t ) &= \Theta_{\rm w} \left[ 1 + \frac{\mu_+}{\mu_- - \mu_+} \erfcx \left( - \mu_+ \sqrt{t} \right) - \frac{\mu_-}{\mu_- - \mu_+} \erfcx \left( - \mu_- \sqrt{t} \right) \right] , \\
    \label{eq:sol-sph-diff}
    \Theta_{\rm d} ( t , \zeta ) &= - \Theta_{\rm w} \frac{R \exp\left( - \frac{\zeta^2}{4 a t}\right)}{R - \zeta} \Bigg[ \frac{\mu_+}{\mu_- - \mu_+} \erfcx \left( \frac{\zeta}{\sqrt{4 a t}} - \mu_+ \sqrt{t} \right) - \frac{\mu_-}{\mu_- - \mu_+} \erfcx \left( \frac{\zeta}{\sqrt{4 a t}} - \mu_- \sqrt{t} \right) \Bigg] .
\end{align}

\section{Short-time asymptotics and thermal boundary layer thickness} \label{sec:asy}

In this section, the short-time asymptotics and the thermal boundary layer thickness are analyzed in non-dimensional form. This analysis is performed for both the planar and spherical configurations under Dirichlet and Neumann boundary conditions. Time is scaled by the characteristic piston time scale $ \tau_{\rm p} $, while spatial dimensions are normalized by the characteristic length scale $ \ell $, yielding the dimensionless time and space variables
\begin{align}
    \label{eq:nondim-var}
    \ct &= \frac{t}{\tau_{\rm p}} , &
    \check{\xi} &= \frac{\xi}{\ell} .
\end{align}
The ratio of the piston time scale to the diffusion time scale directly appears in all solutions and is denoted by $ \varepsilon = \frac{\tau_{\rm p}}{\tau_{\rm d}} $. In the vicinity of the critical point, $ \varepsilon \ll 1 $, meaning that $ \varepsilon $ serves as the fundamental small parameter to distinguish between the short-time and long-time regimes. For the Dirichlet boundary condition, the temperature is non-dimensionalized using the prescribed wall temperature difference $\Theta_{\rm w}$. For the Neumann boundary condition, the natural temperature scale is given by the dimensional coefficient $\ell \frac{\dot{q}_{\rm w}}{\lambda}$. Consequently, the dimensionless temperature is defined as
\begin{align}
    \label{eq:nondim-fun}
    \cTheta &= \frac{\Theta}{\Theta_{\rm w}} , && \text{and} &
    \cTheta &= \frac{\Theta}{\ell \frac{\dot{q}_{\rm w}}{\lambda}} ,
\end{align}
respectively. Since the thermal boundary layer develops in the immediate vicinity of the heated wall, the solutions are analyzed in terms of the wall-normal coordinate $ \zeta $, where $ \zeta = x $ in the planar geometry and $ \zeta = R - r $ in the spherical configuration. Finally, the thermal boundary layer thickness, representing the thermal penetration depth, is evaluated using the classical gradient-based definition applied strictly to the diffusive temperature contribution $ \Theta_{\rm d} $,
\begin{align}
    \delta_{\rm th} ( t ) = \frac{\Theta_{\rm d} ( t , \zeta = 0 )}{- \left. \pdz{\Theta_{\rm d}} \right|_{ \zeta = 0 } } .
\end{align}
Geometrically, this definition of the thermal penetration depth corresponds to the intersection of the wall tangent of the temperature profile with the instantaneous bulk temperature\footnote{Formally, this construction resembles the definition of the time constant in first-order linear systems, where a characteristic measure is obtained from the ratio of a quantity and its initial rate of temporal change.}, thereby providing a characteristic measure of the thickness of the conduction-dominated thermal boundary layer. Note that it is physically essential to isolate $ \Theta_{\rm d} $ and exclude the bulk temperature rise $ \Theta_{\rm b} $ from this evaluation, because the rapid adiabatic compression uniformly elevates the background fluid temperature. Evaluating the penetration depth using the total temperature field would obscure the true thickness of the conduction-dominated zone, as the local wall heating operates on top of a dynamically rising bulk baseline. 

Throughout the subsequent parts of this section, the dimensionless spatial variables utilized for the planar and spherical configurations are explicitly defined as
\begin{align}
    \label{eq:nondim-var-1}
    \cx &= \frac{x}{L} , &
    \cz &= \frac{R - r}{R} = \frac{\zeta}{R} .
\end{align}
All graphical representations presented in this section were generated by direct evaluation of the analytical solutions derived in Sec.~\ref{sec:an-sol} and of the corresponding asymptotic approximations and thermal penetration depths derived in the present section. For all numerical evaluations and plots presented in this section, the single free model parameter is fixed at $ \varepsilon = 0.01 $.

\subsection{Constant heat flux on the wall}

The dimensionless temperature field for the Neumann boundary condition in Cartesian coordinates can be formulated as 
\begin{align}
    \label{eq:dimless-sol-x-2}
    \cTheta ( \ct , \cx ) = \ct + \frac{2}{\sqrt{\pi}} \sqrt{\varepsilon \ct} \exp \left( - \frac{\cx^2}{4 \varepsilon \ct} \right) - \cx \erfc \frac{\cx}{\sqrt{4 \varepsilon \ct}} ,
\end{align}
directly leading to the classical thermal penetration depth 
\begin{align}
    \check{\delta}_{\rm th} ( \ct ) = \frac{2}{\sqrt{\pi}} \sqrt{\varepsilon \ct} .
\end{align}
Although the solution is not fully self-similar due to the explicit appearance of the length scale $ \ell $ in \re{eq:piston}, the similarity variable\footnote{It should be noted that the present definition of the similarity variable deviates from the formulation $ \eta = \frac{x}{\sqrt{4 a t}} $ usually applied in heat transfer.}
\begin{align}
    \eta ( \ct , \cx ) = \frac{\cx}{\check{\delta}_{\rm th} ( \ct )}
\end{align}
proves highly effective. Thereby, \re{eq:dimless-sol-x-2} can be reformulated as
\begin{align}
    \label{eq:dimless-sol-x-2-eta}
    \cTheta ( \ct , \cx ) = \ct + \check{\delta}_{\rm th} ( \ct ) \left[ \exp \left( - \frac{\eta^2 ( \ct , \cx )}{\pi} \right) - \eta( \ct , \cx ) \erfc \frac{\eta ( \ct , \cx )}{\sqrt{\pi}} \right] .
\end{align}

\begin{figure}[!ht]
    \centering
    \begin{center}
        \makebox[\textwidth][c]{%
            \begin{minipage}[c]{.45\textwidth} \centering
                \includegraphics[width=\linewidth]{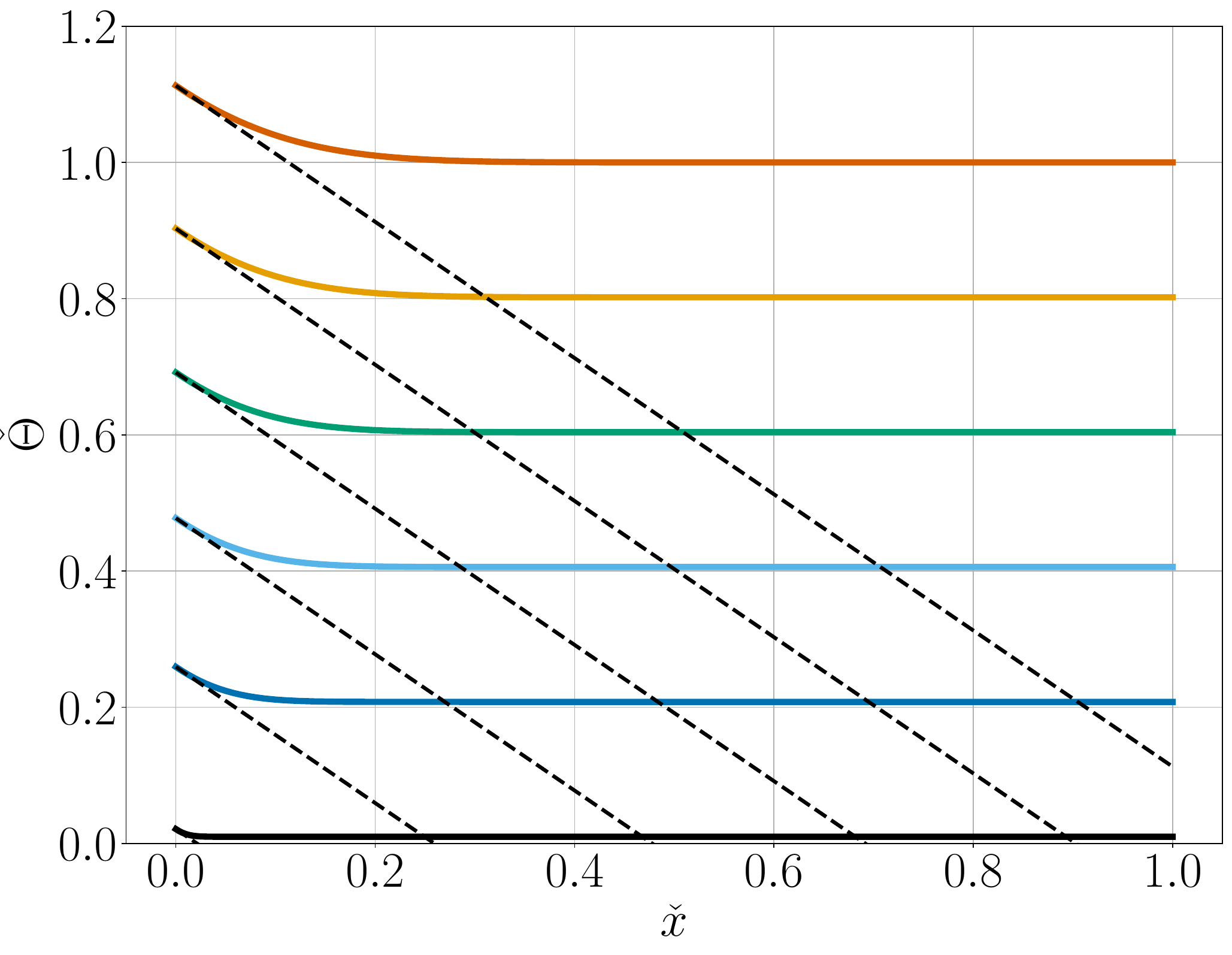}
            \end{minipage}%
            \hspace{.02\textwidth}%
            \begin{minipage}[c]{.32\textwidth} \centering
                \includegraphics[width=\linewidth]{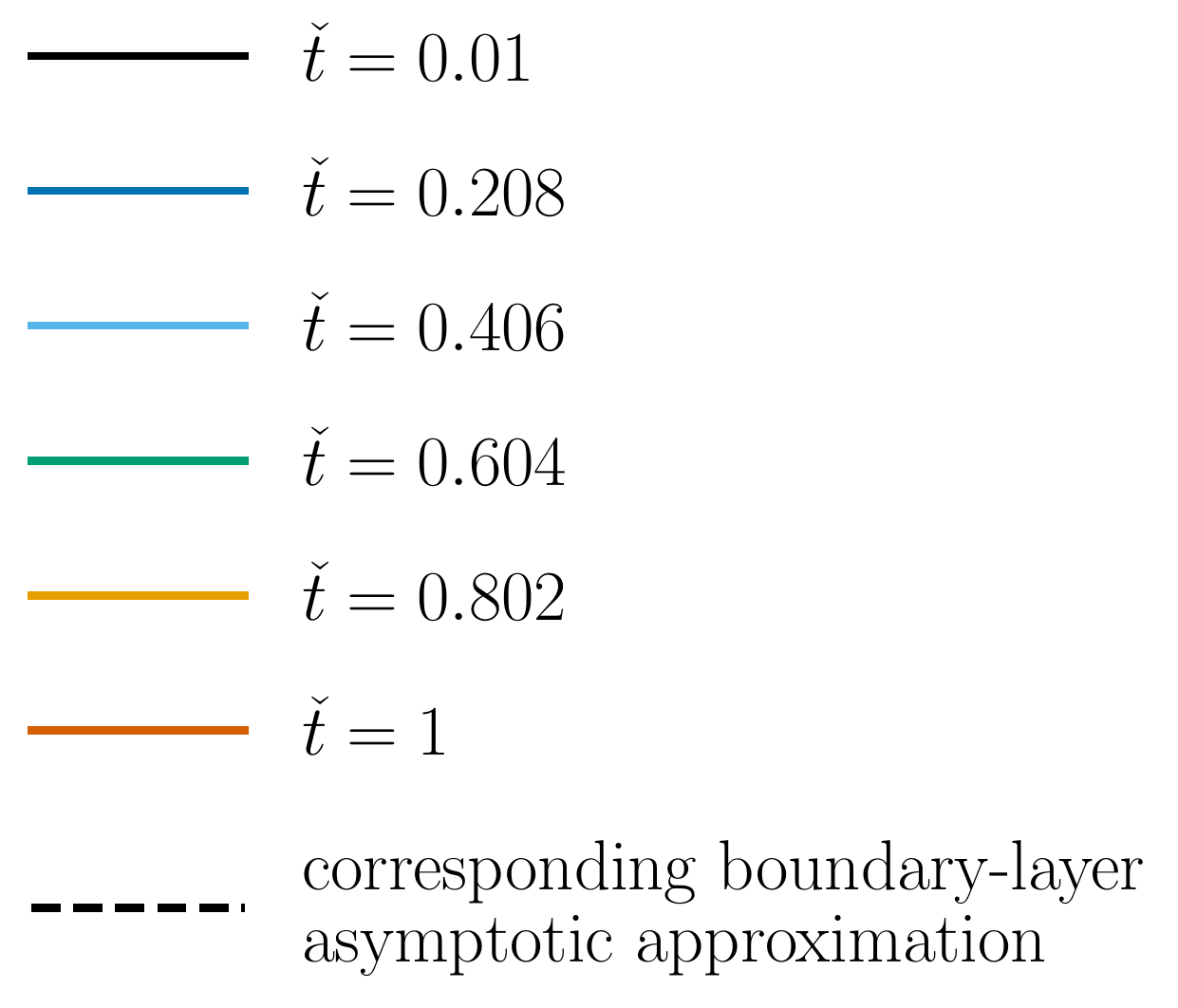}
            \end{minipage}%
            }
    \end{center}
    \caption{Dimensionless temperature distribution evaluated at equal time steps in the interval $\ct \in [0.01, 1]$ under a constant wall heat flux for the planar configuration, where time progresses from the black to the red lines. The solid lines represent the exact analytical solution, and the dashed lines denote the linear boundary asymptotics presented in \re{eq:dimless-sol-x-2-asy}.}
    \label{fig:T-x-2-asy}
\end{figure}

Within the thermal boundary layer, \ie, for $ \cx \ll \check{\delta}_{\rm th} ( \ct ) $, corresponding to $ \eta ( \ct , \cx ) \ll 1 $, the dimensionless temperature field \re{eq:dimless-sol-x-2-eta} can be approximated as
\begin{align}
    \label{eq:dimless-sol-x-2-asy}
    \cTheta ( \ct , \cx ) \approx \ct + \check{\delta}_{\rm th} ( \ct ) \left[ \left( 1 - \frac{\eta^2 ( \ct , \cx )}{\pi} \right) - \eta( \ct , \cx ) \left( 1 - \frac{2}{\pi} \eta ( \ct , \cx ) \right) \right] \approx \ct + \check{\delta}_{\rm th} ( \ct ) \left[ 1 - \eta( \ct , \cx ) \right] = \ct - \cx + \frac{2}{\sqrt{\pi}} \sqrt{\varepsilon \ct} .
\end{align}
The resulting temperature distribution reveals a dynamically shifting linear profile, in which the superposition of three distinct physical phenomena can be recognized. The linear spatial dependence $ - \cx $ ensures flux consistency at the interface, establishing the steady Fourier gradient required to transport the prescribed heat flux into the medium. This time-independent spatial gradient is superimposed on the linearly growing bulk temperature $ \ct $ driven by the global adiabatic compression. Finally, the term $ \frac{2}{\sqrt{\pi}} \sqrt{\varepsilon \ct} $ represents the near-boundary temperature rise caused by local diffusion. This indicates that while the local temperature gradient remains fixed to satisfy the constant heat flux condition, this linear profile shifts upward as time progresses. Furthermore, as the thermal boundary layer grows, the spatial extent of this linear zone continuously expands deeper into the medium, meaning that its far end moves progressively further away from the heated wall. These are illustrated in Fig.~\ref{fig:T-x-2-asy}.

A straightforward scaling check of the similarity variable reveals an important physical constraint regarding the condition $ \eta \ll 1 $. While this inequality is naturally satisfied near the heated boundary, mathematically it can also be fulfilled if $ \sqrt{\varepsilon \ct} \gg 1 $. In dimensional variables, this secondary limit corresponds to $ \sqrt{\frac{t}{\tau_{\rm d}}} \gg 1 $, which represents the long-time diffusion-dominated regime. Since this contradicts our initial assumption that $ t \ll \tau_{\rm d } $, no physical conclusions can be drawn from this secondary limit within the framework of the present model.

Outside the thermal boundary layer, \ie, for $ \cx \gg \check{\delta}_{\rm th} ( \ct ) $, corresponding to $ \eta ( \ct , \cx ) \gg 1 $, the diffusive contributions decay, reducing \re{eq:dimless-sol-x-2-eta} to the uniform bulk temperature
\begin{align}
    \cTheta ( \ct , \cx ) \approx \ct + \check{\delta}_{\rm th} ( \ct ) \left[ \exp \left( - \frac{\eta^2 ( \ct , \cx )}{\pi} \right) - \eta \frac{\exp \left( - \frac{\eta^2 ( \ct , \cx )}{\pi} \right)}{\frac{\eta}{\sqrt{\pi}} \sqrt{\pi}}\right] = \ct .
\end{align}
This limit where $\eta \gg 1$ carries profound physical meaning for the early-stage dynamics. At the very beginning of the process, the thermal boundary layer thickness approaches zero, implying that the condition $ \eta \gg 1 $ holds true throughout virtually the entire fluid volume. In dimensional variables, this short-time regime directly corresponds to $ \sqrt{\frac{t}{\tau_{\rm d}}} \ll 1 $. Since this perfectly aligns with our initial assumption that $ t \ll \tau_{\rm d} $, the physical validity of this limit is strictly consistent with the scope of our model. This reveals that initially the fluid undergoes a spatially homogeneous temperature rise driven by the hidden acoustic process manifesting in the adiabatic compression.

\begin{figure}[!ht]
    \begin{center}
        \makebox[\textwidth][c]{%
            \begin{minipage}[c]{.5\textwidth} \centering
                \includegraphics[width=\linewidth]{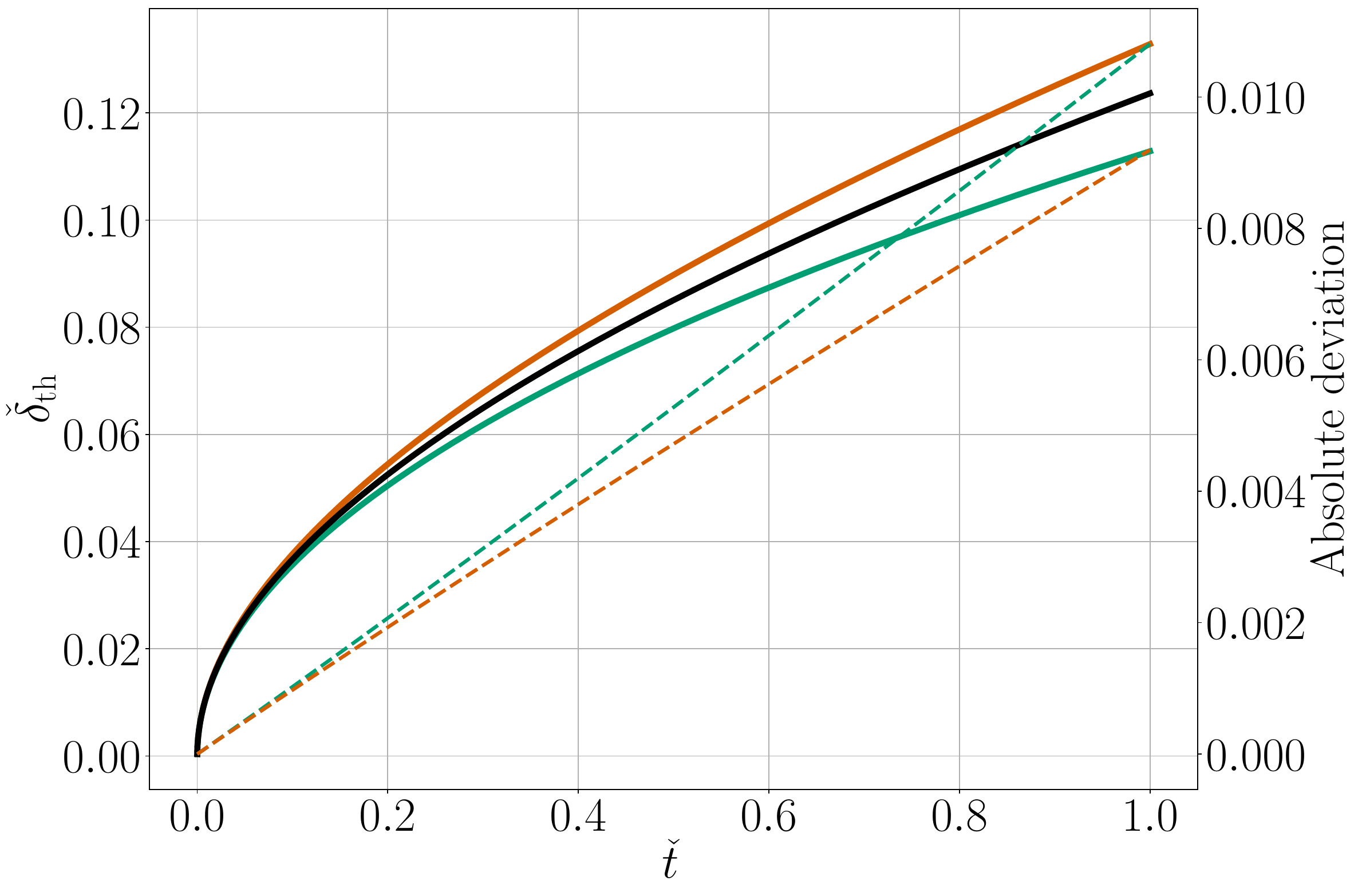}
            \end{minipage}%
            \hspace{.02\textwidth}%
            \begin{minipage}[c]{.35\textwidth} \centering
                \includegraphics[width=\linewidth]{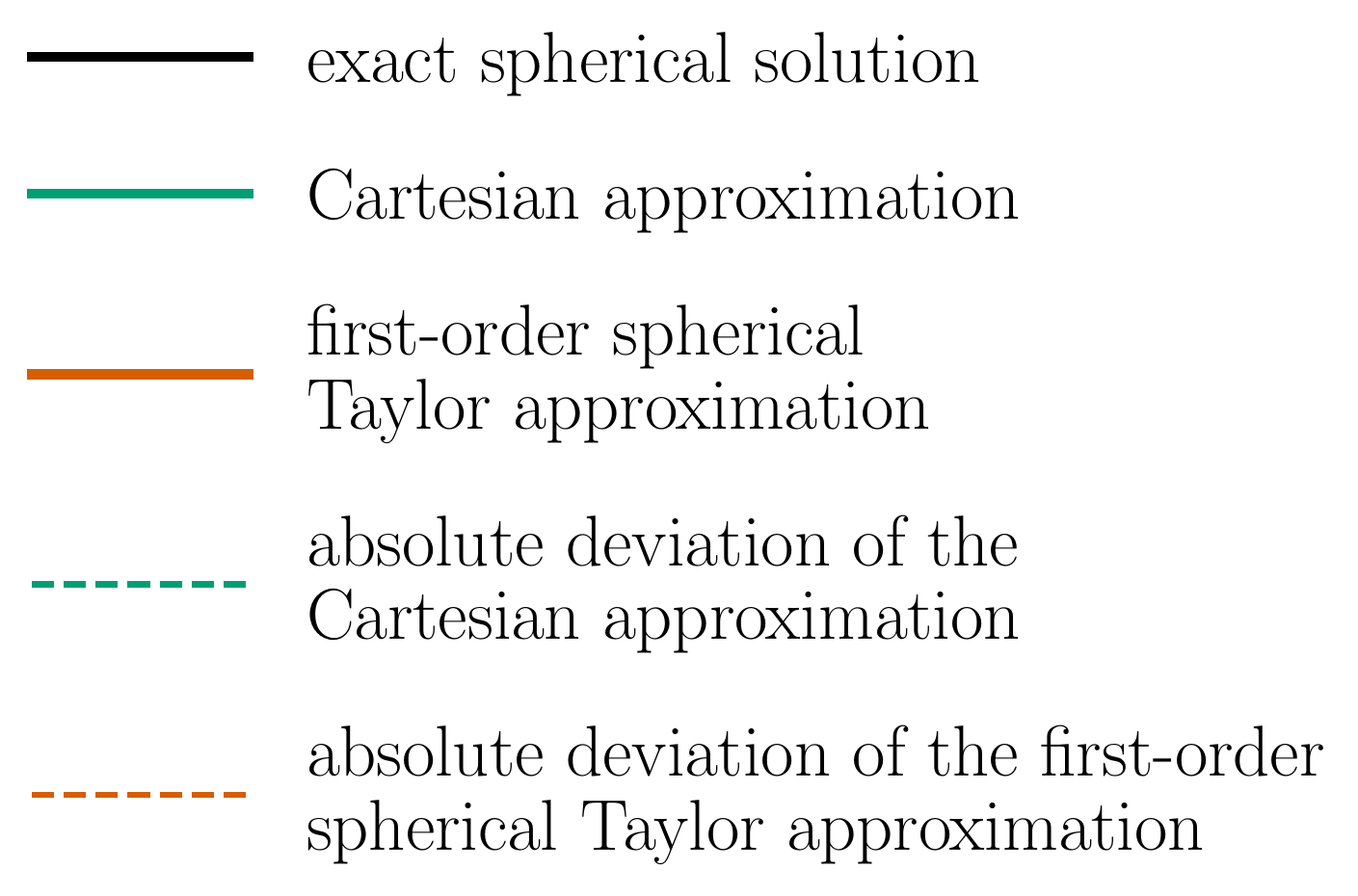}
            \end{minipage}%
            }
    \end{center}
    \caption{Evolution of the dimensionless thermal penetration depth (left axis) and the corresponding absolute deviation from the exact solution (right axis) for the planar and spherical configurations under a constant wall heat flux. On the left axis, the exact spherical solution is represented by the solid black line, the Cartesian approximation by the solid green line, and the first-order spherical Taylor approximation by the solid red line. On the right axis, the dashed green line denotes the absolute deviation of the Cartesian approximation, while the dashed red line represents the absolute deviation of the first-order Taylor approximation.}
    \label{fig:del-2}
\end{figure}

The dimensionless temperature field for the Neumann boundary condition in the spherical geometry is given by
\begin{align}
    \label{eq:dimless-sol-r-2}
    \cTheta ( \ct , \cz ) = \ct + \frac{\exp \left( - \frac{\cz^2}{4 \varepsilon \ct} \right)}{1 - \cz} \left[ \erfcx \left( \frac{\cz}{\sqrt{4 \varepsilon \ct}} - \sqrt{\varepsilon \ct} \right) - \erfcx \frac{\cz}{\sqrt{4 \varepsilon \ct}} \right] ,
\end{align}
which yields the thermal penetration depth
\begin{align}
    \label{dimless-del-r-2}
    \check{\delta}_{\rm th} ( \ct ) = \erfcx \left( - \sqrt{\varepsilon \ct} \right) - 1 = 2 \exp \left( \varepsilon \ct \right) - \erfcx \sqrt{\varepsilon \ct} - 1 .
\end{align}
When $ \sqrt{\varepsilon \ct} \ll 1 $, corresponding to the short-time regime $ t \ll \tau_{\rm d} $, \re{dimless-del-r-2} can be approximated as
\begin{align}
    \check{\delta}_{\rm th} ( \ct ) \approx 2 \left( 1 + \varepsilon \ct \right) - \left( 1 - \frac{2}{\sqrt{\pi}} \sqrt{\varepsilon \ct} \right) - 1 = \frac{2}{\sqrt{\pi}} \sqrt{\varepsilon \ct} \left( 1 + \sqrt{\pi} \sqrt{\varepsilon \ct} \right) .
\end{align}
This Taylor expansion demonstrates that in the leading-order approximation, the spherical configuration perfectly recovers the thermal penetration depth derived for the Cartesian geometry, $ \check{\delta}_{\rm th}^{(1)} ( \ct ) = \frac{2}{\sqrt{\pi}} \sqrt{\varepsilon \ct} $. The first-order correction term reveals that the boundary layer growth is accelerated in the spherical configuration. This physical speeding up is a direct consequence of the converging geometry. As the thermal disturbance propagates inward from the outer heated boundary, the surface area of the concentric fluid shells continuously decreases, leading to a geometric concentration of the heat flux that drives the penetration depth faster into the fluid domain. These developments and the corresponding absolute deviations are illustrated in Fig.~\ref{fig:del-2}.

\begin{figure}[!ht]
    \centering
    \begin{center}
        \makebox[\textwidth][c]{%
            \begin{minipage}[c]{.45\textwidth} \centering
                \includegraphics[width=\linewidth]{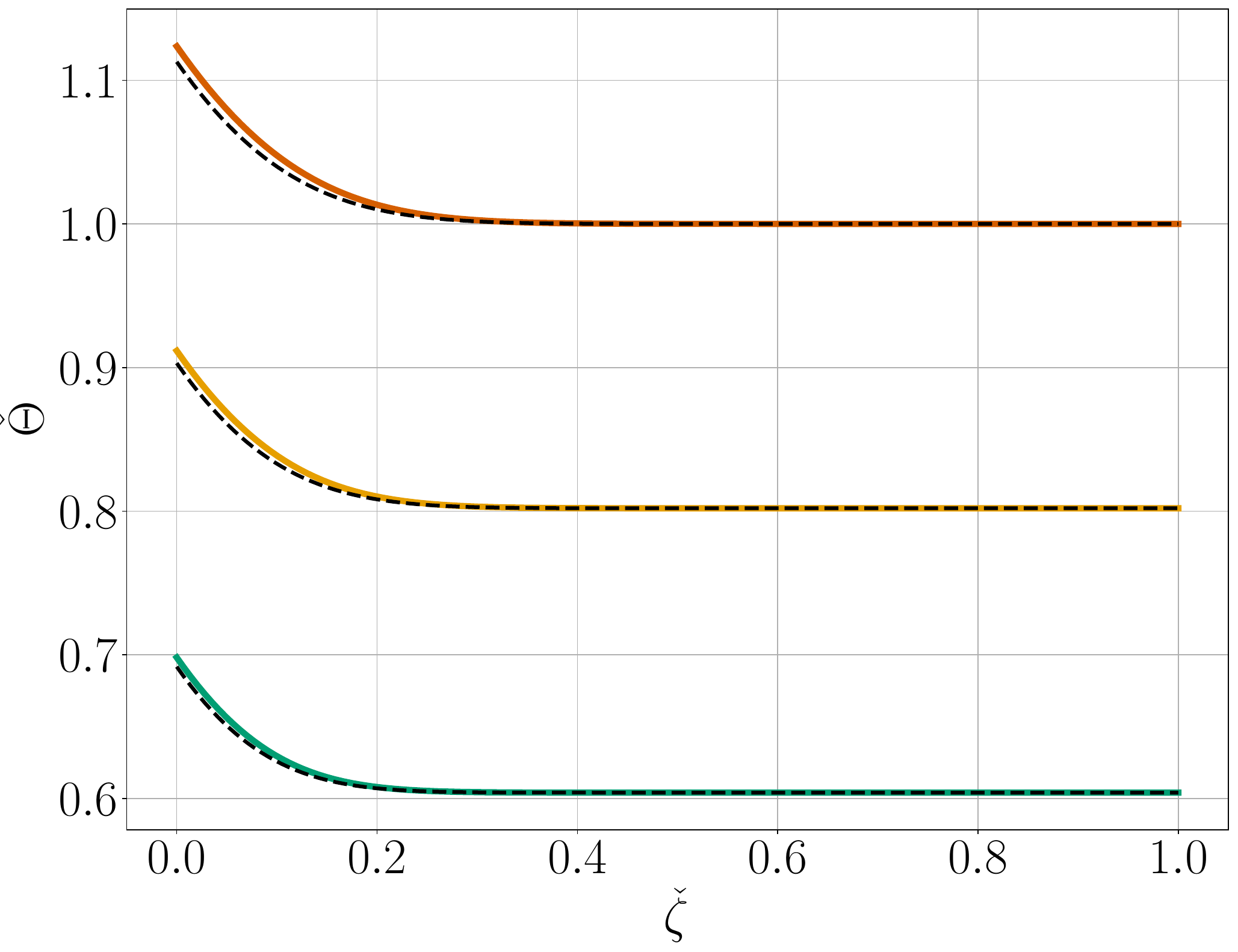}
            \end{minipage}%
            \hspace{.02\textwidth}%
            \begin{minipage}[c]{.32\textwidth} \centering
                \includegraphics[width=\linewidth]{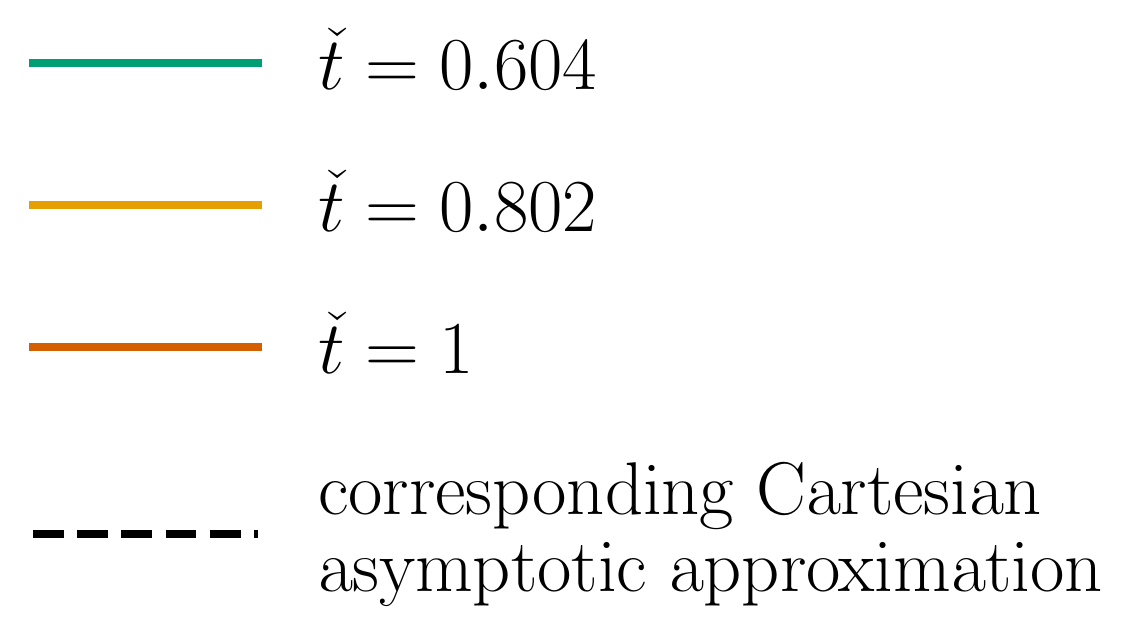}
            \end{minipage}%
            }
    \end{center}
    \caption{Comparison of the dimensionless temperature fields for the planar and spherical configurations under a constant wall heat flux, evaluated at equal time steps within $\ct \in [0.01, 1]$. To focus on the active transient regime, the first three time steps are omitted due to the lack of significant deviation between the geometries. The solid lines represent the exact spherical solution (with time progressing from the green to the red lines), and the dashed lines denote the Cartesian approximation, where the close agreement demonstrates the accuracy of the planar representation within the thermal boundary layer.}
    \label{fig:T-2-x-vs-r}
\end{figure}

Introducing the similarity variable
\begin{align}
    \eta ( \ct , \cz ) = \frac{\cz}{\check{\delta}_{\rm th}^{(1)} ( \ct )}
\end{align}
the spherical solution can be written as
\begin{align}
    \cTheta ( \ct , \cz ) \approx \ct + \frac{\check{\delta}_{\rm th}^{(1)} ( \ct )}{1 - \check{\delta}_{\rm th}^{(1)} ( \ct ) \eta ( \ct , \cz )} \left[ \exp \left( - \frac{\eta^2 ( \ct , \cz )}{\pi} \right) - \eta ( \ct , \cz ) \erfc \frac{\eta ( \ct , \cz )}{\sqrt{\pi}} \right] .
\end{align}
Within the thermal boundary layer, \ie, for $ \cz \ll \check{\delta}_{\rm th}^{(1)} ( \ct ) $, corresponding to $ \eta ( \ct , \cz ) \ll 1 $, this expression can be approximated as
\begin{align}
    \nonumber
    \cTheta ( \ct , \cz ) &\approx \ct + \left( 1 + \check{\delta}_{\rm th}^{(1)} ( \ct ) \eta ( \ct , \cz ) \right) \check{\delta}_{\rm th}^{(1)} ( \ct ) \left[ \exp \left( - \frac{\eta^2 ( \ct , \cz )}{\pi} \right) - \eta ( \ct , \cz ) \erfc \frac{\eta ( \ct , \cz )}{\sqrt{\pi}} \right] \\
    & \approx \ct + \check{\delta}_{\rm th}^{(1)} ( \ct ) \left[ \exp \left( - \frac{\eta^2 ( \ct , \cz )}{\pi} \right) - \eta ( \ct , \cz ) \erfc \frac{\eta ( \ct , \cz )}{\sqrt{\pi}} \right] .
\end{align}
Thereby, near the boundary, the spherical solution reduces to the Cartesian one [cf.\ \re{eq:dimless-sol-x-2-eta}] in the leading order. This close agreement demonstrates that despite the geometric convergence of the spherical domain, the planar approximation remains highly accurate within the thermal boundary layer during the short-time regime. Consequently, the first-order curvature correction discussed above has a minor impact on the local temperature field near the wall, preserving the practical validity of the Cartesian representation throughout the investigated time window. These results are illustrated in Fig.~\ref{fig:T-2-x-vs-r}.

\subsection{Constant wall temperature}

For the Dirichlet boundary condition, the dimensionless temperature field in Cartesian coordinates can be written as
\begin{align}
    \label{eq:dimless-sol-x-1}
    \cTheta ( \ct , \cx ) = 1 - \erfcx \sqrt{\frac{\ct}{\varepsilon}} + \exp \left(- \frac{\cx^2}{4 \varepsilon \ct} \right) \erfcx \left( \frac{\cx}{\sqrt{4 \varepsilon \ct}} + \sqrt{\frac{\ct}{\varepsilon}} \right) ,
\end{align}
which directly yields the dimensionless thermal penetration depth as
\begin{align}
    \check{\delta}_{\rm th} ( \ct ) = \frac{\erfcx \sqrt{\frac{\ct}{\varepsilon}}}{\frac{1}{\sqrt{\pi} \sqrt{\varepsilon \ct}} - \frac{1}{\varepsilon} \erfcx \sqrt{\frac{\ct}{\varepsilon}} } .
\end{align}
To understand the initial transient behavior, we analyze the short-time regime under the condition $ \sqrt{\frac{\ct}{\varepsilon}} \ll 1 $ (implying $ \ct \ll \varepsilon \ll 1$ ), where the scaled complementary error function can be linearized. Under this approximation, the thermal penetration depth simplifies to
\begin{align}
    \nonumber
    \check{\delta}_{\rm th} ( \ct ) &\approx \frac{1 - \frac{2}{\sqrt{\pi}} \sqrt{\frac{\ct}{\varepsilon}}}{\frac{1}{\sqrt{\pi} \sqrt{\varepsilon \ct}} - \frac{1}{\varepsilon} \left( 1 - \frac{2}{\sqrt{\pi}} \sqrt{\frac{\ct}{\varepsilon}} \right)} = \varepsilon \sqrt{\pi} \sqrt{\frac{\ct}{\varepsilon}} \frac{1 - \frac{2}{\sqrt{\pi}} \sqrt{\frac{\ct}{\varepsilon}}}{1 - \sqrt{\pi} \sqrt{\frac{\ct}{\varepsilon}} + 2 \frac{\ct}{\varepsilon}} \\
    & \approx \varepsilon \sqrt{\pi} \sqrt{\frac{\ct}{\varepsilon}} \left( 1 - \frac{2}{\sqrt{\pi}} \sqrt{\frac{\ct}{\varepsilon}} \right) \left(1 + \sqrt{\pi} \sqrt{\frac{\ct}{\varepsilon}} \right) \approx \sqrt{\pi \varepsilon \ct} \left( 1 + \frac{\pi -2}{\sqrt{\pi}} \sqrt{\frac{\ct}{\varepsilon}} \right) .
\end{align}
The leading-order term $ \sqrt{\pi \varepsilon \ct} $ corresponds to classical diffusion, while the first-order correction captures the distortion induced by the bulk heating. 

Conversely, for longer times where $ \sqrt{\frac{\ct}{\varepsilon}} \gg 1 $, the asymptotic representation of the scaled error function can be applied. In this long-time regime, the thermal penetration depth behaves as
\begin{align}
    \label{eq:del-1-x}
    \check{\delta}_{\rm th} ( \ct ) &\approx \frac{\frac{1}{\sqrt{\pi} \sqrt{\frac{\ct}{\varepsilon}}}}{\frac{1}{\sqrt{\pi} \sqrt{\varepsilon \ct}} - \frac{1}{\varepsilon} \frac{1}{\sqrt{\pi} \sqrt{\frac{\ct}{\varepsilon}}}} \rightarrow \infty ,
\end{align}
confirming that as time progresses, the local diffusive perturbation expands in an unbounded domain and eventually fills the entire bulk domain, rendering the short-time semi-infinite assumption invalid.

\begin{figure}[!ht]
    \begin{center}
        \makebox[\textwidth][c]{%
            \begin{minipage}[c]{.5\textwidth} \centering
                \includegraphics[width=\linewidth]{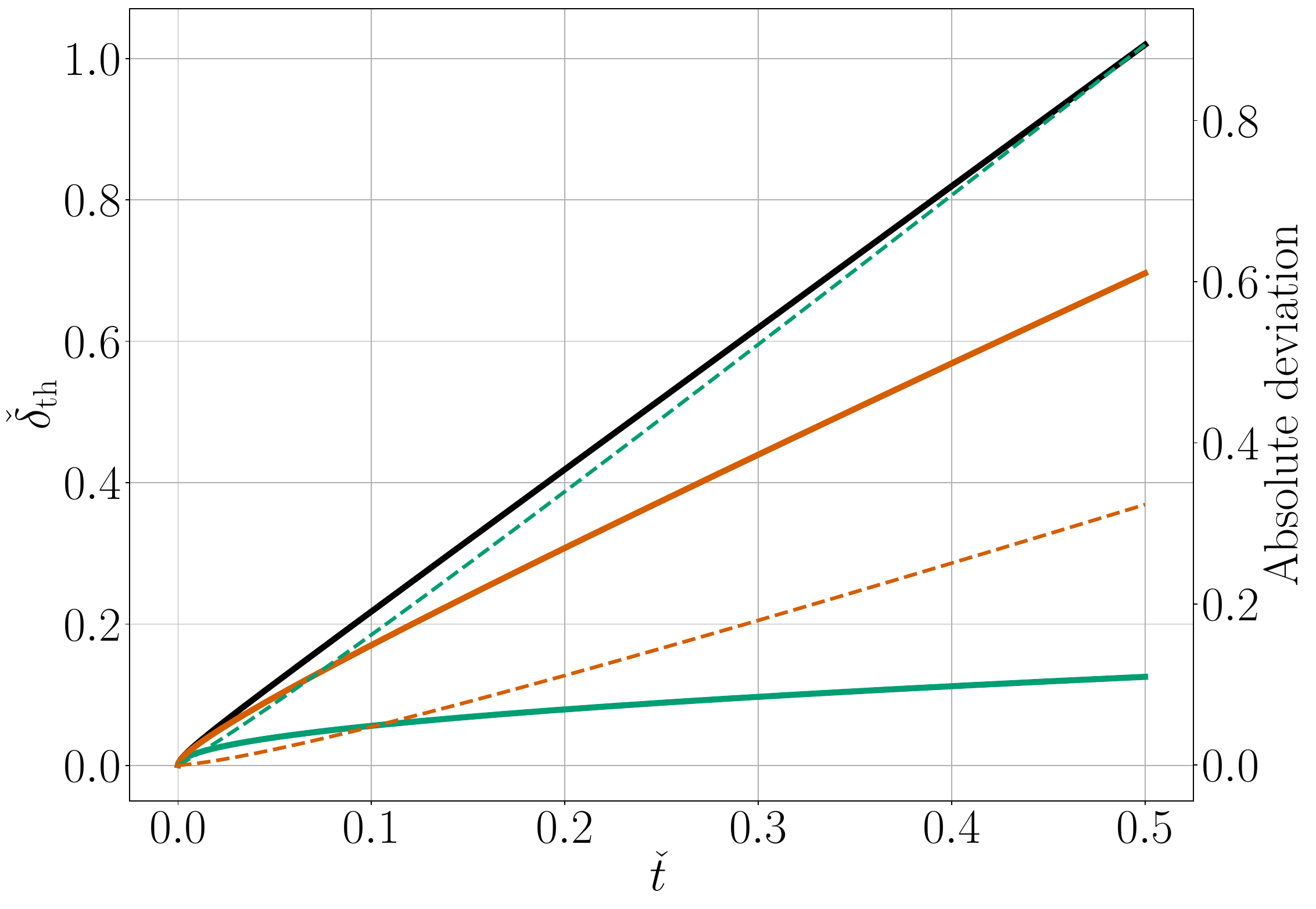}
            \end{minipage}%
            \hspace{.02\textwidth}%
            \begin{minipage}[c]{.35\textwidth} \centering
                \includegraphics[width=\linewidth]{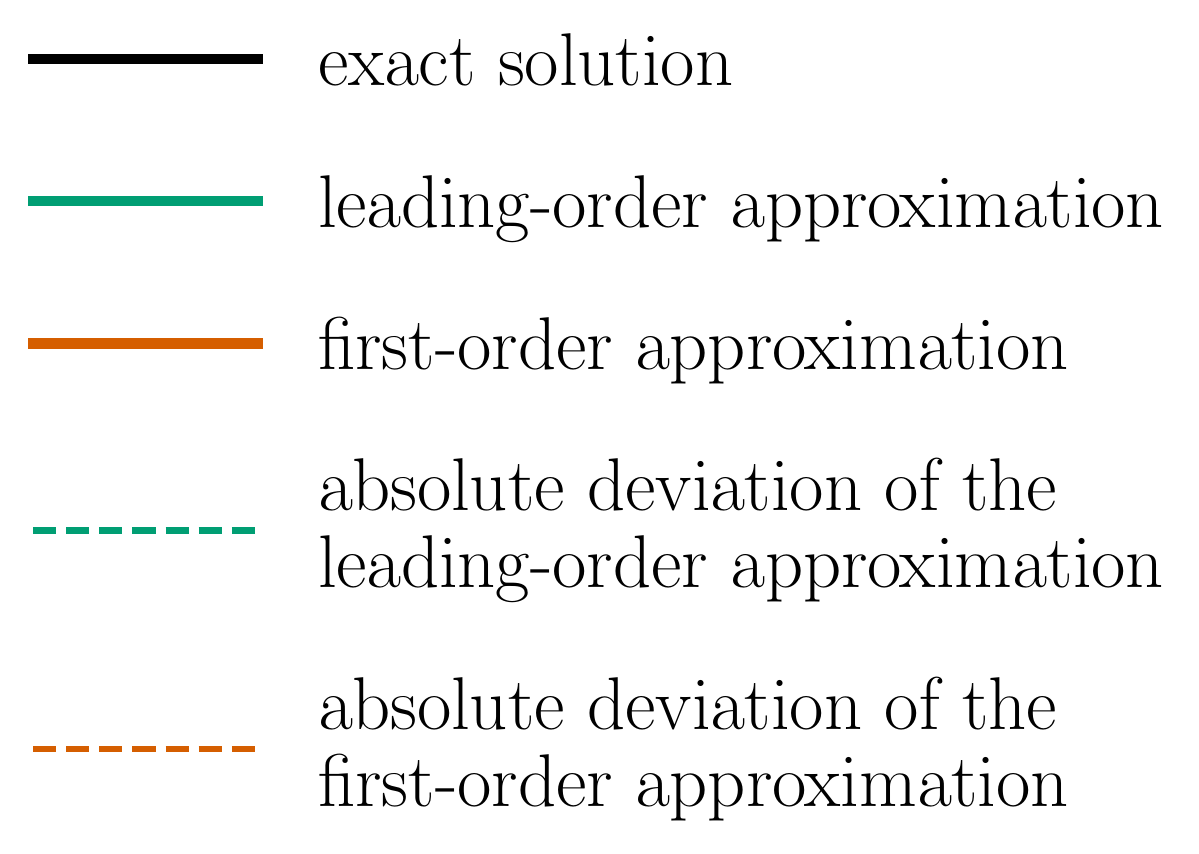}
            \end{minipage}%
            }
    \end{center}
    \caption{Evolution of the dimensionless thermal penetration depth (left axis) and the corresponding absolute deviation from the exact solution (right axis) for the planar configuration under a constant wall temperature. On the left axis, the exact solution is represented by the solid black line, the leading order approximation by the solid green line, and the first-order approximation by the solid red line. On the right axis, the dashed green line denotes the absolute deviation of the leading order approximation, while the dashed red line represents the absolute deviation of the first-order approximation.}
    \label{fig:del-1}
\end{figure}

This physical behavior and the performance of the approximations are demonstrated in Fig.~\ref{fig:del-1}. A key observation is that the exact dimensionless thermal penetration depth grows remarkably fast, reaching a value close to unity around $ \ct = 0.5 $. Physically, this indicates that the thermal boundary layer has already penetrated the entire characteristic length of the domain, meaning that the thermal disturbance has reached the opposite boundary. Once the disturbance crosses this geometric threshold, the semi-infinite medium assumption completely breaks down, and the fluid can no longer be treated as unbounded. Furthermore, the figure illustrates how the absolute deviations of the approximations evolve over time. The leading-order approximation tracks the initial phase well, but its absolute deviation grows steadily as the boundary layer expands. The first-order Taylor expansion provides a significantly lower absolute deviation during the early stages, but it also begins to diverge as the system transitions toward the long-time regime. This confirms that while the short-time expansions are highly precise at the very beginning of the process, their predictive capability is strictly capped by the geometric confinement of the container, which manifests around $ \ct \approx 0.5 $.

To analyze the behavior in the immediate vicinity of the heated boundary, the dimensionless temperature field \re{eq:dimless-sol-x-1} is expanded into a Taylor series around $ \cx = 0 $, which yields
\begin{align}
    \label{eq:T-x-1-near-wall}
    \cTheta ( \ct , \cx ) = 1 - \eta + \frac{\cx}{\varepsilon} \erfcx \sqrt{\frac{\ct}{\varepsilon}} 
\end{align}
with the scaled spatial variable $ \eta = \frac{\cx}{\sqrt{\pi \varepsilon \ct}} $ introduced by the leading-order term of the thermal penetration depth.

\begin{figure}[!ht]
    \begin{center}
        \makebox[\textwidth][c]{%
            \begin{minipage}[c]{.45\textwidth} \centering
                \includegraphics[width=\linewidth]{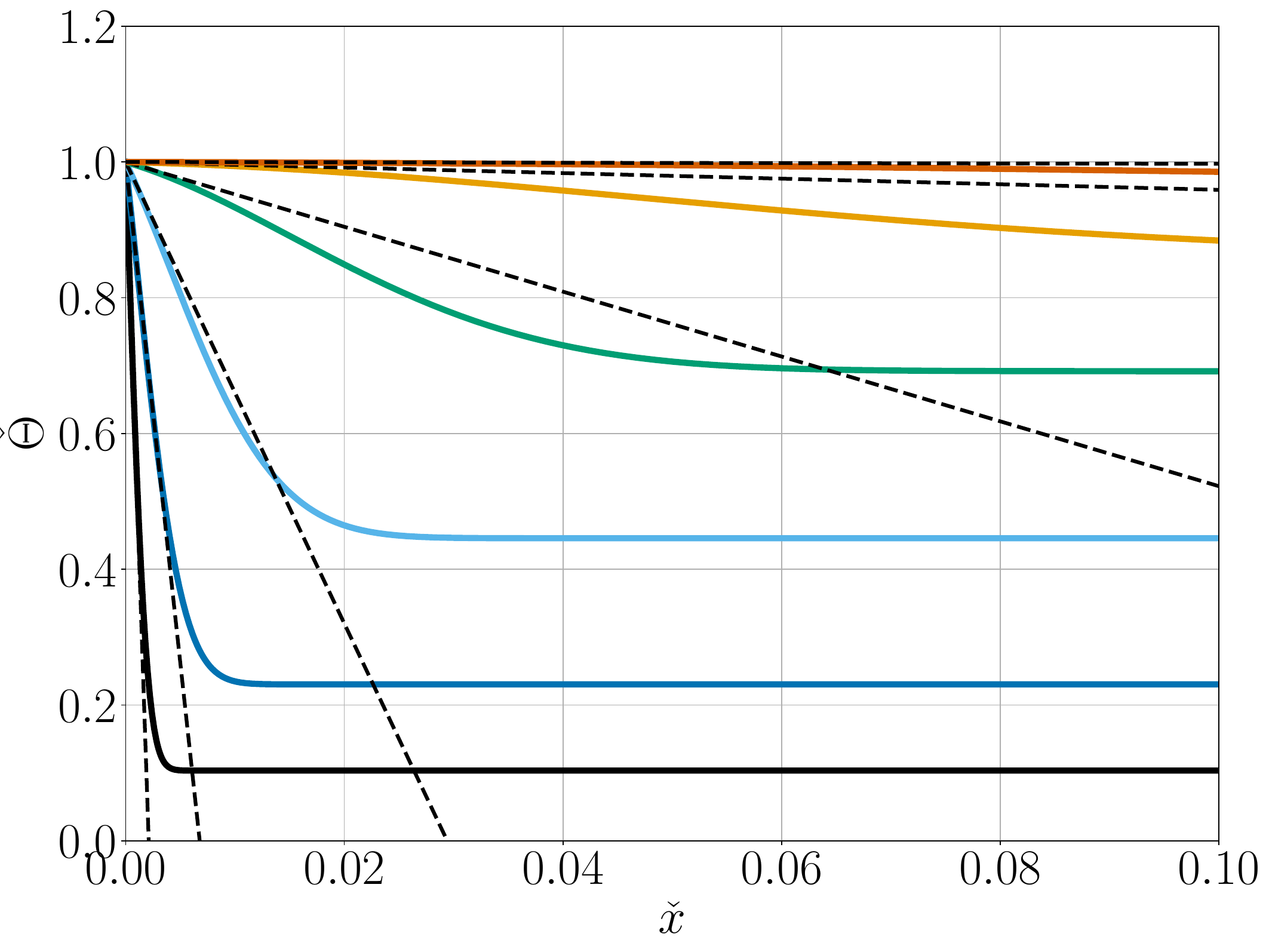}
            \end{minipage}%
            \hspace{.02\textwidth}%
            \begin{minipage}[c]{.32\textwidth} \centering
                \includegraphics[width=\linewidth]{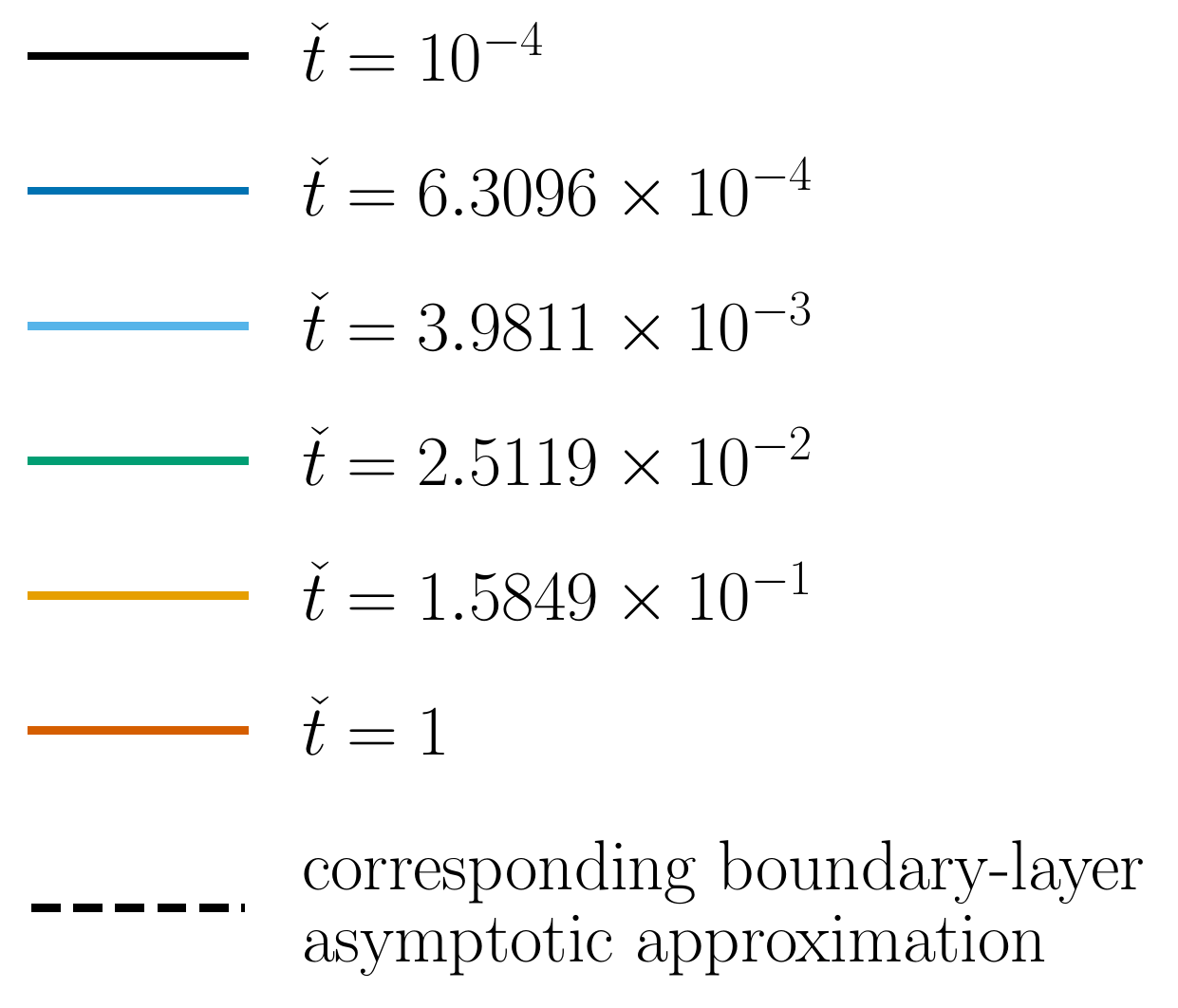}
            \end{minipage}%
            }
    \end{center}
    \caption{Dimensionless temperature distribution evaluated at logarithmically spaced time steps in the interval $ \ct \in [10^{-4}, 1] $ under a constant wall temperature for the planar configuration, where time progresses from the black to the red lines. The solid lines represent the exact analytical solution, and the dashed lines denote the near-wall asymptotics presented in \re{eq:T-x-1-near-wall}.}
    \label{fig:T-x-1-near-wall}
\end{figure}

The structural form of this near-wall expansion reveals that the temperature distribution is governed by the superposition of three distinct physical phenomena. The constant first term explicitly recovers the prescribed Dirichlet boundary condition at the wall. The negative scaled spatial term $ -\eta $ accounts for the classical thermal diffusion. Because $ \eta $ scales inversely with $ \sqrt{\ct} $, the spatial temperature gradient in the physical coordinate system approaches infinity as $ \ct \to 0 $, providing the physical explanation for the vertical drop of the temperature profiles observed at the earliest instances. Finally, the last term represents the dynamic feedback of the piston effect. Driven by the global adiabatic compression that uniformly heats the bulk fluid, this term progressively reduces the thermal driving force between the boundary and the bulk fluid, explaining the rapid temporal flattening of the local spatial gradient near the wall. As demonstrated in Fig.~\ref{fig:del-1}, the thermal boundary layer grows continuously and penetrates the entire finite domain rapidly, causing the fast breakdown of the semi-infinite approximation. The exact analytical solution together with these near-wall approximations is illustrated in Fig.~\ref{fig:T-x-1-near-wall}.

To provide further insight, the historically important solution adopted by Straub \etal and applied in the early Spacelab experiments \cite{straub1995dynamic} is now connected to our asymptotic findings. First, by returning to the dimensional formalism, a possible re-derivation of their formula is presented. It is assumed that the local thermal boundary layer retains its classical diffusion profile but is modulated by the dynamically reduced temperature difference $\Theta_{\rm w} - \Theta_{\rm b}(t)$, yielding the ansatz
\begin{align}
    \Theta_{\rm d} ( t , x ) = \left[ \Theta_{\rm w} - \Theta_{\rm b} ( t ) \right] \erfc \frac{x}{\sqrt{4 a t}} .
\end{align}
Substituting this ansatz into the governing equation of the bulk temperature rise \re{eq:piston-ode} leads to the ordinary differential equation
\begin{align}
    \dt{\Theta_{\rm b}} = \frac{\ell}{\tau_{\rm p}} \frac{\Theta_{\rm w} - \Theta_{\rm b} ( t )}{\sqrt{ \pi a t}} ,
\end{align}
the direct integration of which gives the bulk temperature rise as
\begin{align}
    \Theta_{\rm b} ( t ) = \Theta_{\rm w} \left[ 1 - \exp \left( - \sqrt{\frac{4}{\pi} \frac{\tau_{\rm d}}{\tau_{\rm p}} \frac{t}{\tau_{\rm p}}} \right) \right] .
\end{align}
Consequently, the decoupled diffusive contribution can be formulated as
\begin{align}
    \Theta_{\rm d} ( t , x ) = \Theta_{\rm w} \exp \left( - \sqrt{\frac{4}{\pi} \frac{\tau_{\rm d}}{\tau_{\rm p}} \frac{t}{\tau_{\rm p}}} \right) \erfc \frac{x}{\sqrt{4 a t}} .
\end{align}
Combining these terms yields the total temperature field expressed as
\begin{align}
    \Theta ( t , x ) = \Theta_{\rm w} \left[ 1 - \exp \left( - \sqrt{\frac{4}{\pi} \frac{\tau_{\rm d}}{\tau_{\rm p}} \frac{t}{\tau_{\rm p}}} \right) \erf \frac{x}{\sqrt{4 a t}} \right] ,
\end{align}
which can be rearranged into the form\footnote{It is worth noting that the original formulation presented in \cite{straub1995process} erroneously utilized the complementary error function $ \erfc $ in the final superposition steps instead of the standard error function $ \erf $. This typo leads to a formal violation of the prescribed Dirichlet boundary condition at $ x = 0 $, a mathematical inconsistency corrected here by enforcing the strict definition of the additive field components.}
\begin{align}
    \frac{\Theta_{\rm w} - \Theta ( t , x )}{\Theta_{\rm w}} = \exp \left( - \sqrt{\frac{4}{\pi} \frac{\tau_{\rm d}}{\tau_{\rm p}} \frac{t}{\tau_{\rm p}}} \right) \erf \frac{x}{\sqrt{4 a t}} 
\end{align}
as reported in \cite{straub1995process}. By applying \re{eq:nondim-var} and \re{eq:nondim-fun}, the non-dimensional form of the solution reported by Straub \etal is expressed as
\begin{align}
    \label{eq:straub-nondim}
    \cTheta_{\rm Straub} \left( \ct , \cx \right) &= 1 - \exp \left( - \sqrt{\frac{4}{\pi} \frac{\ct}{\varepsilon}} \right) \erf \frac{\cx}{\sqrt{4 \varepsilon \ct}} .
\end{align}

Now, this historical formulation is directly compared with our exact solution \re{eq:dimless-sol-x-1}. Initially, when $ \sqrt{\frac{\ct}{\varepsilon}} \ll 1 $, the first-order approximation of \re{eq:straub-nondim} is given by
\begin{align}
    \cTheta_{\rm Straub} ( \ct , \cx ) \approx \erfc \frac{\cx}{\sqrt{4 \varepsilon \ct}} + \frac{2}{\sqrt{\pi}} \sqrt{\frac{\ct}{\varepsilon}} \erf \frac{\cx}{\sqrt{4 \varepsilon \ct}} .
\end{align}
Our dimensionless solution \re{eq:dimless-sol-x-1} can be reformulated as
\begin{align}
    \cTheta ( \ct , \cx ) = 1 - \erfcx \sqrt{\frac{\ct}{\varepsilon}} + \exp \left( \sqrt{\frac{\ct}{\varepsilon}} \left( 2 \frac{\cx}{\sqrt{4 \varepsilon \ct}} + \sqrt{\frac{\ct}{\varepsilon}} \right) \right) \erfc \left( \frac{\cx}{\sqrt{4 \varepsilon \ct}} + \sqrt{\frac{\ct}{\varepsilon}} \right) .
\end{align}
Assuming that the scaling variable $\frac{\cx}{\sqrt{4 \varepsilon \ct}}$ remains bounded, the first-order Taylor expansion of this expression yields
\begin{align}
    \cTheta ( \ct , \cx ) \approx \erfc \frac{\cx}{\sqrt{4 \varepsilon \ct}} + \frac{2}{\sqrt{\pi}} \sqrt{\frac{\ct}{\varepsilon}} \left( 1 - \sqrt{\pi} \ierfc \frac{\cx}{\sqrt{4 \varepsilon \ct}} \right).
\end{align}

\begin{figure}[!ht]
    \begin{center}
        \makebox[\textwidth][c]{%
            \begin{minipage}[c]{.45\textwidth} \centering
                \includegraphics[width=\linewidth]{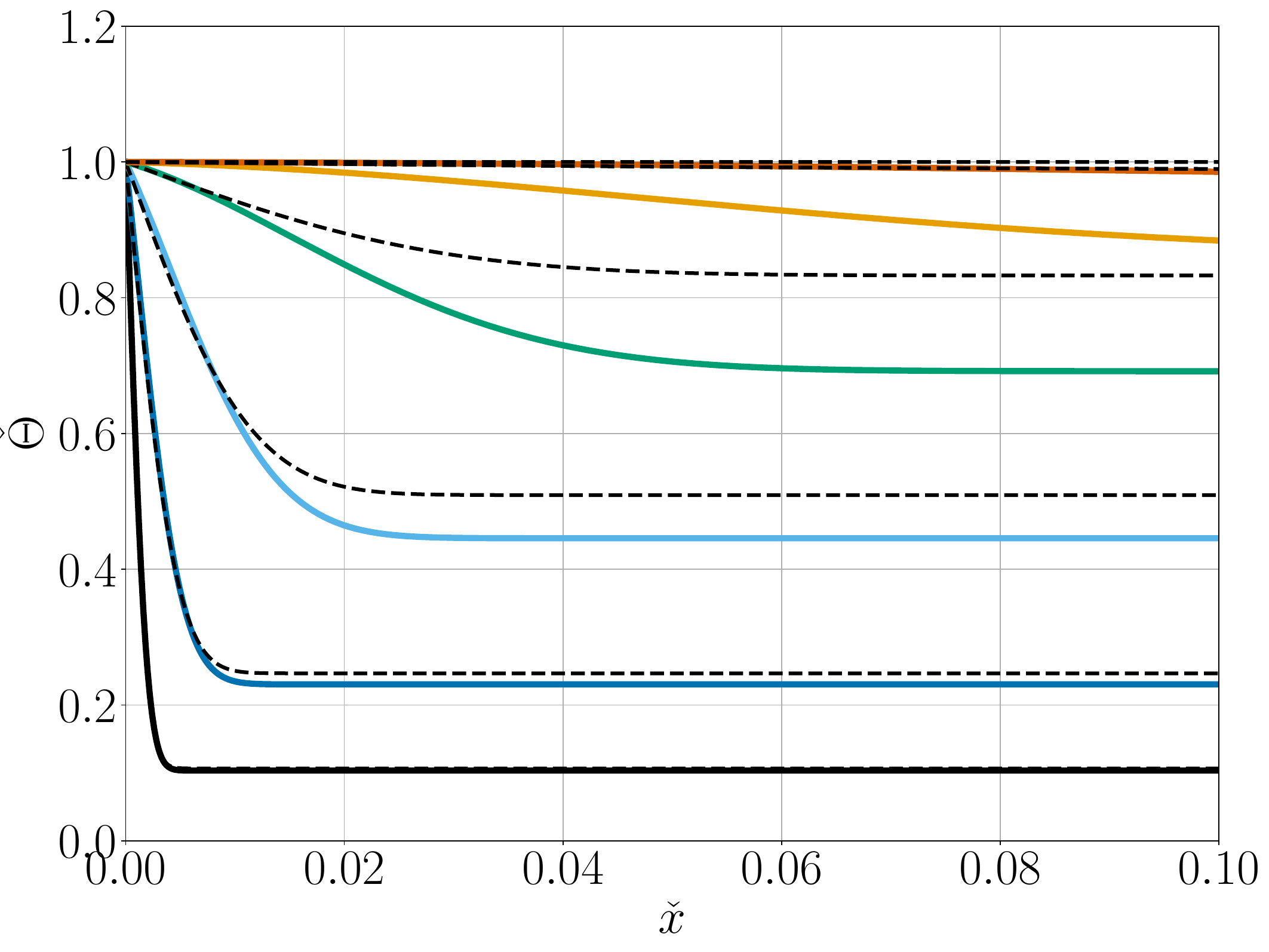}
            \end{minipage}%
            \hspace{.02\textwidth}%
            \begin{minipage}[c]{.32\textwidth} \centering
                \includegraphics[width=\linewidth]{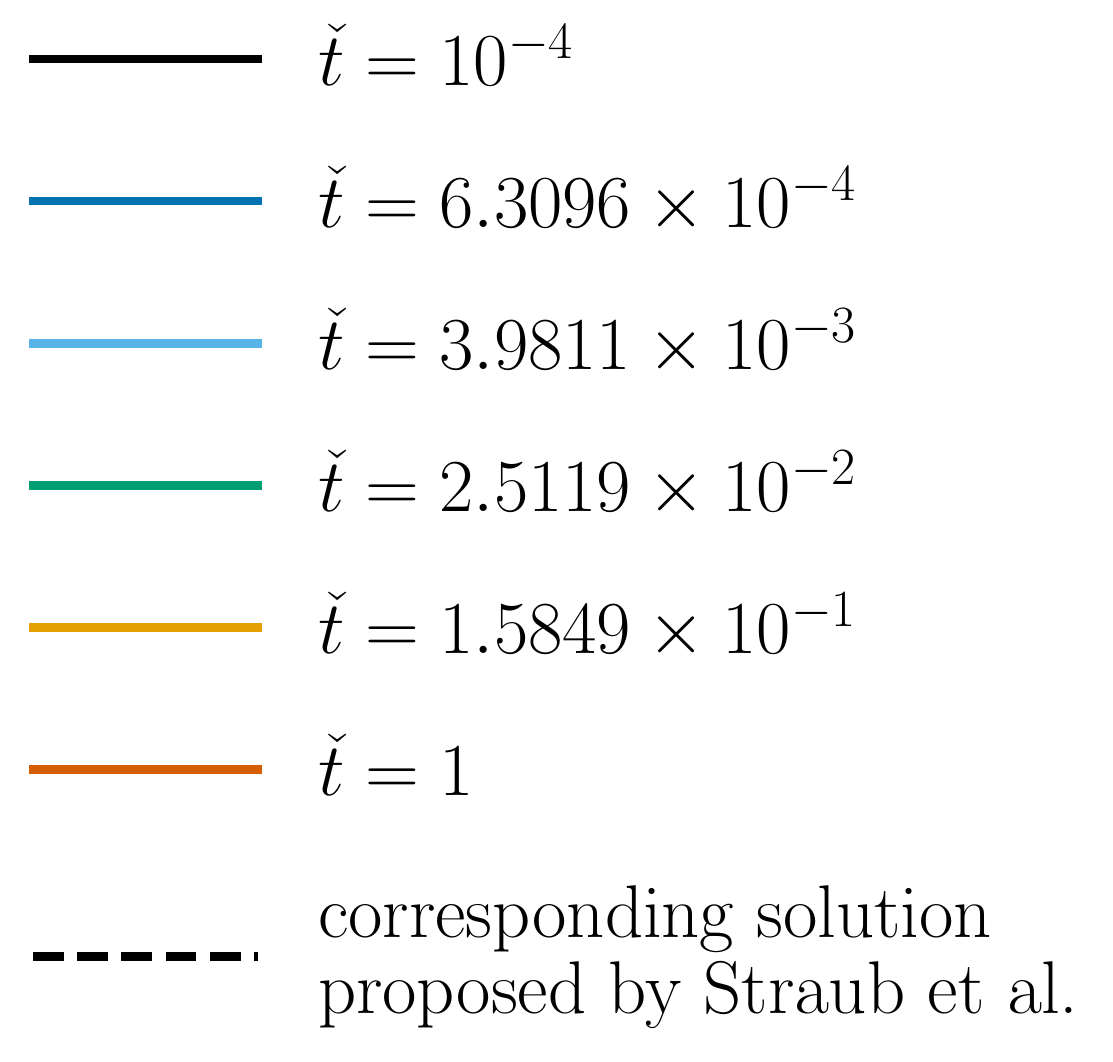}
            \end{minipage}%
            }
    \end{center}
    \caption{Comparison of the dimensionless temperature distribution evaluated at logarithmically spaced time steps in the interval $ \ct \in [10^{-4}, 1] $ under a constant wall temperature for the planar configuration, where time progresses from the black to the red lines. The solid lines represent our exact analytical solution \re{eq:dimless-sol-x-1}, and the dashed lines denote the approximation given in \re{eq:straub-nondim}.}
    \label{fig:T-Straub}
\end{figure}

This asymptotic formalism is clearly supported by the results presented in Fig.~\ref{fig:T-Straub}, which directly compares the exact analytical solution \re{eq:dimless-sol-x-1} with the model reported by Straub \etal \re{eq:straub-nondim}. At the lowest curves representing the very early instances, both formulations run perfectly together because their leading-order structures are identical, verifying that the solution reported by Straub \etal captures the correct initial diffusion state. The reason why the intermediate curves differ significantly lies in the difference between these first-order correction terms, highlighting a fundamental distinction in how the physical coupling is treated. The model reported by Straub \etal considers only a temporal coupling, where the rising bulk temperature purely modulates the boundary driving force over time, but the local boundary layer cannot alter its internal spatial structure to feedback into the bulk dynamics. In this configuration, the correction scales with the $ \erf \frac{\cx}{\sqrt{4 \varepsilon \ct}} $ term. In contrast, our exact formulation incorporates a full spatiotemporal coupling yielding the bracketed term $ 1 - \sqrt{\pi} \ierfc \frac{\cx}{\sqrt{4 \varepsilon \ct}} $. Omitting this dynamic spatial profile change causes the model reported by Straub \etal to underestimate the local thermal resistance, thereby forcing a seemingly accelerated thermalization that manifests as an overshoot with respect to our exact solution.

\begin{figure}[!ht]
    \centering
    \includegraphics[width=.55\textwidth]{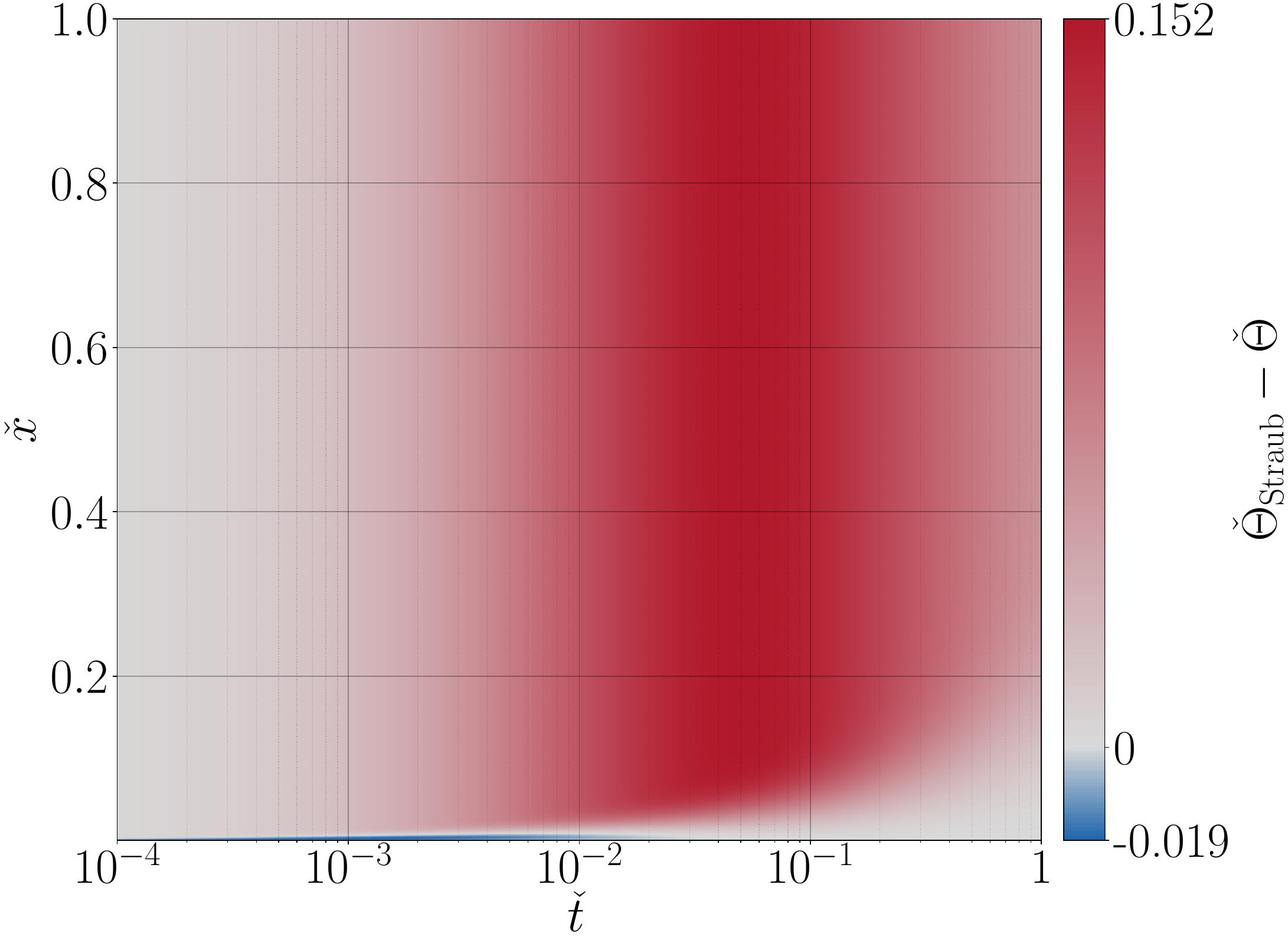}
    \caption{Dimensionless temperature deviation between the solution proposed by Straub \etal and the present exact analytical solution, $ \cTheta_{\rm Straub} ( \ct , \cx ) - \cTheta ( \ct , \cx ) $, evaluated over the dimensionless time interval $ \ct \in [ 10^{-4} , 1 ] $ and the spatial domain $ \cx \in [ 0 , 1 ] $. The deviation remains small during the earliest stage of the transient but subsequently develops into a spatially extended region of overprediction.}
    \label{fig:T-Straub-deviation}
\end{figure}

To quantify the magnitude, spatial distribution, and temporal evolution of this discrepancy, we further investigate the dimensionless temperature deviation $ \cTheta_{\rm Straub} ( \ct , \cx ) - \cTheta ( \ct , \cx ) $. Since both temperature fields are normalized by the prescribed wall-temperature difference, this quantity directly represents the deviation relative to the imposed thermal scale. Positive values indicate overprediction by the formulation of Straub \etal, whereas negative values indicate underprediction. The resulting deviation field is presented in Fig.~\ref{fig:T-Straub-deviation} over the complete investigated spatial domain and across four orders of magnitude in dimensionless time. At the earliest times, the deviation remains small, consistently with the identical leading-order asymptotic behavior of the two formulations. Subsequently, a spatially extended region of positive deviation develops and propagates into the fluid domain. The maximum overprediction reaches approximately $ 0.152 $, corresponding to about $ 15\% $ of the prescribed wall-temperature difference. A comparatively small underprediction also appears near the wall at early dimensionless times, but its magnitude remains below approximately $ 2\% $. Overall, the comparison reveals a dominant overprediction, mainly in the bulk, confirming that the spatially decoupled formulation predicts an excessively rapid thermal response once the first-order spatiotemporal correction to the diffusive temperature field becomes significant.

Now, we present the dimensionless temperature field for the spherical configuration subject to a prescribed wall temperature. Although the resulting closed-form expression is lengthy due to the simultaneous presence of curvature effects and piston-effect coupling, it follows directly from the previously derived dimensional solutions by straightforward substitution and nondimensionalization. Specifically, substituting the roots $ \mu_\pm $ from \re{eq:piston-Tb-s-bc1-mu-r} into the sum of \re{eq:sol-sph-bulk} and \re{eq:sol-sph-diff}, and subsequently introducing the dimensionless variables defined in \re{eq:nondim-var} and \re{eq:nondim-fun}, yields the dimensionless temperature field in the spherical geometry as
\begin{align}
    \nonumber
    \cTheta ( \ct , \cz ) = 1 &- \frac{1}{2} \left( \frac{1}{\sqrt{1 + 4 \varepsilon}} + 1 \right) \erfcx \left( \frac{1}{2} \left( 1 + \sqrt{1 + 4 \varepsilon} \right) \sqrt{\frac{\ct}{\varepsilon}} \right) \\
    \nonumber
    &+ \frac{1}{2} \left( \frac{1}{\sqrt{1 + 4 \varepsilon}} - 1 \right) \erfcx \left( \frac{1}{2} \left( 1 - \sqrt{1 + 4 \varepsilon} \right) \sqrt{\frac{\ct}{\varepsilon}} \right) \\
    \nonumber
    &- \frac{\exp\left( - \frac{\cz^2}{4 \varepsilon \ct}\right)}{1 - \cz} \Bigg[ \frac{1}{2} \left( \frac{1}{\sqrt{1 + 4 \varepsilon}} + 1 \right) \erfcx \left( \frac{\cz}{\sqrt{4 \varepsilon \ct}} + \frac{1}{2} \left( 1 + \sqrt{1 + 4 \varepsilon} \right) \sqrt{\frac{\ct}{\varepsilon}} \right) \\
    \label{eq:dimless-sol-r-1}
    &\hskip 15ex - \frac{1}{2} \left( \frac{1}{\sqrt{1 + 4 \varepsilon}} - 1 \right) \erfcx \left( \frac{\cz}{\sqrt{4 \varepsilon \ct}} + \frac{1}{2} \left( 1 - \sqrt{1 + 4 \varepsilon} \right) \sqrt{\frac{\ct}{\varepsilon}} \right) \Bigg] .
\end{align}
Since $ \varepsilon \ll 1 $, the first-order expansions $ \sqrt{1 + 4 \varepsilon} \approx 1 + 2 \varepsilon $ and $ \frac{1}{\sqrt{1 + 4 \varepsilon}} \approx 1 - 2 \varepsilon $ can be introduced, which simplifies the expression to
\begin{align}
    \nonumber
    \cTheta ( \ct , \cz ) \approx 1 &- ( 1 - \varepsilon ) \erfcx \left( \sqrt{\frac{\ct}{\varepsilon}} + \sqrt{\varepsilon \ct} \right) - \varepsilon \erfcx \left( - \sqrt{\varepsilon \ct} \right) \\
    \label{eq:dimless-sol-r-1-app}
    & + \frac{\exp\left( - \frac{\cz^2}{4 \varepsilon \ct}\right)}{1 - \cz} \left[ (1 - \varepsilon) \erfcx \left( \frac{\cz}{\sqrt{4 \varepsilon \ct}} + \sqrt{\frac{\ct}{\varepsilon}} + \sqrt{\varepsilon \ct} \right) + \varepsilon \erfcx \left( \frac{\cz}{\sqrt{4 \varepsilon \ct}} - \sqrt{\varepsilon \ct} \right) \right] .
\end{align}
Instead of directly analyzing the complicated structure of \re{eq:dimless-sol-r-1}, we rely on the investigation of the simplified form \re{eq:dimless-sol-r-1-app}. The thermal penetration depth is then obtained as
\begin{align}
    \label{eq:del-1-r}
    \check{\delta}_{\rm th} ( \ct ) = \frac{( 1 - \varepsilon) \erfcx \left( \sqrt{\frac{\ct}{\varepsilon}} + \sqrt{\varepsilon \ct} \right) + \varepsilon \erfcx \left( - \sqrt{\varepsilon \ct} \right) }{\frac{1}{\sqrt{\pi} \sqrt{\varepsilon \ct}} - \frac{1}{\varepsilon} \left( 1 + \varepsilon - 2 \varepsilon^2 \right) \erfcx \left( \sqrt{\frac{\ct}{\varepsilon}} + \sqrt{\varepsilon \ct} \right) } ,
\end{align}
which proves to be equivalent to the expression calculated for the planar case up to the zeroth order of $ \varepsilon $ [cf.\ \re{eq:del-1-x}]. This agreement is expected since near the heated boundary, the curvature effect can be linearized to establish a planar-like boundary layer approximation. For sufficiently small $ \varepsilon $, the approximated solution \re{eq:dimless-sol-r-1-app} visually approaches the planar limits. In these limits, the terms where $ \varepsilon $ appears outside the denominators asymptotically vanish as $ \varepsilon \to 0 $. A more precise derivation for this correspondence can be achieved in the short-time regime where $ \sqrt{\varepsilon \ct} \ll 1 $ (corresponding to $ t \ll \tau_{\rm d} $). Performing an asymptotic expansion and neglecting higher-order terms proportional to $ \varepsilon $ yields
\begin{align}
    \nonumber
    \cTheta ( \ct , \cz ) &\approx 1 - \erfcx \sqrt{\frac{\ct}{\varepsilon}} + \exp \left(- \frac{\cz^2}{4 \varepsilon \ct} \right) \erfcx \left( \frac{\cz}{\sqrt{4 \varepsilon \ct}} + \sqrt{\frac{\ct}{\varepsilon}} \right) + \frac{2}{\sqrt{\pi}} \sqrt{\varepsilon \ct} \left[ 1 - \exp \left( - \frac{\cz^2}{4 \varepsilon \ct} \right) \right] \\
    \label{eq:dimless-sol-r-1-app-elegant}
    & \hskip 5ex - 2 \ct \erfcx \sqrt{\frac{\ct}{\varepsilon}} + \left( \cz + 2 \ct \right) \exp \left(- \frac{\cz^2}{4 \varepsilon \ct} \right) \erfcx \left( \frac{\cz}{\sqrt{4 \varepsilon \ct}} + \sqrt{\frac{\ct}{\varepsilon}} \right) .
\end{align}
Comparing this result with \re{eq:dimless-sol-x-1} demonstrates that the leading-order approximation is fully and elegantly recovered by the planar solution. Furthermore, if the condition $ \sqrt{\frac{\ct}{\varepsilon}} \ll 1 $ also holds, the thermal penetration depth can be approximated as
\begin{align}
    \label{eq:del-1-r-app}
    \check \delta_{\rm th} ( \ct ) \approx \sqrt{\pi \varepsilon \ct} \left[ 1 + \frac{\pi - 2}{\sqrt{\pi}} \left( 1 + 2 \ct \right) \sqrt{\frac{\ct}{\varepsilon}} \right] .
\end{align}

\begin{figure}[!ht]
    \begin{center}
        \makebox[\textwidth][c]{%
            \begin{minipage}[c]{.45\textwidth} \centering
                \includegraphics[width=\linewidth]{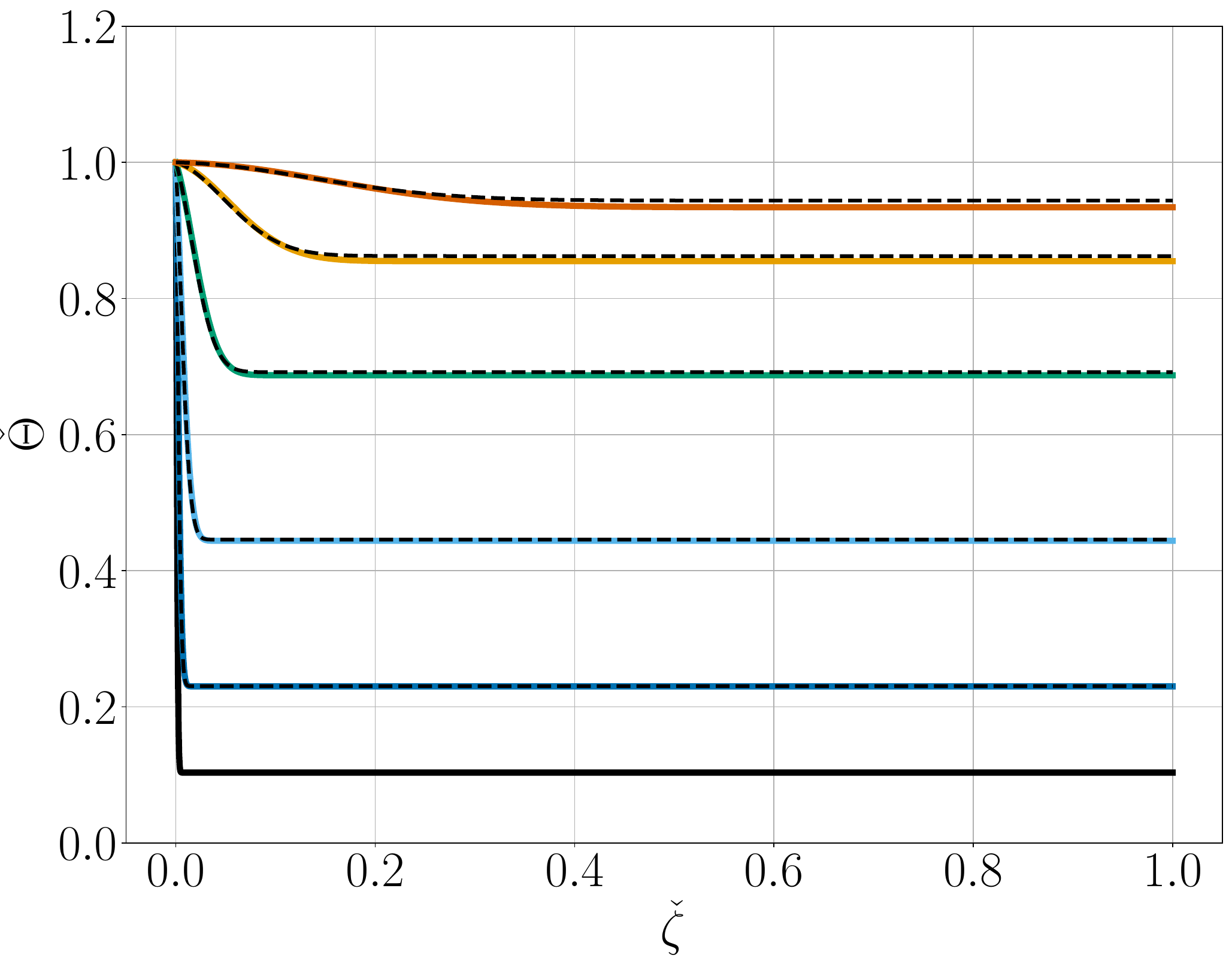}
            \end{minipage}%
            \hspace{.02\textwidth}%
            \begin{minipage}[c]{.32\textwidth} \centering
                \includegraphics[width=\linewidth]{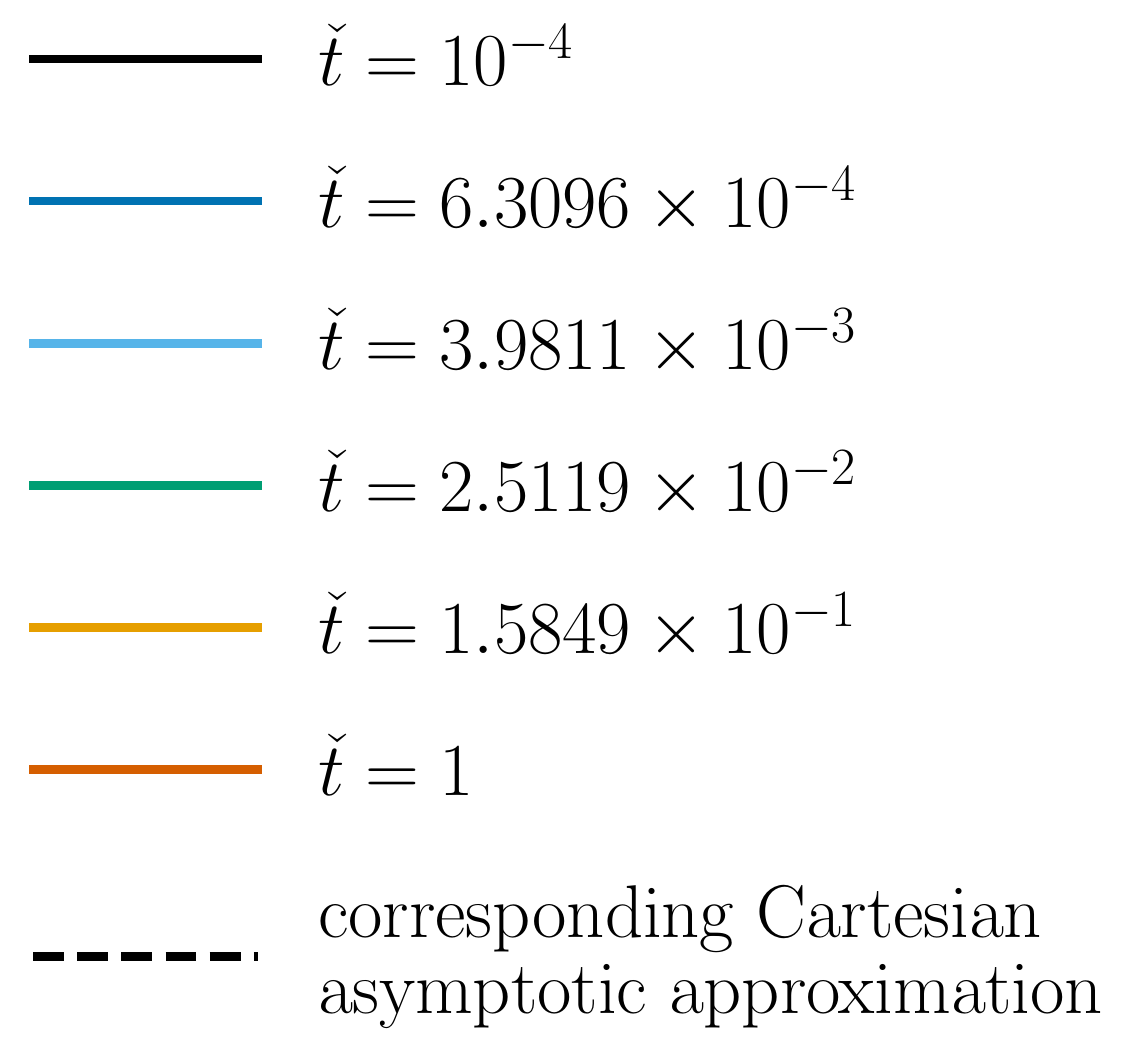}
            \end{minipage}%
            }
    \end{center}
    \caption{Comparison of the dimensionless temperature fields evaluated at logarithmically spaced time steps in the interval $ \ct \in [10^{-4}, 1] $ under a constant wall temperature. The solid lines represent the exact spherical solution (time progresses from the black to the red lines), and the dashed lines denote the Cartesian approximation.}
    \label{fig:T-1-x-vs-r}
\end{figure}

An analysis of the temperature profiles presented in Fig.~\ref{fig:T-1-x-vs-r} demonstrates excellent agreement between the exact spherical solution and the planar approximation. This close correspondence is particularly striking during the short-time transients within the thermal boundary layer, where the curvature effects are minimal. Consequently, these results rigorously justify that near the heated boundary, the curvature can be successfully linearized, validating the accuracy of the planar boundary layer representation for the short-time response.

\begin{figure}[!ht]
    \begin{center}
        \makebox[\textwidth][c]{%
            \begin{minipage}[c]{.5\textwidth} \centering
                \includegraphics[width=\linewidth]{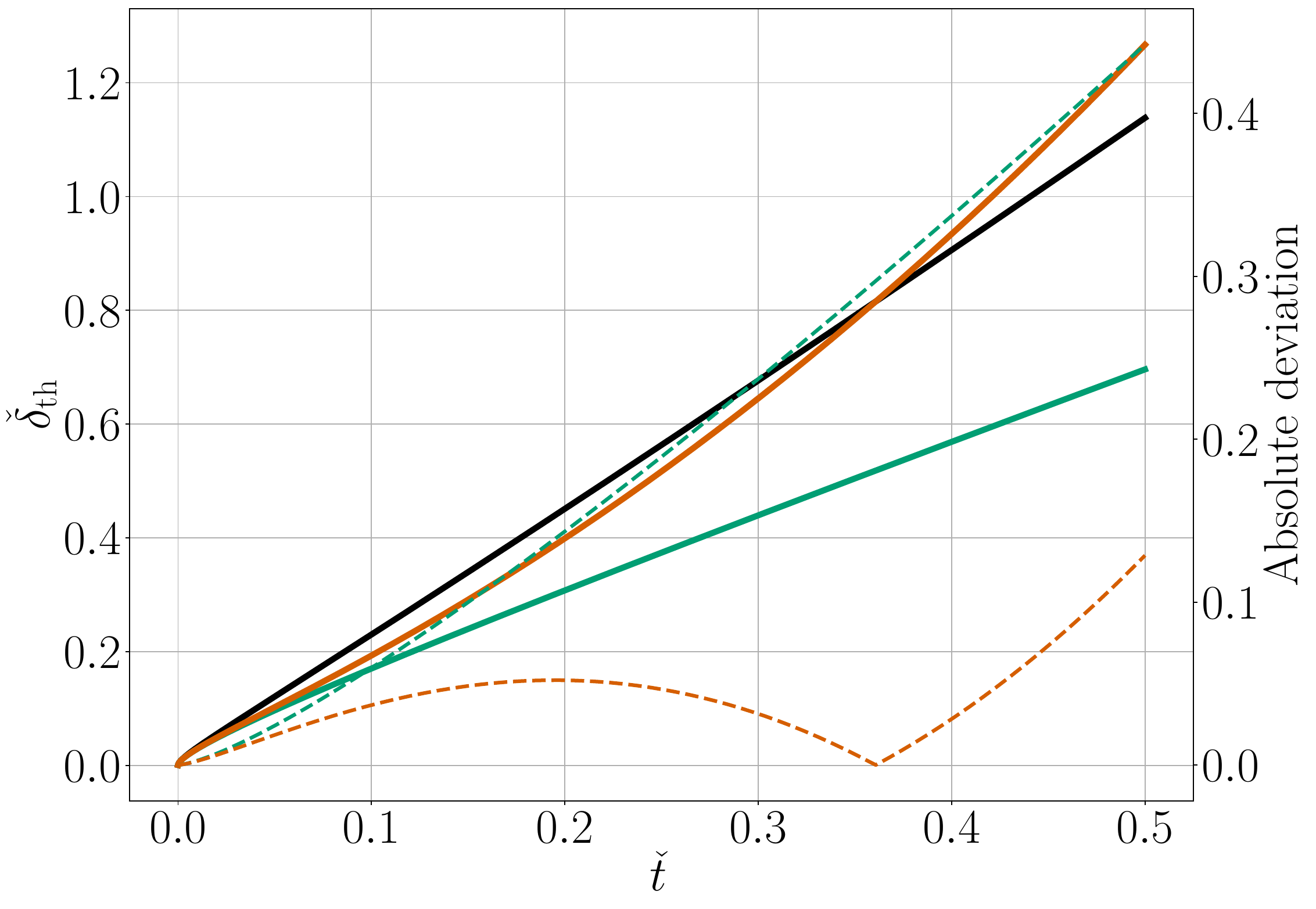}
            \end{minipage}%
            \hspace{.02\textwidth}%
            \begin{minipage}[c]{.35\textwidth} \centering
                \includegraphics[width=\linewidth]{del-2-legend.pdf}
            \end{minipage}%
            }
    \end{center}
    \caption{Evolution of the dimensionless thermal penetration depth (left axis) and the corresponding absolute deviation from the exact solution (right axis) for the spherical configuration under a constant wall temperature. On the left axis, the exact spherical solution is represented by the solid black line, while the leading-order and first-order approximations are shown by the solid green and orange lines, respectively. On the right axis, the corresponding absolute deviations are represented with dashed lines.}
    \label{fig:del-1-r}
\end{figure}

The temporal evolution of the thermal penetration depth presented in Fig.~\ref{fig:del-1-r} confirms that incorporating higher-order terms in the asymptotic expansion significantly extends the accuracy of the approximation over longer intervals. While the leading-order planar solution begins to deviate early from the exact spherical curve, the first-order correction captures the geometric curvature successfully, tracking the exact behavior with minimal absolute error across the entire investigated dimensionless time range.

\begin{figure}[!ht]
    \centering
    \begin{center}
        \makebox[\textwidth][c]{%
            \begin{minipage}[c]{.45\textwidth} \centering
                \includegraphics[width=\linewidth]{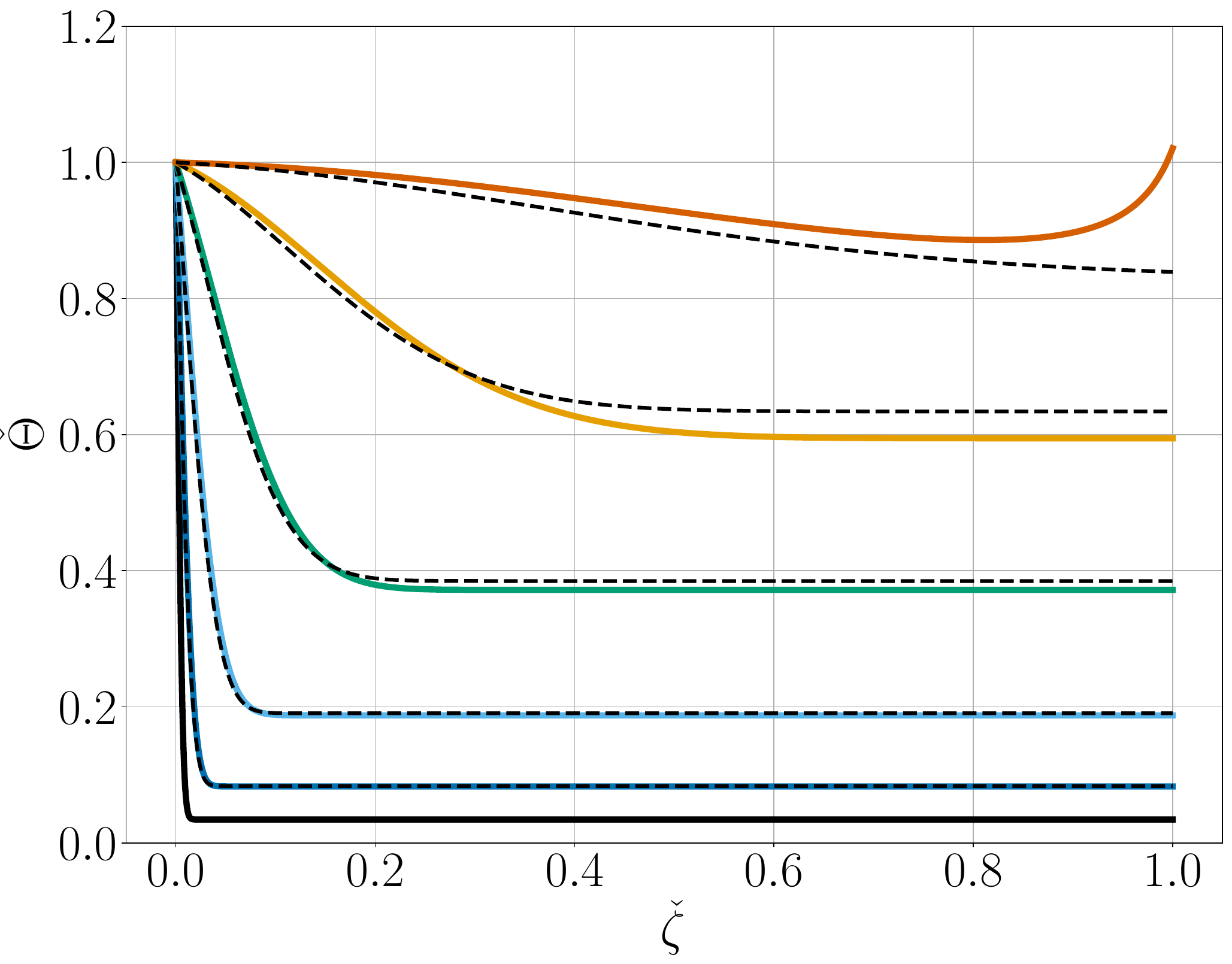}
            \end{minipage}%
            \hspace{.02\textwidth}%
            \begin{minipage}[c]{.32\textwidth} \centering
                \includegraphics[width=\linewidth]{T-x-1-x-vs-r-legend.pdf}
            \end{minipage}%
            }
    \end{center}
    \caption{Comparison of the dimensionless temperature fields evaluated at logarithmically spaced time steps in the interval $\ct \in [10^{-4}, 1]$ under a constant wall temperature with $ \varepsilon = 0.1 $. The solid lines represent the exact spherical solution, where time progresses from the black to the red lines, and the dashed lines denote the Cartesian approximation. Note that the final red profile represents a regime where the solution is no longer physically valid due to the thermal front reaching the center of the sphere.}
    \label{fig:T-1-x-vs-r-mod}
\end{figure}

Finally, Fig.~\ref{fig:T-1-x-vs-r-mod} highlights the absolute mathematical importance of the smallness of the small parameter $\varepsilon$ in maintaining the fidelity of the asymptotic framework. In this figure, a comparison of the temperature profiles evaluated at a larger small parameter of $ \varepsilon = 0.1 $ is presented, demonstrating more pronounced structural deviations between the two configurations. At the final dimensionless time step ($ \ct = 1 $), the exact spherical solution exhibits a distinct upward curvature near the center of the domain ($ \cz \to 1 $). This behavior indicates that the boundary perturbation has fully reached the core of the sphere and renders the model physically invalid at this terminal stage.

\subsection{Summary and physical interpretation of the asymptotic results}

For all configurations considered above, the thermal penetration depth recovers the classical diffusive behavior to leading order and therefore scales as $ \sqrt{t} $. This common behavior reflects the fact that, at the earliest stages of the process, thermal transport remains confined to a thin region adjacent to the heated wall and is governed primarily by local diffusion. The effects of piston-effect-induced bulk heating, the imposed boundary condition, and the curvature of the spherical domain enter through the subsequent correction terms.

Under a prescribed wall heat flux, the planar penetration depth retains the classical diffusive form, whereas the spherical result contains an additional curvature correction. Specifically, the leading-order dimensionless penetration depth is proportional to $ \sqrt{\varepsilon \check{t}} $, and its relative curvature correction is also of order $ \sqrt{\varepsilon \check{t}} $. After multiplication by the leading term, the first additive correction is therefore proportional to $ \varepsilon \check{t} $. Physically, the prescribed heat flux maintains a constant wall-normal temperature gradient, while the penetration depth increases as diffusion transports the thermal disturbance farther into the fluid. The curvature correction remains perturbatively small as long as $ \sqrt{\varepsilon \check{t}} \ll 1 $, consistently with the short-time condition $ t \ll \tau_{\rm d} $.

Under a prescribed wall temperature, the dimensionless penetration depth also follows the classical diffusive scaling $ \sqrt{\varepsilon \check{t}} $ to leading order. However, its first relative correction is proportional to $ \sqrt{\frac{\check{t}}{\varepsilon}} $, so that the corresponding first additive correction is proportional to $ \check{t} $. In contrast to the prescribed-flux case, this correction is already present in the planar configuration and originates from the dynamically changing thermal driving force. As the piston effect raises the spatially uniform bulk temperature, the instantaneous temperature difference between the wall and the bulk continuously decreases, modifying the temperature gradient that determines the penetration depth. The factor $ \frac{1}{\sqrt{\varepsilon}} $ in the relative correction indicates that this feedback becomes increasingly significant near the critical point. Consequently, the asymptotic expansion requires the more restrictive condition $ \check{t} \ll \varepsilon $, and deviations from the leading-order diffusive behavior become appreciable earlier than under a prescribed wall heat flux.

The relationship between the planar and spherical geometries is governed by the ratio of the growing penetration depth to the radius. At sufficiently short times, the thermal layer is much thinner than the radius of the sphere, and the spherical boundary is locally indistinguishable from a plane. Consequently, the planar and spherical solutions coincide to leading order. As the thermal penetration depth increases, however, the inward-propagating disturbance encounters concentric fluid shells of continuously decreasing surface area. The transferred heat is therefore distributed over progressively smaller fluid volumes, producing a geometric concentration that modifies the local temperature gradient and accelerates the inward thermal penetration relative to the planar configuration. Consequently, curvature effects become increasingly important as the thermal penetration depth grows relative to the radius of the sphere, leading to progressively larger deviations from the planar approximation.

\section{Transient response considering the thermal inertia of the container wall}
\label{sec:spacelab}

When modeling an experimental configuration, the idealization of an instantly prescribed constant wall temperature or a strictly constant heat flux may result in a predicted temperature response that differs strongly from the actually measured values.

Regarding the microgravity experiments performed during the Spacelab D-2 mission \cite{straub1995dynamic}, the heat pulse experiments\footnote{In a heat pulse experiment, the duration of the boundary heating is orders of magnitude shorter than all other characteristic time scales, as numerically verified in subsection~\ref{subsec:spacelab}.} were conducted in a thin-walled spherical tank made of copper and coated with a thin gold layer to minimize radiation heat losses. The heating wire was glued on the outer surface of this sphere along a single equator. Since copper is an excellent heat conductor, the wall temperature proved to be almost perfectly homogeneous during the experiments. The numerical solutions presented by Straub \etal applied the measured wall temperature directly as a boundary condition, thereby achieving excellent agreement with the experimental data \cite{straub1995dynamic}. However, when no experimentally measured boundary values are available, this procedure cannot be implemented. In such cases, a comprehensive model of the experimental setup can provide deeper insight. The heating wire warms up due to the electric current flowing through it, which heats the outer surface of the tank. The inner surface of the shell, heated by conduction through the wall, then transfers heat to the fluid layer in contact with it. A detailed analysis of the setup reveals that the heat capacity of the wire is negligible relative to that of the wall, but the latter is comparable to the heat capacity of the supercritical fluid filling. Furthermore, the thin-walled spherical copper shell can be treated as a lumped homogeneous body because its internal diffusion time scale is negligible even compared to the short duration of the boundary heating.

In what follows, we formulate the conjugate boundary condition \cite{perelman1961conjugated,dorfman2009conjugate} characterizing the energy storage capacity of the wall, which delays and dampens the thermal footprint of the heater. The post-acoustic temperature response in the fluid domain is then presented by applying this derived effective boundary condition. Finally, the obtained exact analytical solution is validated against the experimental data gathered during the Spacelab D-2 mission.

\subsection{Formulation of the conjugate boundary problem and the effective boundary condition}

\begin{figure}[!ht]
    \centering
    \includegraphics[width=.35\textwidth]{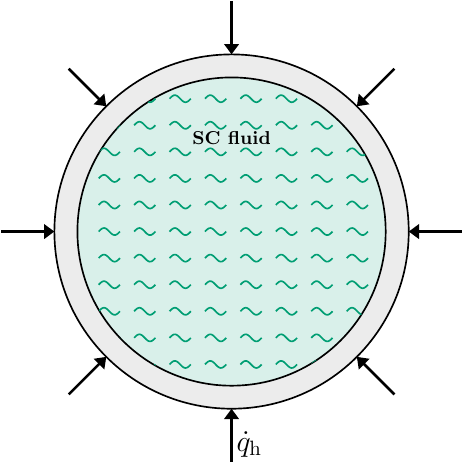}
    \caption{Schematic of the experimental setup used in the microgravity experiments during the Spacelab D2 mission.}
    \label{fig:D2-schematic}
\end{figure}

As previously noted, the heat capacity of the heater wire is neglected, therefore the heating power $ \dot{Q}_{\rm h} $ emitted by the wire directly heats the sphere shell, which is treated as a lumped capacitance body. The schematic representation of the experimental setup is presented in Fig.~\ref{fig:D2-schematic}. Accordingly, the energy balance of the shell can be formulated as
\begin{align}
    \label{eq:wall-e-bal}
    m_{\rm w} c_{\rm w} \dt{\Theta_{\rm w}} = \dot{Q}_{\rm h} - \dot{Q}_{\rm w} ,
\end{align}
where $ m_{\rm w} $ and $ c_{\rm w} $ represent the mass and the specific heat capacity of the wall, respectively, while $ \dot{Q}_{\rm w} $ denotes the heating power transferred from the wall to the fluid layer in contact with it.

Conjugate conditions at the solid--fluid interface provide continuity of the temperature and heat flux fields, expressed as
\begin{align}
    \label{eq:conj-bc-1}
    \Theta_{\rm w} ( t ) &= \Theta ( t , R ) , \\
    \label{eq:conj-bc-2}
    \lambda \left. \pdr{\Theta} \right|_{r = R} &= \dot{q}_{\rm w} ,
\end{align}
where $ R $ denotes the inner radius of the spherical shell. For a thin-walled shell with wall thickness $ \delta_{\rm w} $ satisfying $ \delta_{\rm w} \ll R $, the outer and inner surface areas can be approximated as equal. For simplicity, the inner radius is chosen as the only characteristic dimension of the system. Therefore, the mass of the shell can be approximated as $ m_{\rm w} \approx \qrho_{\rm w} 4 \pi R^2 \delta_{\rm w} $. Consequently, the solid--fluid heat flux can be derived from \re{eq:wall-e-bal} as
\begin{align}
    \label{eq:qw-mcdT}
    \dot{q}_{\rm w} = \dot{q}_{\rm h} - \qrho_{\rm w} c_{\rm w} \delta_{\rm w} \dt{\Theta_{\rm w}} ,
\end{align}
with the heater heat flux defined as $ \dot{q}_{\rm h} = \frac{\dot{Q}_{\rm h}}{4 \pi R^2} $ and the solid--fluid heat flux as $ \dot{q}_{\rm w} = \frac{\dot{Q}_{\rm w}}{4 \pi R^2} $. Note that both heat current densities are evaluated relative to the inner surface of the shell. Substituting \re{eq:qw-mcdT} into the conjugate boundary condition \re{eq:conj-bc-2} and applying the temperature continuity constraint \re{eq:conj-bc-1} yields the effective boundary condition
\begin{align}
    \label{eq:effective-bc}
    \left. \pdr{\Theta} \right|_{r = R} = \frac{\dot{q}_{\rm h}}{\lambda} - \frac{\qrho_{\rm w} c_{\rm w} \delta_{\rm w}}{\lambda} \left. \pdt{\Theta} \right|_{r = R} ,
\end{align}
acting directly on the fluid domain. This effective boundary condition can be recognized as a modified Neumann boundary condition where the ideal constant heat flux is dynamically dampened and delayed due to the heat capacity of the container. The coefficient of the time derivative of the temperature can be rearranged as
\begin{align}
    \label{eq:therm-vel}
    \frac{\qrho_{\rm w} c_{\rm w} \delta_{\rm w}}{\lambda} = \frac{\qrho_{\rm w} c_{\rm w}}{\qrho c_p} \frac{\delta_{\rm w}}{a} = \frac{\qrho_{\rm w} c_{\rm w}}{\qrho c_p} \sqrt{\frac{a_{\rm w}}{a}} \frac{\delta_{\rm w}}{\sqrt{a_{\rm w}}} \frac{1}{\sqrt{a}} = \sqrt{\frac{\lambda_{\rm w} \qrho_{\rm w} c_{\rm w}}{\lambda \qrho c_p}} \sqrt{\frac{\tau_{\rm w}}{a}} = \frac{\qB \sqrt{\tau_{\rm w}}}{\sqrt{a}} ,
\end{align} 
where $ a_{\rm w} = \frac{\lambda_{\rm w}}{\qrho_{\rm w} c_{\rm w}} $ is the thermal diffusivity of the solid wall and $ \tau_{\rm w} = \frac{\delta_{\rm w}^2}{a_{\rm w}} $ is the corresponding diffusion time scale of the shell, furthermore, $ \qB = \sqrt{\frac{\lambda_{\rm w} \qrho_{\rm w} c_{\rm w}}{\lambda \qrho c_p}} $ denotes the involved thermal effusivity ratio. This representation highlights that the thermal contact and the corresponding energy buffering at the solid--fluid interface can be characterized by an emerging effective characteristic thermal velocity $ \mathsf{v}_{\rm th} = \frac{\sqrt{a}}{\qB \sqrt{\tau_{\rm w}}} = \frac{\sqrt{a_{\rm w} a}}{\qB \delta_{\rm w}} $. Nevertheless, to simplify the subsequent calculations, we retain the original unified form of the coefficient presented in \re{eq:therm-vel}.

\subsection{Temperature response for the effective boundary condition in the post-acoustic approximation}

Substituting the temperature field decomposition \re{eq:bulk+diff} into \re{eq:effective-bc}, the effective boundary condition on the diffusive contribution
\begin{align}
    \label{eq:BC-tank}
    \left. \pdr{\Theta_{\rm d}} \right|_{r = R} = \frac{\dot{q}_{\rm h}}{\lambda} - \frac{\qB \sqrt{\tau_{\rm w}}}{\sqrt{a}} \left[ \dt{\Theta_{\rm b}} ( t ) + \left. \pdt{\Theta_{\rm d}} \right|_{r = R} \right] 
\end{align}
is obtained. By utilizing \re{eq:piston-ode}, this relation can be reformulated as
\begin{align}
    \label{eq:BC-eff}
    \left( 1 + \qB \sqrt{\frac{\tau_{\rm w}}{\varepsilon \tau_{\rm p}}} \right) \left. \pdr{\Theta_{\rm d}} \right|_{r = R}  =  \frac{\dot{q}_{\rm h}}{\lambda} - \frac{\qB \sqrt{\tau_{\rm w}}}{\sqrt{a}} \left. \pdt{\Theta_{\rm d}} \right|_{r = R} ,
\end{align}
with $ \varepsilon = \frac{\tau_{\rm p}}{\tau_{\rm d}} $. The localized Cartesian inward radial coordinate transformation via \re{eq:transf-r-2-x} and \re{eq:transf-r-2-zeta} applied to the diffusion equation \re{eq:piston-pde} and the effective boundary condition \re{eq:BC-eff} yields
\begin{align}
    \pdt{\vartheta} &= a \ppd{\vartheta}{\zeta} , \\
    \left. \pdt{\vartheta} \right|_{\zeta = 0} - \sqrt{a} \frac{1 + \qB \sqrt{\frac{\tau_{\rm w}}{\varepsilon \tau_{\rm p}}}}{\qB \sqrt{\tau_{\rm w}}} \left. \pdz{\vartheta} \right|_{\zeta = 0} - \frac{1 + \qB \sqrt{\frac{\tau_{\rm w}}{\varepsilon \tau_{\rm p}}}}{\qB \sqrt{\tau_{\rm d} \tau_{\rm w}}} \vartheta ( t , 0 ) &= \frac{\sqrt{a} R}{\lambda \qB \sqrt{\tau_{\rm w}}} \dot{q}_{\rm h} .
\end{align}
Via Laplace transformation, these equations become
\begin{align}
    \label{eq:Str-diff-L}
    s \htheta ( s , \zeta ) &= a \ppd{\htheta}{\zeta} , \\
    \label{eq:Str-BC-L}
    s \htheta ( s , 0 ) - \sqrt{a} \frac{1 + \qB \sqrt{\frac{\tau_{\rm w}}{\varepsilon \tau_{\rm p}}}}{\qB \sqrt{\tau_{\rm w}}} \left. \pdz{\htheta} \right|_{\zeta = 0} - \frac{1 + \qB \sqrt{\frac{\tau_{\rm w}}{\varepsilon \tau_{\rm p}}}}{\qB \sqrt{\tau_{\rm d} \tau_{\rm w}}} \htheta ( s , 0 ) &= \frac{1}{s} \frac{\sqrt{a} R}{\lambda \qB \sqrt{\tau_{\rm w}}} \dot{q}_{\rm h} .
\end{align}
The solution of \re{eq:Str-diff-L} is given in \re{eq:piston-pde-s-bc1-sol-r}, while the boundary condition \re{eq:Str-BC-L} determines the complex frequency-dependent amplitude   
\begin{align}
    \label{eq:Str-A}
    A (s) = \frac{\frac{\sqrt{a} R}{\lambda \qB \sqrt{\tau_{\rm w}}} \dot{q}_{\rm h}}{s \left( s + \frac{1 + \qB \sqrt{\frac{\tau_{\rm w}}{\varepsilon \tau_{\rm p}}}}{\qB \sqrt{\tau_{\rm w}}} \sqrt{s} - \frac{1 + \qB \sqrt{\frac{\tau_{\rm w}}{\varepsilon \tau_{\rm p}}}}{\qB \sqrt{\tau_{\rm d} \tau_{\rm w}}} \right)} = \frac{c_1}{\sqrt{s}} + \frac{c_2}{s} + \frac{c_3}{\sqrt{s} - \mu_+} + \frac{c_4}{\sqrt{s} - \mu_-}
\end{align}
expressed via partial fraction decomposition applying the roots
\begin{align}
    \mu_\pm = - \frac{1}{2} \frac{1 + \qB \sqrt{\frac{\tau_{\rm w}}{\varepsilon \tau_{\rm p}}}}{\qB \sqrt{\tau_{\rm w}}} \left( 1 \pm \sqrt{1 + \frac{4 \qB \sqrt{\tau_{\rm w}}}{\sqrt{\tau_{\rm d}} \left( 1 + \qB \sqrt{\frac{\tau_{\rm w}}{\varepsilon \tau_{\rm p}}}\right)}} \right)
\end{align}
of the denominator from \re{eq:Str-A} and the coefficients
\begin{align}
    c_2 &= \frac{\frac{\sqrt{a} R}{\lambda \qB \sqrt{\tau_{\rm w}}} \dot{q}_{\rm h}}{\mu_+ \mu_-} , &
    c_1 &= c_2 \sqrt{\tau_{\rm d}} , &
    c_3 &= \frac{1 - \sqrt{\tau_{\rm d}} \mu_-}{\mu_- - \mu_+} c_2 , &
    c_4 &= - \frac{1 - \sqrt{\tau_{\rm d}} \mu_+}{\mu_- - \mu_+} c_2 .
\end{align}
Note that $ c_1 + c_3 + c_4 = 0 $. Accordingly, the Laplace-transformed field reads as
\begin{align}
    \label{eq:Str-theta-L}
    \htheta ( s , \zeta ) = - R^2 \frac{\dot{q}_{\rm h}}{\lambda \left( 1 + \qB \sqrt{\frac{\tau_{\rm w}}{\varepsilon \tau_{\rm p}}} \right)} \exp\left( - \sqrt{\frac{s}{a}} \zeta \right) \left( \frac{\sqrt{\tau_{\rm d}}}{\sqrt{s}} + \frac{1}{s} + \frac{1 - \sqrt{\tau_{\rm d}} \mu_-}{\mu_- - \mu_+} \frac{1}{\sqrt{s} - \mu_+} - \frac{1 - \sqrt{\tau_{\rm d}} \mu_+}{\mu_- - \mu_+} \frac{1}{\sqrt{s} - \mu_-} \right) ,
\end{align}
and the diffusive contribution is obtained via inverse transformation and \re{eq:transf-r-2-x} as 
\begin{align}
    \nonumber
    \Theta_{\rm d} ( t , \zeta ) = &- R \frac{\dot{q}_{\rm h}}{\lambda \left( 1 + \qB  \sqrt{\frac{\tau_{\rm w}}{\varepsilon \tau_{\rm p}}} \right)} \frac{ R \exp\left( - \frac{\zeta^2}{4 a t} \right)}{R - \zeta} \Bigg[ \erfcx{\frac{\zeta}{\sqrt{4 a t}}} \\
    & + \mu_+ \frac{1 - \sqrt{\tau_{\rm d}} \mu_-}{\mu_- - \mu_+} \erfcx \left( \frac{\zeta}{\sqrt{4 a t}} - \mu_+ \sqrt{t} \right) - \mu_- \frac{1 - \sqrt{\tau_{\rm d}} \mu_+}{\mu_- - \mu_+} \erfcx \left( \frac{\zeta}{\sqrt{4 a t}} - \mu_- \sqrt{t} \right) \Bigg] .
\end{align}
The bulk temperature rise contribution is determined from \re{eq:piston-ode} via the transformed temperature field $ \vartheta ( t, \zeta ) $. Applying \re{eq:transf-r-2-x} and \re{eq:transf-r-2-zeta} yields
\begin{align}
    \dt{\Theta_{\rm b}} = - \frac{1}{\tau_{\rm p}} \left( \frac{1}{R} \vartheta ( t , 0 ) + \left. \pdz{\vartheta} \right|_{\zeta = 0} \right) ,
\end{align}
the Laplace transform of which is
\begin{align}
    \label{eq:Str-Tb-L}
    s \hTheta_{\rm b} (s) = - \frac{1}{\tau_{\rm p}} \left( \frac{1}{R} \htheta ( s , 0 ) + \left. \pdz{\htheta} \right|_{\zeta = 0} \right) .
\end{align}
Substituting \re{eq:Str-theta-L} into \re{eq:Str-Tb-L}, the bulk temperature rise in the Laplace domain becomes
\begin{align}
    \hTheta_{\rm b} (s) = \frac{1}{\tau_{\rm p} R} \left( \sqrt{\tau_{\rm d}} \frac{A (s)}{\sqrt{s}} - \frac{A (s)}{s} \right) .
\end{align}
Applying partial fraction decomposition for the expressions $ \frac{A ( s )}{\sqrt{s}} $ and $ \frac{A ( s )}{s} $ inside the bracket again yields
\begin{align}
    \frac{A ( s )}{\sqrt{s}} &= - \left( \frac{c_3}{\mu_+} + \frac{c_4}{\mu_-} \right) \frac{1}{\sqrt{s}} + \frac{c_1}{s} + \frac{c_2}{\sqrt{s}^3} + \frac{c_3}{\mu_+ \left( \sqrt{s} - \mu_+ \right)} + \frac{c_4}{\mu_- \left( \sqrt{s} - \mu_- \right)} , \\
    \frac{A ( s )}{s} &= - \left( \frac{c_3}{\mu_+^2} + \frac{c_4}{\mu_-^2} \right) \frac{1}{\sqrt{s}} - \left( \frac{c_3}{\mu_+} + \frac{c_4}{\mu_-} \right) \frac{1}{s} + \frac{c_1}{\sqrt{s}^3} + \frac{c_2}{s^2} + \frac{c_3}{\mu_+^2 \left( \sqrt{s} - \mu_+ \right)} + \frac{c_4}{\mu_-^2 \left( \sqrt{s} - \mu_- \right)} .
\end{align}
A lengthy but straightforward algebraic rearrangement via the inverse Laplace transformation results in the final bulk temperature rise
\begin{align}
    \nonumber
    \Theta_{\rm b} ( t ) &= R \frac{\dot{q}_{\rm h}}{\lambda \left( 1 + \qB \sqrt{\frac{\tau_{\rm w}}{\varepsilon \tau_{\rm p}}} \right)} \frac{1}{\tau_{\rm p}} \Bigg\{ \left( t - \tau_{\rm d} \right) - \frac{1}{\mu_+} \frac{1 - \sqrt{\tau_{\rm d}} \mu_-}{\mu_- - \mu_+} \left[ 1 - \left( 1- \sqrt{\tau_d} \mu_+ \right) \erfcx \left( - \mu_+ \sqrt{t} \right) \right] \\
    & \hskip 2ex + \frac{1}{\mu_-} \frac{1 - \sqrt{\tau_{\rm d}} \mu_+}{\mu_- - \mu_+} \left[ 1 - \left( 1- \sqrt{\tau_d} \mu_- \right) \erfcx \left( - \mu_- \sqrt{t} \right) \right] \Bigg\} .
\end{align}

\subsection{Comparison with the Spacelab D-2 microgravity measurements}
\label{subsec:spacelab}

Now, the exact analytical solution corresponding to the effective boundary condition \re{eq:effective-bc} is compared directly with the microgravity experimental data gathered during the Spacelab D-2 mission. The experimental configuration consists of a thin-walled copper sphere shell with an outer diameter of $ 20 $~mm and a wall thickness of $ \delta_{\rm w} = 0.4 $~mm, yielding an inner radius of $ R = 9.6 $~mm. The thermophysical properties of the copper are characterized by a mass density of $ \qrho_{\rm w} = 8950 \ {\rm \frac{kg}{m^3}} $, a specific heat capacity of $c_{\rm w} = 385 \ \rm{\frac{J}{kg \, K}} $, and a thermal conductivity of $ \lambda_{\rm w} = 400 \ {\rm \frac{W}{m \, K}} $. Based on these parameters, the internal thermal diffusion time scale of the solid shell is evaluated as $ \tau_{\rm w} = \frac{\delta_{\rm w}^2}{a_{\rm w}} = 0.0014 $~s. 

During the experiment, the electrical boundary heating lasts  $ \tau_{\rm h} = 10 $~s with a heating power of $ \dot{Q}_{\rm h} = 3.85 $~mW. A comparison of the time scales reveals that the diffusion time of the copper wall is orders of magnitude shorter than the duration of the heating pulse, namely $ \tau_{\rm w} \ll \tau_{\rm h} $. This stark multi-scale separation justifies our core physical assumption, confirming that the high-conductivity copper shell behaves as a lumped capacitance body with an almost perfectly homogeneous spatial temperature distribution throughout the thermal transient. To incorporate the finite duration constant heating into the continuous exact analytical solution, the linear superposition technique is applied in the explicit form
\begin{align}
    \Theta_{\rm D2} ( t , \zeta ) = \Theta ( t , \zeta ) - H ( t - \tau_{\rm h} ) \Theta ( t - \tau_{\rm h} , \zeta ) ,
\end{align}
where $ H $ denotes the Heaviside step function.

The experimental fluid examined under critical density conditions during the Spacelab D-2 mission is sulfur hexafluoride (SF$_6$). The measurements reported by Straub \etal \cite{straub1995dynamic} encompass two distinct supercritical fluid states evaluated at different temperature distances from the critical point, namely at $ T_0 = T_{\rm c} + 4.75 $~K and $ T_0 = T_{\rm c} + 0.1 $~K, where $ T_{\rm c} = 318.72 $~K is the critical temperature of SF$_6$. The necessary thermophysical properties for both cases were gathered from the NIST Chemistry WebBook \cite{lemmon2024thermophysical} database and are summarized in Tab.~\ref{tab:sf6-properties}. It is worth noting that for the second case, the material parameters were evaluated at a slightly adjusted distance of $ T_0 = T_{\rm c} + 0.15 $~K because the extreme proximity of the $ 0.1 $~K state causes several essential thermophysical properties to be undefined in the NIST database due to critical singularities. 
\begin{table}[!ht]
    \begin{center}
        \caption{Thermophysical properties of sulfur hexafluoride obtained from the NIST database \cite{lemmon2024thermophysical} for the two examined supercritical states, presented with five significant digits.}
        \label{tab:sf6-properties}
        \begin{tabular}{ c | c | c | c | c }
            $ \left( T_0 - T_{\rm c} \right) \ / \ {\rm K} $ & $ c_p \ / \ {\rm \frac{J}{kg \, K}} $ & $ c_v  \ / \ {\rm \frac{J}{kg \, K}} $ & $ \qas \ / \ {\rm \frac{m}{s}} $ & $ \lambda \ / \ {\rm \frac{W}{m \, K}}$ \\
            \hline
            0.15 & 404550 & 1053.5 & 61.841 & 0.12035 \\
            4.75 & 8978.8 & 829.62 & 74.836 & 0.044203  
        \end{tabular}
    \end{center}
\end{table}

Analyzing these two configuration states is highly instructive because, as summarized in Table~\ref{tab:sf6-calculated}, approaching the critical point triggers strong variations in the specific heat capacity ratio $ \gamma $ and the thermal diffusivity $ a $, which represent the two independent thermophysical parameters determining the realized process according to the post-acoustic approximation. As the system moves from the $T_{\rm c} + 4.75 $~K state closer to the critical point, the heat capacity ratio $ \gamma $ diverges significantly, while the thermal diffusivity approaches zero. In parallel, the isentropic speed of sound also approaches zero [cf.\ Table~\ref{tab:sf6-properties}]. Consequently, the acoustic time scale $ \tau_{\rm a} $ increases slightly, but this change remains minor for the given experimental geometry. In contrast, the thermal diffusion time scale $ \tau_{\rm d} $ changes drastically, swelling to a massive value due to the critical slowing down of the pure conduction process. Shifting between these limits, the intermediate piston time scale $ \tau_{\rm p} $ behaves exactly as anticipated by the post-acoustic reduction, positioning itself firmly between the fast acoustic and the slow diffusive regimes.

Comparing these scales to the actual experimental duration shows that the heating duration $ \tau_{\rm h} = 10 $~s is orders of magnitude shorter than both the piston and the diffusion time scales, mathematically confirming that the setup operates strictly as a heat pulse experiment. Simultaneously, because this pulse duration is several orders of magnitude longer than the acoustic time scale ($ \tau_{\rm h} \gg \tau_{\rm a} $), the post-acoustic approximation remains entirely valid throughout the process. Finally, the small parameter $ \varepsilon $ drops dramatically from $ 0.033934 $ down to $ 0.00087031 $ as the critical point is approached. This sharp decrease provides a justification that directly supports the validity of our multi-scale asymptotic expansions established in Sec.~\ref{sec:asy}.
\begin{table}[!ht]
    \begin{center}
        \caption{Calculated specific heat capacity ratio, characteristic time scales and the small parameter for the two examined supercritical states, presented with five significant digits.}
        \label{tab:sf6-calculated}
        \begin{tabular}{ c | c | c | c | c | c | c }
            $ \left( T_0 - T_{\rm c} \right) \ / \ {\rm K} $ & $ \gamma $ & $ a = \frac{\lambda}{\qrho c_p} \ / \ {\rm \frac{mm^2}{s}} $ & $ \tau_{\rm a} = \frac{R}{\qas} \ / \ {\rm s} $ & $ \tau_{\rm p} = \frac{R^2}{3 (\gamma -1 ) a} \ / \ {\rm s} $ & $ \tau_{\rm d} = \frac{R^2}{a} \ / \ {\rm s} $ & $ \varepsilon = \frac{\tau_{\rm p}}{\tau_{\rm d}}$ \\
            \hline
            0.15 & 384.01 & 0.00040076 & 0.00015524 & 200.13 & 229958 & 0.00087031 \\
            4.75 & 10.823 & 0.0066321 & 0.00012828 & 471.56 & 13896 & 0.033934
        \end{tabular}
    \end{center}
\end{table}

\begin{figure}[!ht]
    \centering
    \includegraphics[height=.37\textwidth]{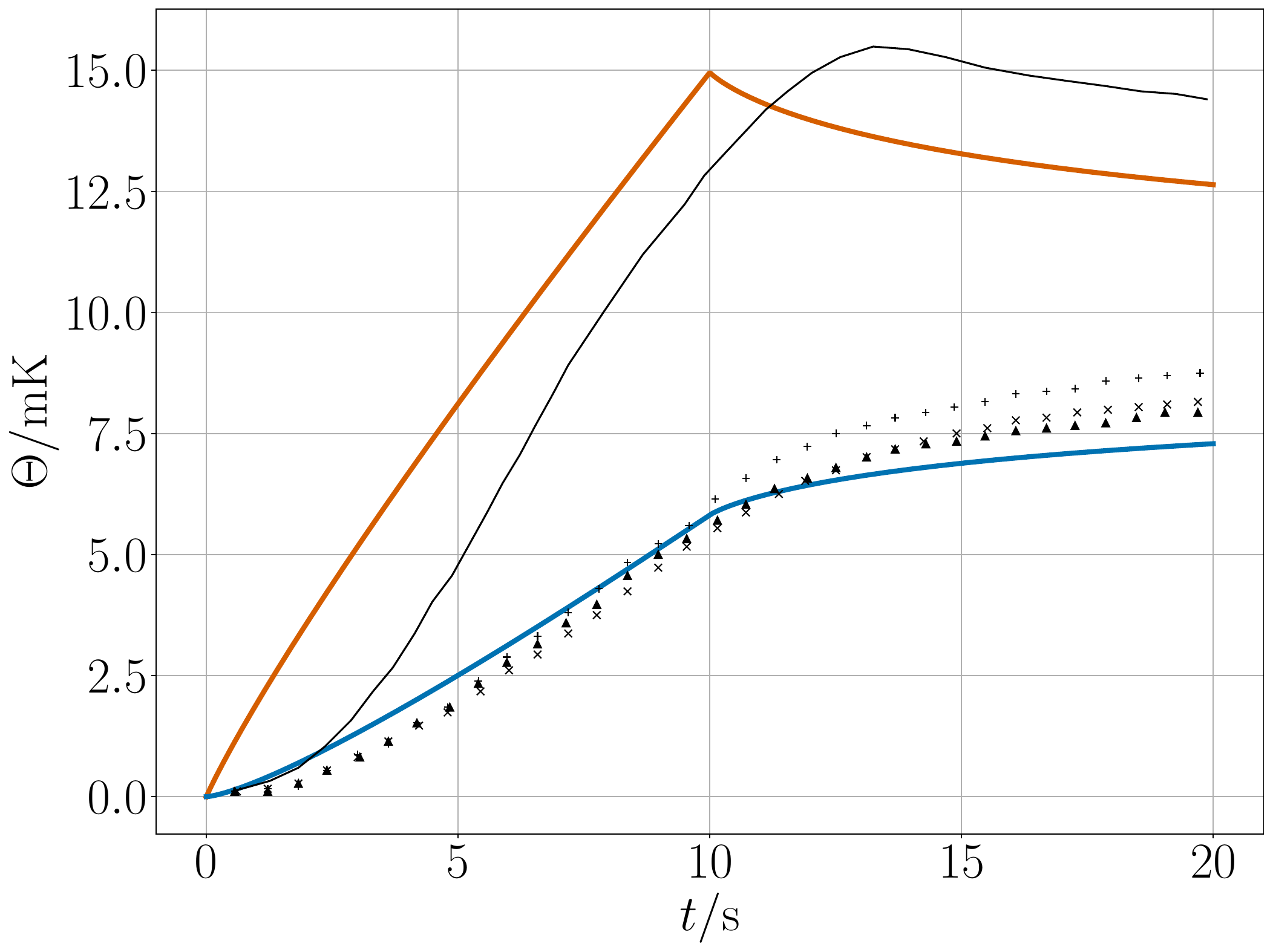}
    \hfill
    \includegraphics[height=.37\textwidth]{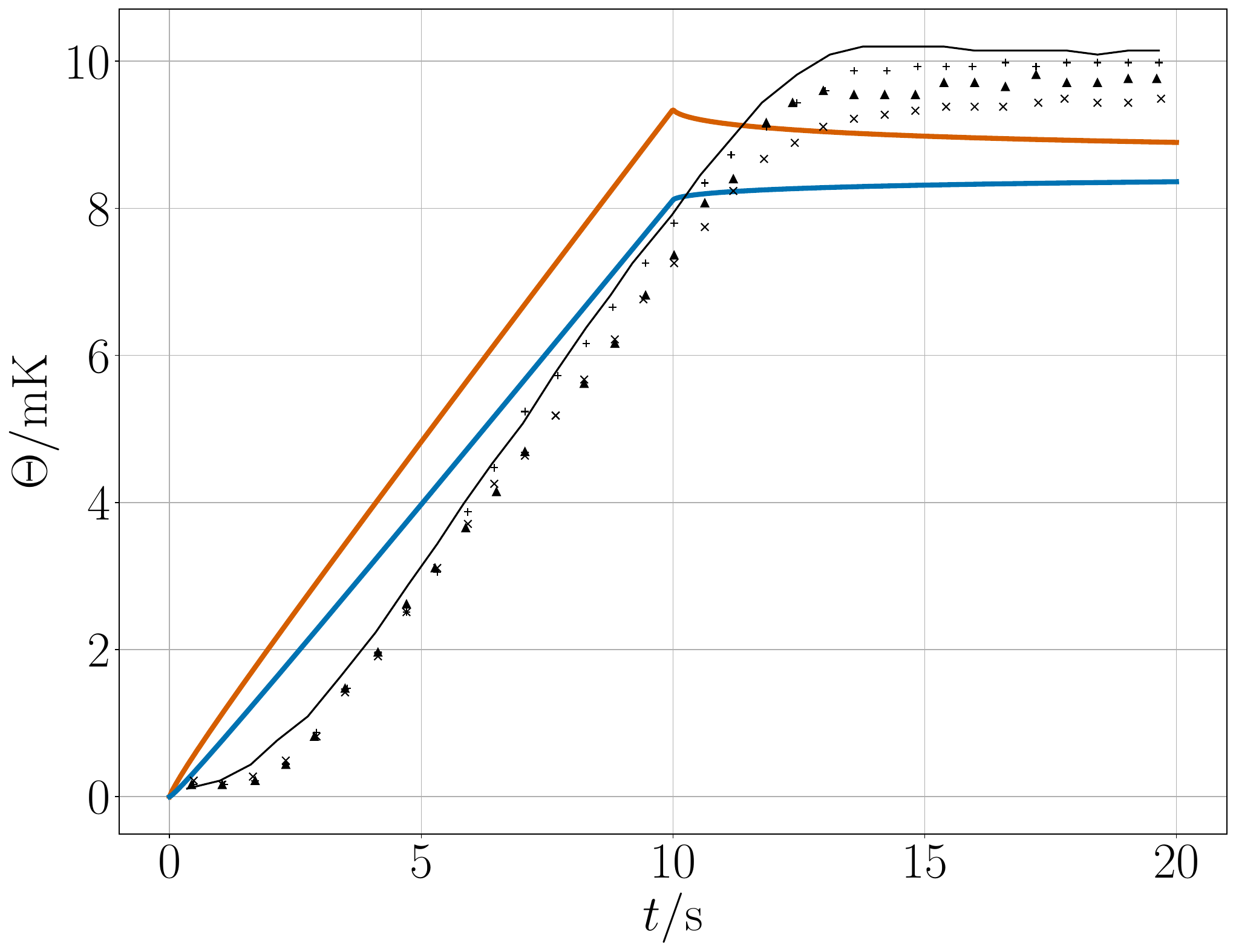} \\
    \includegraphics[width=0.8\textwidth]{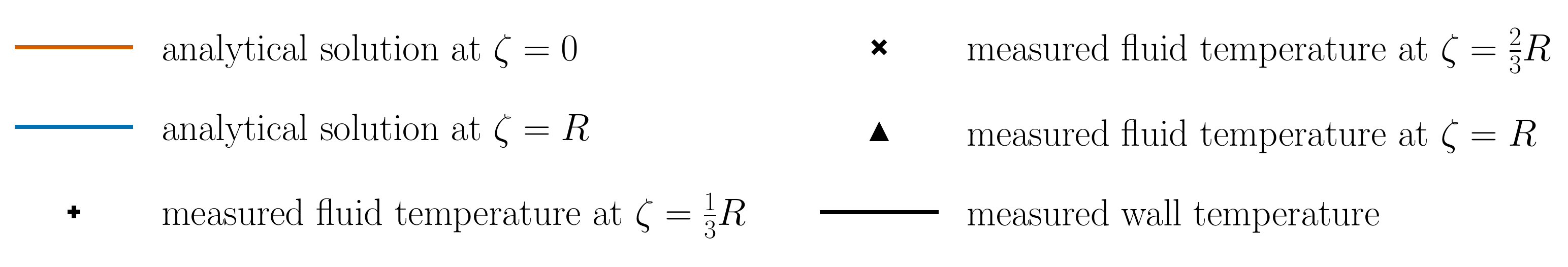}
    \caption{Comparison of the exact analytical temperature response with the Spacelab D-2 microgravity experimental data for supercritical sulfur hexafluoride. \emph{Left:} State further away from the critical point evaluated at $ T_0 = T_{\rm c} + 4.75 $~K. \emph{Right:} State near the critical point evaluated at $ T_0 = T_{\rm c} + 0.15 $~K. The discrete black symbols represent the measured fluid temperatures at different radial positions measured from the wall, where $ + $ markers denote $ \nicefrac{R}{3}$, crosses denote $ \nicefrac{2R}{3} $, and triangles represent the center of the sphere. The solid black line is the measured wall temperature profile reported by Straub \etal \cite{straub1995dynamic}. The solid red and blue lines represent the fluid temperature at the wall and in the center of the sphere, respectively, computed from the exact analytical solution for the effective boundary condition.}
    \label{fig:T-SLD2}
\end{figure}

The performance and predictive capability of the developed exact analytical solution for the conjugate boundary condition are evaluated against the Spacelab D-2 microgravity experimental data across both fluid states in Fig.~\ref{fig:T-SLD2}. A prominent physical feature observed simultaneously in both configurations is that the experimental fluid temperatures measured at the distinct spatial locations—namely at $ \nicefrac{R}{3} $, $ \nicefrac{2R}{3} $, and the sphere center—run almost perfectly together. This spatial uniformity provides a striking experimental confirmation of our post-acoustic framework, proving that the acoustically driven adiabatic compression acts as an instantaneous, spatially homogeneous volumetric heat source throughout the entire bulk domain. Our analytical solution captures this uniform bulk heating with high fidelity, tracking the overall slope and the maximum temperature amplitude of the bulk fluid points effectively for both the $ T_{\rm c} + 4.75 $~K and $ T_{\rm c} + 0.15 $~K states.

The specific physical mechanisms driving the temporal variations between these two distinct temperature differences from the critical point can be verified quantitatively by examining the calculated values of the thermal effusivity ratio $ \qB $ and the effective thermal velocity $ \mathsf{v}_{\rm th} $. Further away from the critical point ($ T_{\rm c} + 4.75 $~K), the effusivity ratio is high, $ \qB \approx 68.4 $, and the inverse thermal velocity reaches a substantial value of $ \nicefrac{1}{\mathsf{v}_{\rm th}} \approx 31181 \ {\rm \frac{s}{m}} $. This large inertia coefficient mathematically explains the prolonged, soft S-shaped time lag visible in the left plot of Fig.~\ref{fig:T-SLD2}, as the copper shell acts as a dominant thermal buffer that temporarily stores the incoming heat flux. In contrast, as the system approaches the critical point ($ T_{\rm c} + 0.15 $~K), the divergence of the fluid specific heat capacity causes the effusivity ratio to drop by a factor of ten to $ \qB \approx 6.18 $, while the inertia coefficient decreases to $ \nicefrac{1}{\mathsf{v}_{\rm th}} \approx 11452 \ {\rm \frac{s}{m}} $. Consequently, the characteristic thermal velocity across the interface nearly triples, rising from $ 0.032 \ {\rm \frac{mm}{s}} $ up to $ 0.087 \ {\rm \frac{mm}{s}} $. This massive reduction in the wall boundary thermal resistance allows the heat flux to penetrate the fluid domain much faster, forcing the temperature response in the right plot to rise sharply from the very beginning and visually demonstrating the thermodynamic origin of the critical speeding up of the system.

On the other hand, deviation is observable in both configurations regarding the temporal evolution of the wall temperature. While our exact analytical solution captures the general trend, it slightly precedes the experimental wall profile during the heating phase and does not reproduce the continued temperature rise observed after the shutdown of the heating at $ \tau_{\rm h} = 10 $~s. The physical origin of this thermal lag stems from the multilayer construction of the experimental apparatus. Although the heat capacity of the heating wire itself is negligible, the wire was attached to the copper shell via an adhesive glue layer, which reported a finite thermal conductivity of approximately $ 1 \ {\rm \frac{W}{m \, K}} $ \cite{straub1995dynamic}. This intermediate glue layer introduces an additional contact thermal resistance that delays the heat penetration into the copper shell. Incorporating this localized insulation mechanism would require extending our effective boundary condition to include second-order temporal derivatives to resolve the dual-layer solid transport. While such an extension is mathematically possible, the present first-order formulation already fulfills the primary objective of this study by providing a closed-form analytical description that captures the bulk fluid thermalization and the essential time scales of the coupled transient response with sufficient accuracy.

\section{Conclusions}

In this study, the post-acoustic approximation of the piston effect was analyzed. By applying the method of multiple scales, the fast acoustic transients and the slow thermal diffusion processes were systematically separated, providing a validation for the zero-flow thermalization model originally proposed by Boukari \etal \cite{boukari1990critical}. Furthermore, instead of using the classical integro-differential equation formulation, a boundary-coupled diffusion equation was derived.

The primary theoretical and practical contributions of this work are summarized through the following focal points:
\begin{itemize}
    \item Derive exact, closed-form analytical solutions to the post-acoustic approximation of the piston effect for both planar and spherical geometries under Dirichlet and Neumann boundary conditions.
    \item Analyze the thermal penetration depth and short-time asymptotic behavior for all four configurations investigated above. Our analysis proved that in the short-time regime and near the heated boundary, the spherical and planar solutions coincide to leading order, making the planar approximation adequate. A comparison of our exact solutions with that proposed by Straub \etal \cite{straub1995process} shows that, although the latter correctly captures the initial temporal evolution, its use of a spatially decoupled formulation leads to an overprediction of the thermal response at longer times.
    \item Incorporate a dynamic conjugate boundary condition to analyze the Spacelab D-2 microgravity experiments, where the exact analytical solution demonstrates good agreement with the experimental bulk temperature data.
\end{itemize}

The derived solutions are valid in the short-time regime, before the thermal disturbance reaches the opposite boundary or, in the spherical configuration, the center of the domain. At later times, the semi-infinite approximation breaks down and the finite-domain problem can be continued by a Fourier-type eigenfunction expansion, using the present short-time solution as a matched initial state. In that regime, however, the response becomes increasingly governed by global diffusion and finite-size equilibration rather than by a localized boundary-layer piston-effect mechanism alone. Therefore, the present solutions are intended to provide the closed-form short-time branch of the full transient response, where the piston-effect-driven boundary–bulk coupling is most clearly exposed.

The exact analytical solutions developed in this framework provide a powerful benchmark for validating complex numerical solvers, while offering direct physical insights, including a detailed parameter-dependence analysis of the piston effect. Furthermore, extending the presented asymptotic framework to higher order terms of the small parameter remains an interesting future direction, specifically to investigate the onset and behavior of the secondary fluid flows that are expected to emerge over the slow diffusive time scale.

More broadly, coupled transport problems (including thermoacoustics, thermoelasticity, non-Fourier heat conduction, and related fields) may involve several interacting physical mechanisms and widely separated characteristic time scales associated with diffusion, relaxation, and wave-like responses, among other phenomena. Their mathematical treatment requires a broad range of approaches, including analytical and asymptotic methods, similarity transformations, Laplace-domain solution techniques as well as carefully designed numerical schemes \cite{takacs2025piston,shahid2022numerical,askar2023photo,takacs2024thermodynamically,abouelregal2025modified,carr2025modelling}. Beyond providing solutions to the governing equations, these mathematical approaches facilitate the physical interpretation of the interactions among the underlying transport mechanisms and their characteristic time scales.

\section*{Acknowledgments}

The author wishes to express his gratitude to Tamás Fülöp for the insightful discussions throughout the development of this work, which significantly shaped the content of this study. Special thanks are also extended to Zsolt Szabó for his guideline in multi-scale analysis. The author is grateful to Róbert Kovács for his helpful comments and constructive remarks.

The research was supported by the Sustainable Development and Technologies National Programme of the Hungarian Academy of Sciences (FFT NP FTA) and by the Hungarian Scientific Research Fund under grant agreements NKKP STARTING24 149487 and NKKP Advanced 150038.

\appendix
\section{Inverse Laplace transforms of the applied expressions} \label{sec:app-A}

Let $ \hat{f} ( s ) = \mathcal{L} \{ f( t ) \}( s ) $ denote the Laplace transform of the function $ f ( t ) $, where its inverse is represented by $ f ( t ) = \mathcal{L}^{-1} \{ \hat{f} ( s ) \} ( t ) $. Table~\re{tab:inv-lap} summarizes the analytical Laplace transform pairs utilized throughout the derivations, sourced from the standard monograph of Carslaw and Jaeger \cite{carslaw1959conduction}, where $\mu$ and $k$ denote real constants.
\begin{table}[!ht]
    \begin{center}
        \caption{The inverse Laplace transforms of the appearing expressions.}
        \label{tab:inv-lap}
        \begin{tabular}{ c | c }
            $ \hat{f} ( s ) $ & $ f ( t ) $ \\
            \hline
            $ \frac{1}{\sqrt{s} + \mu} $ & $ \frac{1}{\sqrt{\pi t}} - \mu \erfcx \left( \mu \sqrt{t} \right)$ \\
            $ \frac{\exp \left( - k \sqrt{s} \right)}{\sqrt{s}} $ & $ \frac{\exp \left( - \frac{k^2}{4 t} \right)}{\sqrt{\pi t}} $ \\
            $ \frac{\exp \left( - k \sqrt{s} \right)}{s} $ & $ \erfc \frac{k}{2 \sqrt{t}} $ \\
            $ \frac{\exp \left( - k \sqrt{s} \right)}{\sqrt{s} + \mu} $ & $ \exp \left( - \frac{k^2}{4 t} \right) \left[ \frac{1}{\sqrt{\pi t}} - \mu \erfcx \left( \frac{k}{2 \sqrt{t}} + \mu \sqrt{t} \right) \right] $
        \end{tabular}
    \end{center}
\end{table}

\bibliographystyle{unsrt}
\bibliography{bibs_merged}

@book{bender1999advanced,
  title={Advanced mathematical methods for scientists and engineers: {A}symptotic methods and perturbation theory},
  author={Bender, Carl M and Orszag, Steven A},
  volume={1},
  year={1999},
  publisher={Springer}
}

@article{boukari1990critical,
  title={Critical speeding up in pure fluids},
  author={Boukari, Hac{\`{e}}ne and Shaumeyer, J. N. and Briggs, Matthew E. and Gammon, Robert W.},
  journal={Physical Review A},
  volume={41},
  number={4},
  pages={2260},
  year={1990},
  publisher={APS}
}

@article{carles1998effect,
  title={The effect of bulk viscosity on temperature relaxation near the critical point},
  author={Carl{\`{e}}s, Pierre},
  journal={Physics of Fluids},
  volume={10},
  number={9},
  pages={2164--2176},
  year={1998},
  publisher={AIP Publishing}
}

@article{carles2005two,
  title={Two typical time scales of the piston effect},
  author={Carl{\`e}s, Pierre and Dadzie, Kokou},
  journal={Physical Review E—Statistical, Nonlinear, and Soft Matter Physics},
  volume={71},
  number={6},
  pages={066310},
  year={2005},
  publisher={APS}
}

@article{carles2006thermoacoustic,
  title={Thermoacoustic waves near the liquid-vapor critical point},
  author={Carl{\`{e}}s, Pierre},
  journal={Physics of Fluids},
  volume={18},
  number={12},
  year={2006},
  publisher={AIP Publishing}
}

@article{carles2010brief,
    title = {A brief review of the thermophysical properties of supercritical fluids},
    journal = {The Journal of Supercritical Fluids},
    volume = {53},
    number = {1},
    pages = {2--11},
    year = {2010},
    author = {Carl{\`{e}}s, Pierre},
    issn = {0896-8446}
}

@book{carslaw1959conduction,
  author    = {Carslaw, Horatio Scott and Jaeger, John Conrad},
  year      = {1959},
  title     = {Conduction of Heat in Solids},
  edition   = {2},
  publisher = {Clarendon Press},
  location  = {Oxford}
}

@article{chen2022asymptotic,
  title={Asymptotic analysis of boundary thermal-wave process near the liquid--gas critical point},
  author={Chen, Lin and Zhang, Rui and Kanda, Yuki and Basu, Dipankar N and Komiya, Atsuki and Chen, Haisheng},
  journal={Physics of Fluids},
  volume={34},
  number={3},
  year={2022},
  publisher={AIP Publishing}
}

@article{daniarta2022thermodynamic,
  title={Thermodynamic efficiency of subcritical and transcritical power cycles utilizing selected {ACZ} working fluids},
  author={Daniarta, Sindu and Imre, Attila R and Kolasi{\'n}ski, Piotr},
  journal={Energy},
  volume={254},
  pages={124432},
  year={2022}
}

@book{degroot1962nonequilibrium,
  title={Non-equilibrium thermodynamics},
  author={de~Groot, Sybren Ruurds and Mazur, Peter},
  year={1962},
  address = {Amsterdam},
  publisher={Dover Publications}
}

@inproceedings{dobson2017supercritical,
  title = {{Supercritical Geothermal Systems - A Review of Past Studies and Ongoing Research Activities}},
  author = {Dobson, Patrick and Asanuma, Hiroshi and Huenges, Ernst and Poletto, Flavio and Reinsch, Thomas and Sanjuan, Bernard},
  booktitle = {{42nd Workshop on Geothermal Reservoir Engineering}},
  address = {Stanford, CA, United States},
  series = {Proceedings '' 42nd Workshop on Geothermal Reservoir Engineering''},
  year = {2017}
}

@book{dorfman2009conjugate,
  title={Conjugate problems in convective heat transfer},
  author={Dorfman, Abram S},
  year={2009},
  publisher={CRC Press}
}

@article{garrabos1998relaxation,
  title={Relaxation of a supercritical fluid after a heat pulse in the absence of gravity effects: Theory and experiments},
  author={Garrabos, Y. and Bonetti, M. and Beysens, D. and Perrot, F. and Fr{\"{o}}hlich, T. and Carl{\`{e}}s, P. and Zappoli, B.},
  journal={Physical Review E},
  volume={57},
  number={5},
  pages={5665},
  year={1998},
  publisher={APS}
}

@article{grigull1964prinzip,
  title={Das {P}rinzip von {L}e {C}hatelier und {B}raun},
  author={Grigull, Ulrich},
  journal={International Journal of Heat and Mass Transfer},
  volume={7},
  number={1},
  pages={23--31},
  year={1964},
  publisher={Elsevier}
}

@book{gyarmati1970nonequilibrium,
  title={Non-equilibrium Thermodynamics},
  subtitle="Field Theory and Variational Principles",
  author={Gyarmati, Istv{\'{a}}n},
  year={1970},
  publisher={Springer-Verlag},
  address="Berlin Heidelberg",
  isbn = "978-3-642-51069-4, 978-3-642-51067-0",
  issn = "0173-0274"
}

@article{hasan2012thermoacoustic,
  title = {Thermoacoustic transport in supercritical fluids at near-critical and near-pseudo-critical states},
  author = {Hasan, Nusair and Farouk, Bakhtier},
  journal = {The Journal of Supercritical Fluids},
  volume = {68},
  pages = {13--24},
  year = {2012},
  issn = {0896-8446}
}

@article{imre2019anomalous,
  year = 2019,
  month = {feb},
  publisher = {Periodica Polytechnica Budapest University of Technology and Economics},
  volume = {63},
  number = {2},
  pages = {276--285},
  author = {Imre, Attila R. and Groniewsky, Axel and Gy{\"{o}}rke, G{\'{a}}bor and Katona, Adrienn and Velmovszki, D{\'{a}}vid},
  title = {Anomalous Properties of Some Fluids -- with High Relevance in Energy Engineering -- in Their Pseudo-critical ({W}idom) Region},
  journal = {Periodica Polytechnica Chemical Engineering}
}

@incollection{lemmon2024thermophysical,
  address = {Gaithersburg MD, 20899, USA},
  author = {Lemmon, Eric W. and Bell, Ian H. and Huber, Marcia L. and McLinden, Mark O.}, 
  booktitle = {{NIST} {C}hemistry {W}eb{B}ook, {NIST} {S}tandard {R}eference {D}atabase {N}umber 69},
  editor = {Linstrom, P. J. and Mallard, W. G.},
  publisher = {National Institute of Standards and Technology},
  year = {2024},
  title = {Thermophysical Properties of Fluid Systems},
  note = {https://webbook.nist.gov/chemistry/fluid/ (retrieved June 24, 2025)}
}

@article{longmire2022onset,
  title={Onset of heat transfer deterioration caused by pseudo-boiling in {CO}$_2$ laminar boundary layers},
  author={Longmire, N. and Banuti, D. T.},
  journal={International Journal of Heat and Mass Transfer},
  volume={193},
  pages={122957},
  year={2022},
  publisher={Elsevier}
}

@book{matolcsi2004ordinary,
  author    = "Matolcsi, Tam{\'{a}}s",
  title     = "Ordinary thermodynamics",
  subtitle  = "Nonequilibrium homogeneous processes",
  year      = "2004",
  publisher = "Aka\-d{\'{e}}\-mi\-ai Kiad{\'{o}} (Publishing House of the Hungarian Academy of Sciences)",
  address   = "Budapest",
  isbn      = "9789630581707"
}

@article{onuki1990fast,
  title={Fast adiabatic equilibration in a single-component fluid near the liquid-vapor critical point},
  author={Onuki, Akira and Hao, Hong and Ferrell, Richard A},
  journal={Physical Review A},
  volume={41},
  number={4},
  pages={2256},
  year={1990},
  publisher={APS}
}

@article{onuki2007thermoacoustic,
  title={Thermoacoustic effects in supercritical fluids near the critical point: Resonance, piston effect, and acoustic emission and reflection},
  author={Onuki, Akira},
  journal={Physical Review E—Statistical, Nonlinear, and Soft Matter Physics},
  volume={76},
  number={6},
  pages={061126},
  year={2007},
  publisher={APS}
}

@article{perelman1961conjugated,
  title={On conjugated problems of heat transfer},
  author={Perelman, TL},
  journal={International Journal of Heat and Mass Transfer},
  volume={3},
  number={4},
  pages={293--303},
  year={1961},
  publisher={Elsevier}
}

@article{rahman2020design,
  title = {Design concepts of supercritical water-cooled reactor ({SCWR}) and nuclear marine vessel: A review},
  author = {Mohammad Mizanur Rahman and Ji Dongxu and Nusrat Jahan and Massimo Salvatores and Jiyun Zhao},
  journal = {Progress in Nuclear Energy},
  volume = {124},
  pages = {103320},
  year = {2020},
  issn = {0149-1970}
}

@article{reinsch2017utilizing,
  title={Utilizing supercritical geothermal systems: a review of past ventures and ongoing research activities},
  author={Reinsch, Thomas and Dobson, Patrick and Asanuma, Hiroshi and Huenges, Ernst and Poletto, Flavio and Sanjuan, Bernard},
  journal={Geothermal Energy},
  volume={5},
  number={1},
  pages={1--25},
  year={2017},
  publisher={Springer}
}

@article{shen2011thermoacoustic,
  title={Thermoacoustic waves along the critical isochore},
  author={Shen, B and Zhang, P},
  journal={Physical Review E—Statistical, Nonlinear, and Soft Matter Physics},
  volume={83},
  number={1},
  pages={011115},
  year={2011},
  publisher={APS}
}

@article{straub1995dynamic,
  title={Dynamic temperature propagation in a pure fluid near its critical point observed under microgravity during the {G}erman {S}pacelab {M}ission {D}-2},
  author={Straub, J. and Eicher, L. and Haupt, A.},
  journal={Physical Review E},
  volume={51},
  number={6},
  pages={5556},
  year={1995},
  publisher={APS}
}

@article{straub1995process,
  title={The process of heat and mass transport at the critical point of pure fluids},
  author={Straub, J and Eicher, L and Haupt, A},
  journal={International journal of thermophysics},
  volume={16},
  number={5},
  pages={1051--1058},
  year={1995},
  publisher={Springer}
}

@article{takacs2024leading,
  author   = {Tak{\'a}cs, Don{\'a}t M. and F{\"u}l{\"o}p, Tam{\'a}s and Imre, Attila R.},
  title    = {Leading elliptic relationship for supercritical fluids in the Widom region},
  journal  = "The Journal of Supercritical Fluids",
  year     = {2024},
  volume   = "208",
  pages    = {106216},
  numpages = {9}
}

@article{takacs2025piston,
  title={The piston effect in supercritical fluids investigated via a reversible--irreversible vector field splitting-based explicit time integration scheme},
  author={Tak{\'a}cs, Don{\'a}t M and F{\"u}l{\"o}p, Tam{\'a}s and Kov{\'a}cs, R{\'o}bert and Sz{\"u}cs, M{\'a}ty{\'a}s},
  journal={Physics of Fluids},
  volume={37},
  number={7},
  year={2025},
  publisher={AIP Publishing}
}

@article{theofanous2002boiling1,
  title={The boiling crisis phenomenon: {P}art {I}: nucleation and nucleate boiling heat transfer},
  author={Theofanous, T. G. and Tu, J. P. and Dinh, A. T. and Dinh, Truc-Nam},
  journal={Experimental Thermal and Fluid Science},
  volume={26},
  number={6--7},
  pages={775--792},
  year={2002},
  publisher={Elsevier}
}

@article{theofanous2002boiling2,
  title={The boiling crisis phenomenon: {P}art {II}: dryout dynamics and burnout},
  author={Theofanous, T. G. and Dinh, Truc-Nam and Tu, J. P. and Dinh, A. T.},
  journal={Experimental Thermal and Fluid Science},
  volume={26},
  number={6-7},
  pages={793--810},
  year={2002},
  publisher={Elsevier}
}

@ARTICLE{toth2025initial,
  author     = {Tóth, K. and Szücs, M.},
  title      = {Initial state dependence of thermo-mechanical coupling in heat conduction near the liquid-vapor critical point},
  journal    = {Journal of Computational and Applied Mechanics},
  volume     = {20},
  number     = {2},
  pages      = {71-92},
  year       = {2025}
}

@article{wu2022review,
  title = {A review of existing {S}uper{C}ritical {W}ater reactor concepts, safety analysis codes and safety characteristics},
  author = {Wu, Pan and Ren, Yanhao and Feng, Min and Shan, Jianqiang and Huang, Yanping and Yang, Wen},
  journal = {Progress in Nuclear Energy},
  volume = {153},
  pages = {104409},
  year = {2022},
  issn = {0149-1970}
}

@article{zappoli1990anomalous,
  title = {Anomalous heat transport by the piston effect in supercritical fluids under zero gravity},
  author = {Zappoli, B. and Bailly, D. and Garrabos, Y. and Le Neindre, B. and Guenoun, P. and Beysens, D.},
  journal = {Physical Review A},
  volume = {41},
  number = {4},
  pages = {2264--2267},
  year = {1990},
  month = {Feb},
  publisher = {American Physical Society}
}

@article{zappoli2003nearcritical,
  title = {Near-critical fluid hydrodynamics},
  author = {Zappoli, Bernard},
  journal = {Comptes Rendus M{\'{e}}canique},
  volume = {331},
  number = {10},
  pages = {713--726},
  year = {2003},
  issn = {1631-0721}
}

@book{zappoli2015heat,
  title={Heat transfers and related effects in supercritical fluids},
  author={Zappoli, Bernard and Beysens, Daniel and Garrabos, Yves},
  year={2015},
  publisher={Springer}
}

@article{zhang2023comparison,
  title={Comparison study of fluid thermal boundary-bulk behaviors in the close-to-critical region under different property trends},
  author={Zhang, Rui and Chen, Lin},
  journal={Physics of Fluids},
  volume={35},
  number={8},
  year={2023},
  publisher={AIP Publishing}
}

@article{shahid2022numerical,
  title={Numerical computation of magnetized bioconvection nanofluid flow with temperature-dependent viscosity and Arrhenius kinetic},
  author={Shahid, A and Huang, HL and Bhatti, Muhammad Mubashir and Marin, Marin},
  journal={Mathematics and Computers in Simulation},
  volume={200},
  pages={377--392},
  year={2022},
  publisher={Elsevier}
}

@article{askar2023photo,
  title={Photo-thermoelasticity heat transfer modeling with fractional differential actuators for stimulated nano-semiconductor media},
  author={Askar, Sameh and Abouelregal, Ahmed E and Marin, Marin and Foul, Abdelaziz},
  journal={Symmetry},
  volume={15},
  number={3},
  pages={656},
  year={2023},
  publisher={MDPI}
}

@article{abouelregal2025modified,
  title={A modified spatiotemporal nonlocal thermoelasticity theory with higher-order phase delays for a viscoelastic micropolar medium exposed to short-pulse laser excitation: AE Abouelregal et al.},
  author={Abouelregal, Ahmed E and Marin, Marin and {\"O}chsner, Andreas},
  journal={Continuum Mechanics and Thermodynamics},
  volume={37},
  number={1},
  pages={15},
  year={2025},
  publisher={Springer}
}

@article{carr2025modelling,
  title={Modelling and analysis of laser flash experiments using the Cattaneo heat equation},
  author={Carr, Elliot J},
  journal={Applied Mathematical Modelling},
  pages={116727},
  year={2025},
  publisher={Elsevier}
}

@article{takacs2024thermodynamically,
  title={Thermodynamically extended symplectic numerical simulation of viscoelastic, thermal expansion and heat conduction phenomena in solids},
  author={Tak{\'a}cs, Don{\'a}t M and Pozs{\'a}r, {\'A}ron and F{\"u}l{\"o}p, Tam{\'a}s},
  journal={Continuum Mechanics and Thermodynamics},
  volume={36},
  number={3},
  pages={525--538},
  year={2024},
  publisher={Springer}
}

\end{document}